\documentclass[aps,prd,twocolumn,showpacs,floats,floatfix,letterpaper,nofootinbib,superscriptaddress,]{revtex4-1}
 
\usepackage{hyperref}
\usepackage{amssymb,amsmath,latexsym,mathrsfs}
\usepackage{graphicx,subfigure}
\usepackage{epsfig}
\usepackage{varioref,xr-hyper}
\usepackage{color}
\usepackage{multirow}
\usepackage{array}
\hypersetup{
     colorlinks   = true,
     citecolor    = blue,
     linkcolor = magenta
     }
     
\begin{document}

\title{Studying the precision of ray tracing techniques with Szekeres models}
\author{S. M. Koksbang}
\email{koksbang@phys.au.dk}
\affiliation{Department of Physics and Astronomy, Aarhus University, 8000 Aarhus C, Denmark}
\author{S. Hannestad}
\affiliation{Department of Physics and Astronomy, Aarhus University, 8000 Aarhus C, Denmark}
\begin{abstract}
The simplest standard ray tracing scheme employing the Born and Limber approximations and neglecting lens-lens coupling is used for computing the convergence along individual rays in mock N-body data based on Szekeres swiss cheese and onion models. The results are compared with the exact convergence computed using the exact Szekeres metric combined with the Sachs formalism. A comparison is also made with an extension of the simple ray tracing scheme which includes the Doppler convergence. The exact convergence is reproduced very precisely as the sum of the gravitational and Doppler convergences along rays in Lemaitre-Tolman-Bondi models. This is not the case when the models are based on non-symmetric Szekeres models. For such models, there is a significant deviation between the exact and ray traced paths and hence also the corresponding convergences.
\footnote{\color{red} Please note that this arXiv-version does not correspond to the published version of the paper. An erratum has been published in PRD, giving corrections to errors in section \ref{sec:onion_LTB_results} and appendix \ref{app:curved}. In this arXiv-verion, the errors have been corrected and new results added and discussed in the appropriate sections (the abstract, section \ref{sec:onion_LTB_results}, the conclusion and appendix \ref{app:Doppler_curved}). In addition, a discussion of the relation between the results of section \ref{sec:onion_sz} and the new results of section \ref{sec:onion_LTB_results} has been included at the beginning of section \ref{sec:onion_sz}.
\newline\indent
The published version of the paper can be found as the previous arXiv-version. Please see  \cite{erratum} for the published erratum.}

\end{abstract}


\maketitle

\section{introduction}
With the possible near-future exception of observations based on gravitational waves, all astrophysical observations are based on light. This makes understanding light propagation a crucial element of both theoretical and observational cosmology. The light that is observed in astrophysical observations has propagated through vast regions of the Universe to reach us. During its propagation, the light feels the exact, local spacetime and not some "average" spacetime described by a Friedmann-Lemaitre-Robertson-Walker (FLRW) model. What effects the local inhomogeneities of spacetime have on light propagation and how important these effects are is still up for debate. For redshift-distance relations it has however been shown that averaging over {\em many} light rays will reduce observations approximately to what one would see if the light had simply traveled through the averaged universe model (see {\em e.g.} \cite{Syksy_av1, Syksy_av2, safety_in_numbers, nikolaos,walls,not_statistical, statistics, dallas_cheese, tardis}). For models with vanishing backreaction this implies that averaging over many light rays will yield results approximately corresponding to FLRW results. (For effects of non-vanishing backreaction on light propagation, see {\em e.g.} \cite{tardis}. See {\em e.g.} \cite{bc1, bc2, bc3, bc4} for introductions and reviews on cosmic backreaction.) The number of geodesics needed to obtain such results can be quite large though. In addition, some observables, such as reduced shear and CMB temperature fluctuations, are not suitable for the needed averaging. It is thus important to study the effects of inhomogeneities on light propagation so that the gained knowledge can be used when interpreting especially high precision observations.
\newline\newline
Much work has gone into using perturbation theory to study the effects of inhomogeneities on lensing and redshift-distance relations (see {\em e.g.} \cite{umeh,ido1, ido2,VT, CMB_distance, lensing_bias, weak_2} for some recent examples). Another approach is to use exact, inhomogeneous solutions to Einstein's equations. Among the most realistic, exact solutions to Einstein's equations which contain dynamical structures are the quasi-spherical Szekeres models \cite{Szekeres} including their spherically symmetric limit, the Lemaitre-Tolman-Bondi (LTB) models \cite{LTB1, LTB2, LTB3}. 
Light propagation through these models has been vastly studied, especially with the purpose of studying the effects of inhomogeneities on CMB and supernova observations (see {\em e.g.} \cite{statistics, dallas_cheese, tardis,early_CMB1, early_CMB2, bolejko_cmb, Alnes, first_ltb, bolejko_noelle, giant_void,krasinski_acc,bolejko_andersson, Bolejko_supernova, growth, first_dallas, Enqvist, onion, swiss_cheese_LTB, testing_WMAP, mimick, Havard, onion2, marra, light_cone, Szybka, Tetradis, Tetradis2, mapping_DL, Nambu, lensing_effects, dallas_first, early_anisotropy, distortion, single, drift, bog, Wessel1, Wessel2} for some examples)
\footnote{Other models breaking either the assumption of homogeneity or isotropy have also been used to study light propagation. Examples are \cite{Bianchi, Stephani} concerning light propagation in Bianchi and Stephani models respectively. In this work, the focus will be on the Szekeres models, as the Szekeres structures are more comparable to structures in typical N-body simulations.}.
With the exception of onion models (see {\em e.g.} \cite{onion}), these models are double or triple structure models (see {\em e.g.} \cite{tripple_structure}) and are thus not individually useful for realistic studies of light propagation over large distances. A possible method for  overcoming this issue is to combine few-structure models to build multiple-structure swiss cheese models (first introduced in \cite{cheese}). These models have a high degree of complexity and are thus very important and useful. However, they suffer from simplicities such as a lack of interaction between individual structures ("holes") and often the holes in the cheese are made up of only a few specific inhomogeneous models representing structure formation on a limited scale interval. Perhaps because swiss cheese models have these insufficiencies, universe models based on the output from Newtonian N-body simulations are by many considered the most realistic models of the real universe - despite their lack of relativistic effects such as cosmic backreaction \cite{bcNewt1, bcNewt2}. The standard cosmological setting for studying light propagation through an inhomogeneous universe is thus the output from Newtonian N-body simulations. The caveat is that Newtonian theory does not include a metric. This is problematic as the metric is crucial for studying relativistic light propagation and weak lensing. To circumvent the problem, it is standard to introduce the perturbed FLRW metric in the Newtonian gauge and use this metric to obtain approximation schemes for tracing rays through N-body simulations and computing the corresponding shear, convergence etc.. Work  ({\em e.g.} \cite{improve1, improve2, improve3}) has been done with the goal to improve the standard ray tracing schemes which are based on several approximations besides the use of the perturbed FLRW metric. However, it is difficult to obtain exact quantifications of the impact of these approximations and thus to determine to what extent the suggested improvements are actually worth their trouble.
\newline\indent
Exact, inhomogeneous solutions to Einstein's field equations, such as the Szekeres models, are excellent tools for studying the validity and precision of results obtained by using mainstream approximation schemes. Such models have for instance been used to study the possibility of a giant void being the reason for the CMB cold spot \cite{void1,void2,void3}, to study the precision of single structure thin lensing \cite{single}, and to illustrate the dominance of Doppler convergence at low redshifts \cite{bright_side}.
\newline\newline
In this work, quasi-spherical Szekeres swiss cheese and onion models will be used to study the precision of the simplest, standard ray tracing scheme used for obtaining the convergence along individual geodesics. The convergence is the most basic cosmologically interesting quantity obtainable from ray tracing and can be used to compute {\em e.g.} distance measures. The gravitational convergence can also be used to obtain the shear (see {\em e.g.} \cite{shear_from_conv}), so once the convergence has been obtained, many observables can be studied (see {\em e.g.}  \cite{Schneider1, Schneider2}  for some examples). Hence, obtaining the convergence along light rays is the typical goal of ray tracing.
\newline\indent
The precision of the convergence obtained from standard ray tracing schemes will here be studied by using the direct comparison approach introduced in \cite{dig_selv}. Using this approach, Szekeres models are reproduced as (mock) N-body data. Standard ray tracing approximation schemes can then be combined with the N-body data to obtain the convergence along light rays. The approximate results hereby obtained are compared to the exact results obtained by studying exact light propagation in the exact Szekeres spacetime. The comparisons made in the following show that in the special case of LTB single void/swiss cheese models, the ray tracing method reproduces exact ray paths and corresponding convergences very well. However, in the more general case where the single void models are non-symmetric, as well as in the case of onion models, the ray tracing scheme does not reproduce the exact results.

\section{Quasi-spherical Szekeres models}\label{sec:models}
This section gives a brief introduction to the quasi-spherical Szekeres models including a description of the particular models used in this work. See {\em e.g.} \cite{tripple_structure, wormhole} for more details on the Szekeres models and the procedure for constructing particular models.
\newline\newline
\begin{figure*}
\centering
\subfigure[m = 2]{
\includegraphics[scale = 0.4]{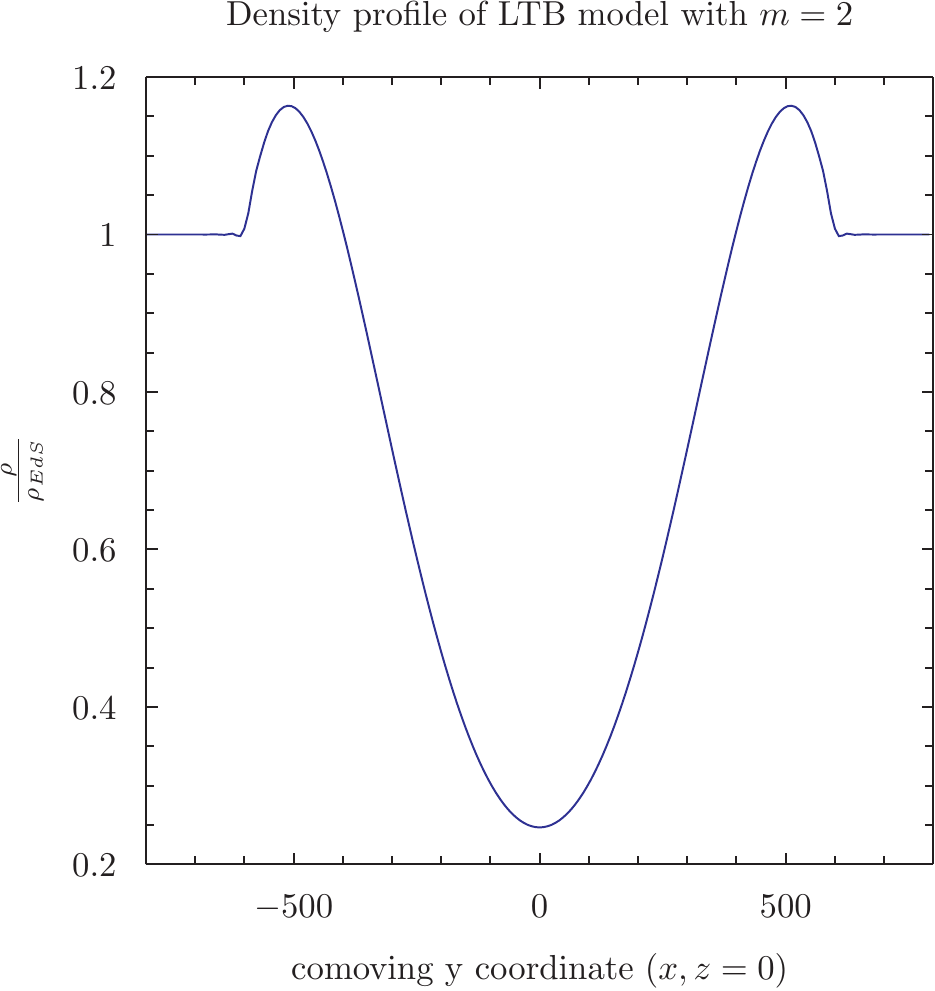}
}
\subfigure[m = 4]{
\includegraphics[scale = 0.4]{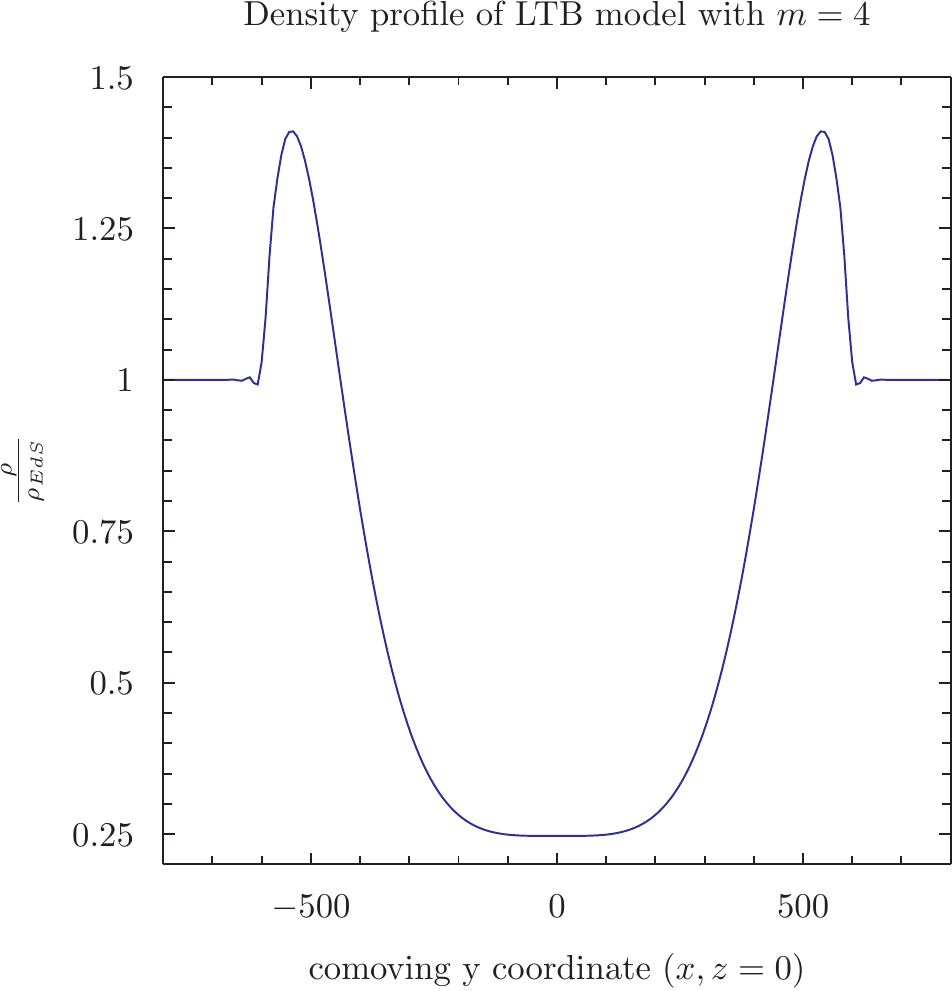}
}
\subfigure[m = 6]{
\includegraphics[scale = 0.4]{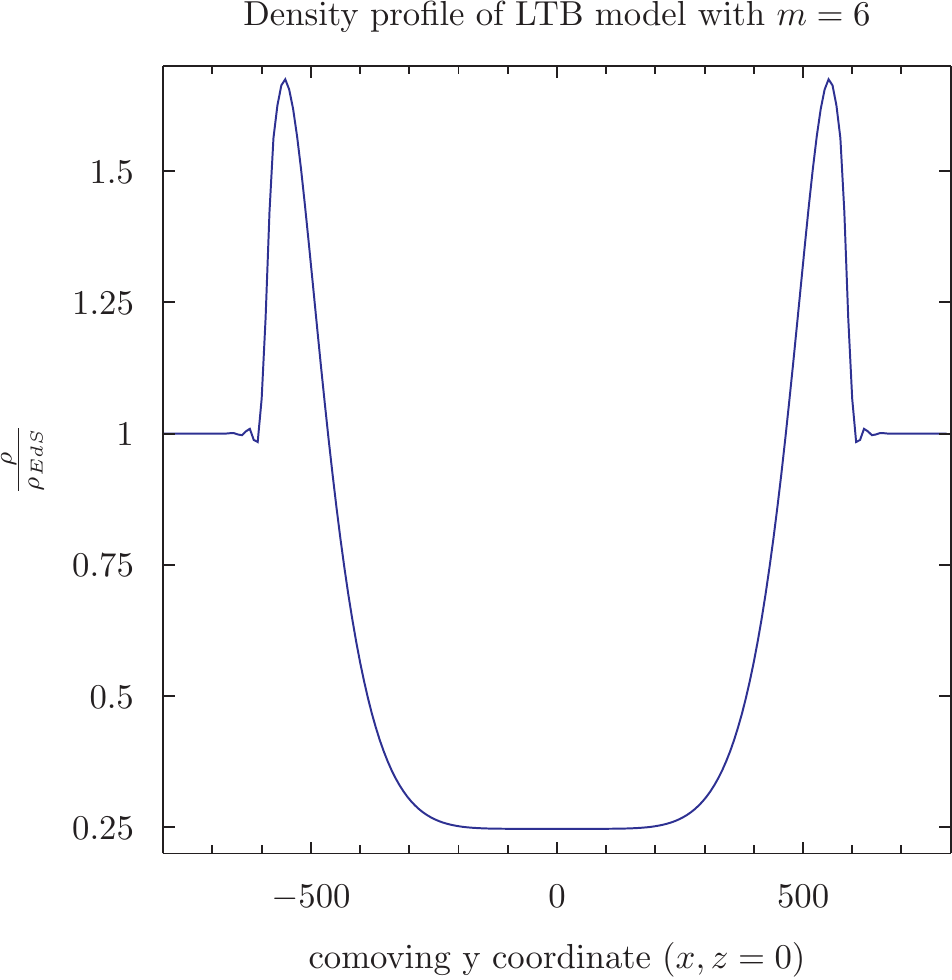}
}\par
\subfigure[m = 2]{
\includegraphics[scale = 0.4]{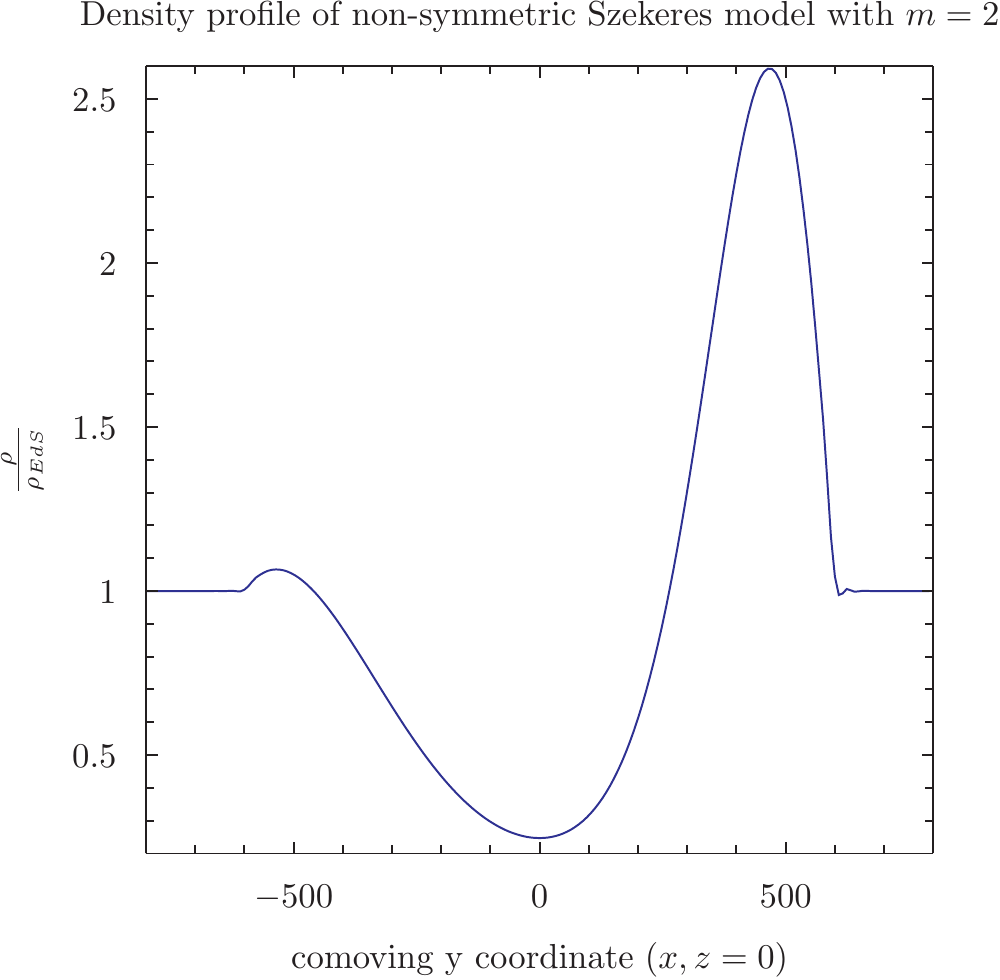}
}
\subfigure[m = 4]{
\includegraphics[scale = 0.4]{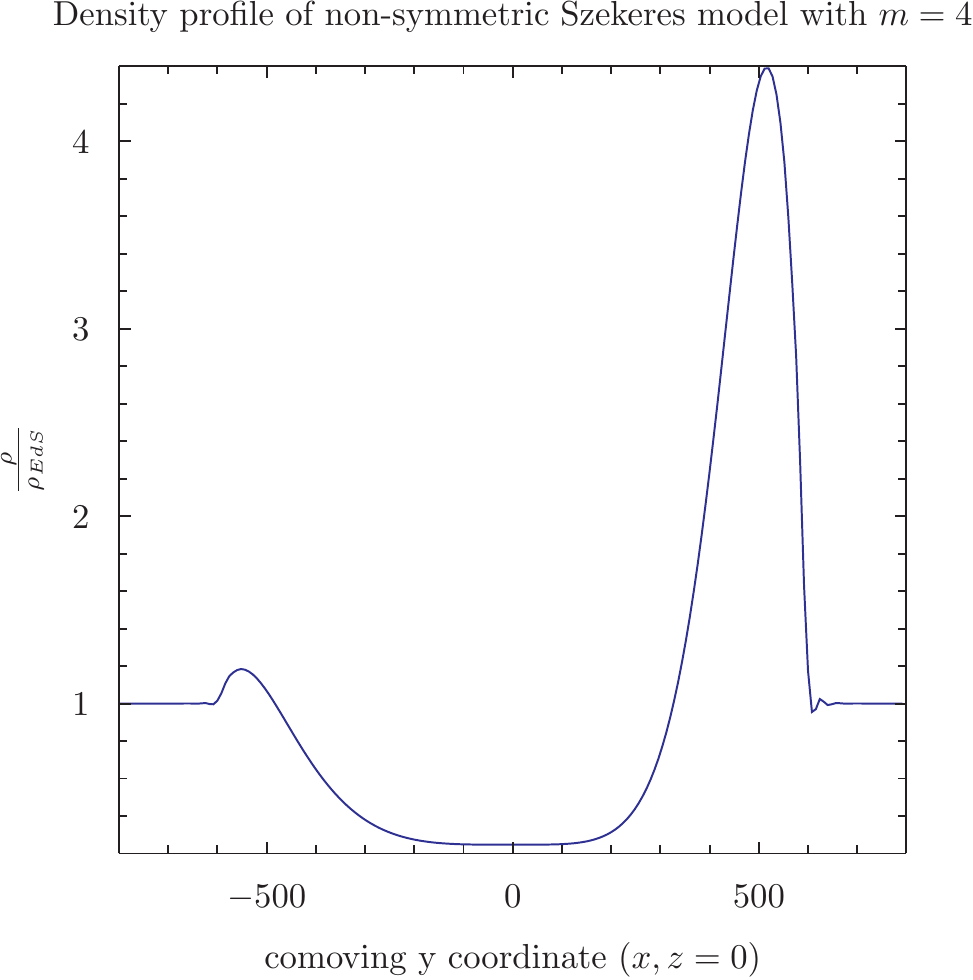}
}
\subfigure[m = 6]{
\includegraphics[scale = 0.4]{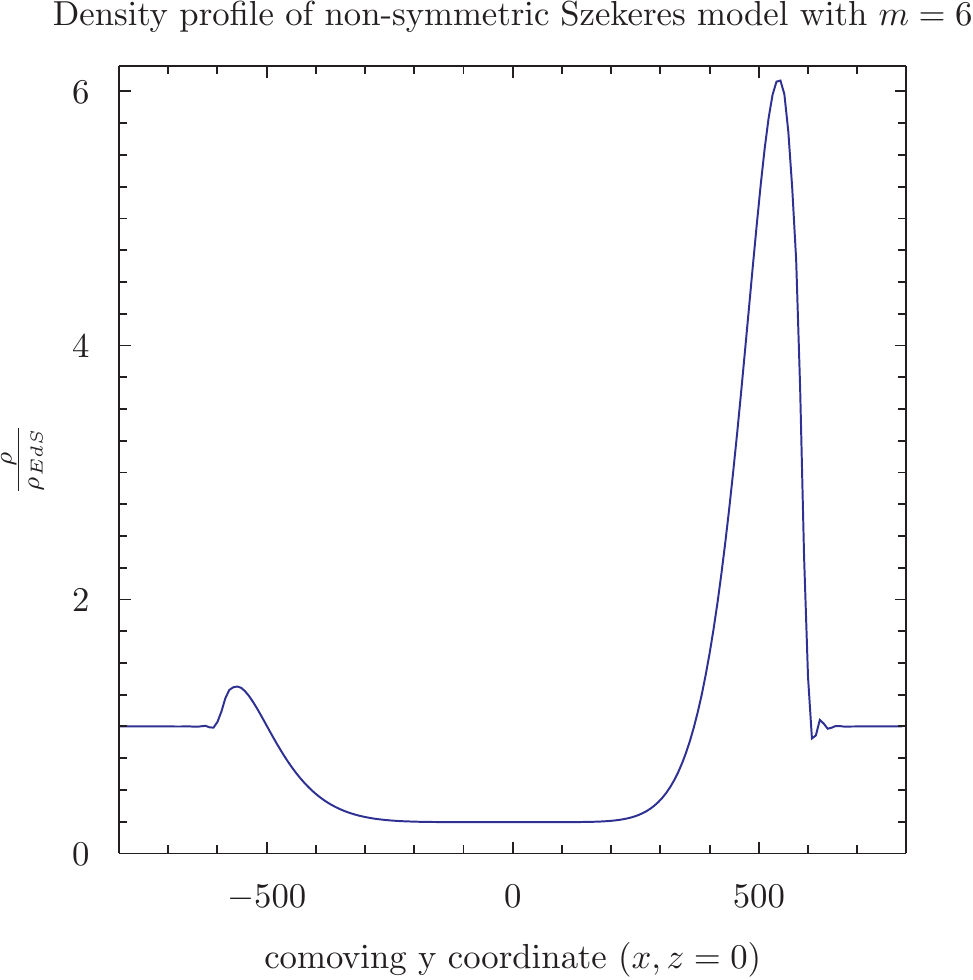}
}
\caption{Present time 1D density profiles of LTB and non-symmetric Szekeres models with $r_b = 60\text{Mpc}$ and varied $m$. For the non-symmetric models, the profiles are shown along the direction of highest degree of anisotropy. The comoving coordinates are normalized at present time in units of $0.1\text{Mpc}$.}
\label{fig:densities}
\end{figure*}

The quasi-spherical Szekeres models are a family of exact, inhomogeneous solutions to Einstein's equations. Using stereographic coordinates, the spacetime of the model can be described by the line element:
\begin{equation}\label{line_element}
\begin{split}
ds^{2} = -c^2dt^{2} +\frac{\left(\Phi_{,r}(t,r)-\Phi(t,r)\frac{E_{,r}(r,p,q)}{E(r,p,q)}\right)^2}{1-k(r)}dr^2 +\\
\frac{\Phi(t,r)^2}{E(r,p,q)^2}(dp^2+dq^2)
\end{split}
\end{equation}
Subscripted commas followed by one or more coordinates or indices indicate partial derivatives with respect to the implied coordinate(s).
\newline\indent
Inserting the line element into Einstein's equations for a dust universe with a cosmological constant yields the following two useful relations:
\begin{equation}\label{eq:friedmann}
\frac{1}{c^2}\Phi_{,t}^2 = \frac{2M}{\Phi}-k+\frac{1}{3c^2}\Lambda \Phi^2
\end{equation}
\begin{equation}\label{eq:density}
\rho = \frac{2M_{,r}-6M\frac{E_{,r}}{E}}{c^2\beta \Phi^2(\Phi_{,r}-\Phi\frac{E_{,r}}{E})},\, \, \, \, \, \, \, \, \beta = 8\pi G_N/c^4
\end{equation}
The function $M = M(r)$ appearing in these equations is a temporal integration constant depending on the radial coordinate and corresponds to the effective gravitational mass at comoving radial coordinate $r$.
\newline\indent
$E$ is given by $E = \frac{1}{2S}(p^2+q^2)-\frac{p P}{S} - \frac{qQ}{S}+\frac{P^2+Q^2+S^2}{2S}$, where $S, P$ and $Q$ are continuous but otherwise arbitrary functions of $r$. The quasi-spherical Szekeres models reduce to LTB models when $P,Q$ and $S$ are constant functions.
\newline\newline
The LTB model has two free functions and an extra degree of freedom from the covariance of the radial coordinate. A quasi-spherical Szekeres model has the three dipole functions as extra free functions and are specified by choosing dipole functions to remove the spherical symmetry of an underlying LTB model.
\newline\indent
The models studied here all have a vanishing cosmological constant. The coordinate covariance of $r$ is in the models removed by setting $\Phi(t_0,r) = r$, where $t_0$ is the present time according to an Einstein-de Sitter (EdS) model with reduced Hubble parameter $h :=H_0/(100\text{km}/\text{s}/\text{Mpc}) =0.7$. One of the free functions of the LTB models is chosen to be the time of the Big Bang which is set equal to zero, {\em i.e.} all the studied models have $\Phi(t = 0, r) = 0$. The last free function of the LTB models is chosen to be $k(r)$ which is specified below for the different models which will be studied.

\subsection{Single void/Swiss cheese models}\label{sec:cheese_models}
The swiss cheese models studied here are based on either LTB or non-symmetric Szekeres single void models. The final specification of the LTB single void models is made by defining $k(r)$ as:
\begin{equation}\label{eq:k}
k(r) = \left\{ \begin{array}{rl}
-r^2k_{\text{max}}\left(\left(\frac{r}{r_b} \right)^m -1 \right)^2  &\text{if} \,\, r<r_b \\
0 &\mbox{ otherwise}
\end{array} \right.
\end{equation}
The models studied here have $k_{\text{max}}=8\cdot10^{-10}$. The parameters $m$ and $r_b$ are varied in order to study the importance of void size and shape. The specific values of $m$ and $r_b$ used here are $2,4,6$ and $30\text{Mpc},60\text{Mpc},100\text{Mpc}$ respectively. Present time density profiles of the models with $m = 60\text{Mpc}$ are shown in the top row of figure \ref{fig:densities}.
To construct non-symmetric Szekeres models, the LTB models are modified by defining
\footnote{The constant term, $-n$, in $Q$ has no physical significance but is included in order to simplify a technical detail regarding ray tracing through the origin of single void models.}
$Q = n(e^{-\sigma r}-1)$ with $n = 2.3$ and $\sigma = \frac{0.002}{0.1\text{Mpc}}$. 1D present time density profiles of the non-symmetric models are shown in the bottom row of figure \ref{fig:densities}.

\subsection{Onion models}
\begin{figure*}[htb!]
\centering
\subfigure[model $O1_{\text{ltb}}$]{
\includegraphics[scale = 0.5]{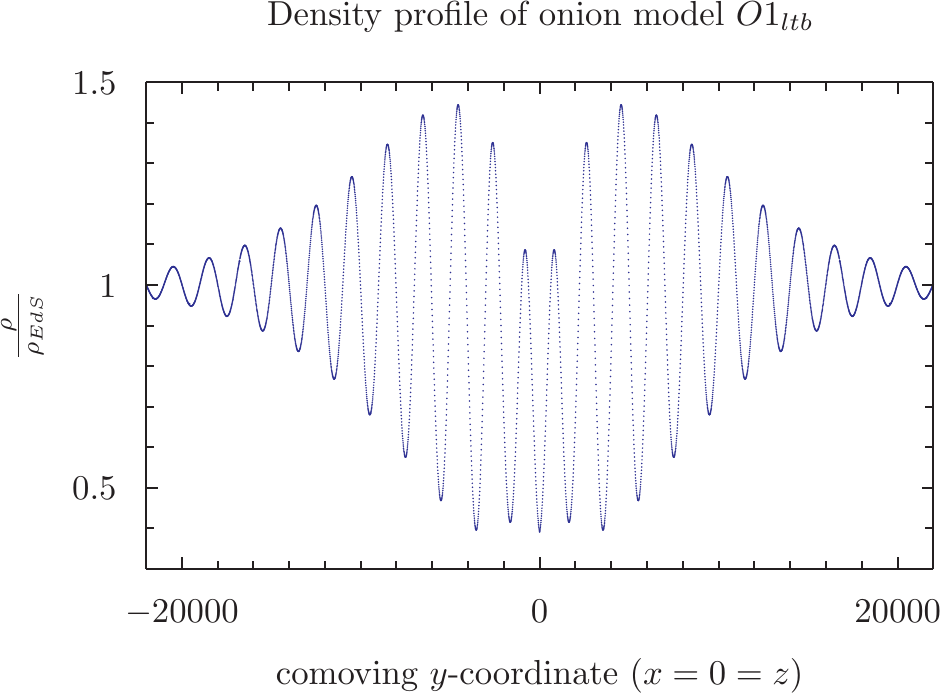}
}
\subfigure[model $O2_{\text{ltb}}$]{
\includegraphics[scale = 0.5]{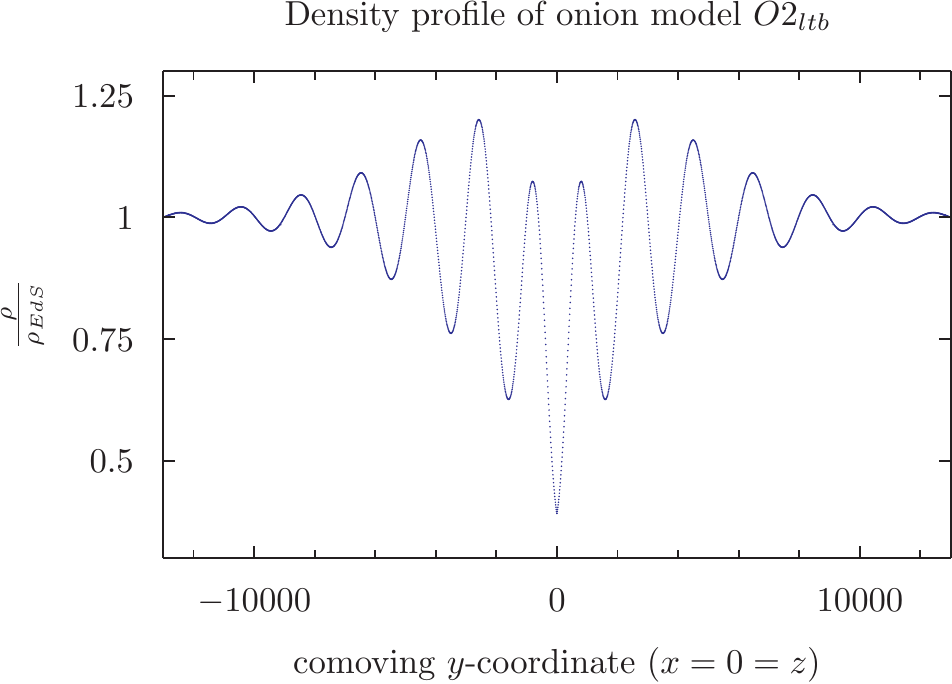}
}
\subfigure[model $O3_{\text{ltb}}$]{
\includegraphics[scale = 0.5]{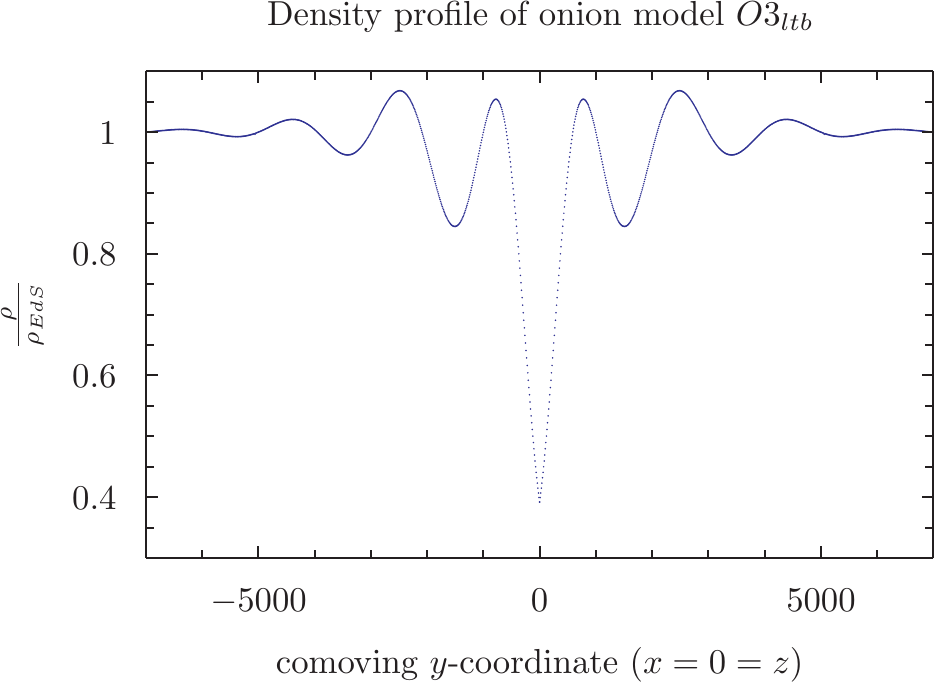}
}\par
\subfigure[model $O1_{\text{sz}}$]{
\includegraphics[scale = 0.5]{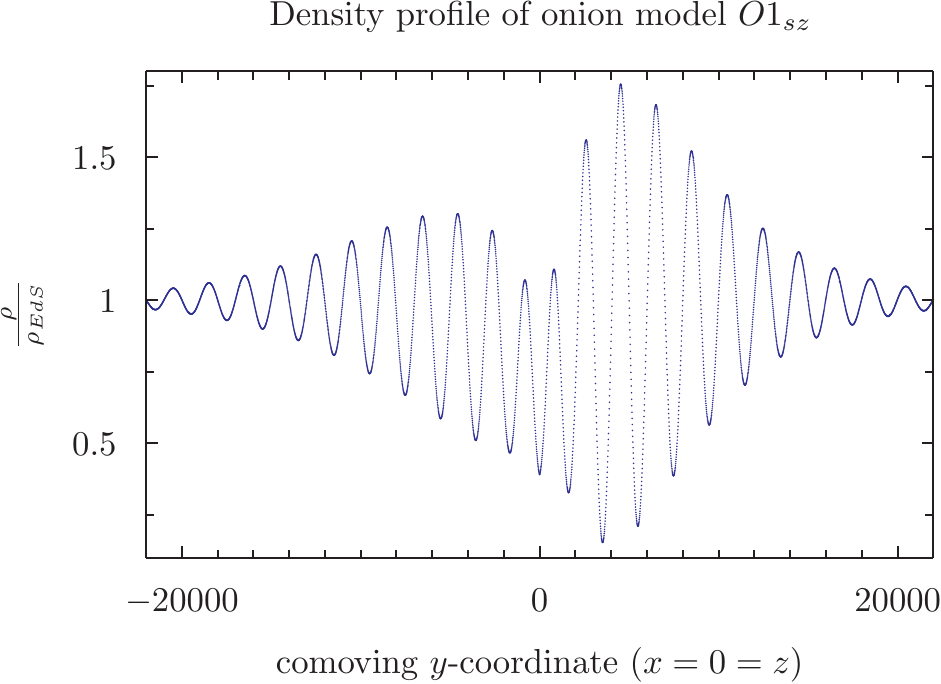}
}
\subfigure[model $O2_{\text{sz}}$]{
\includegraphics[scale = 0.5]{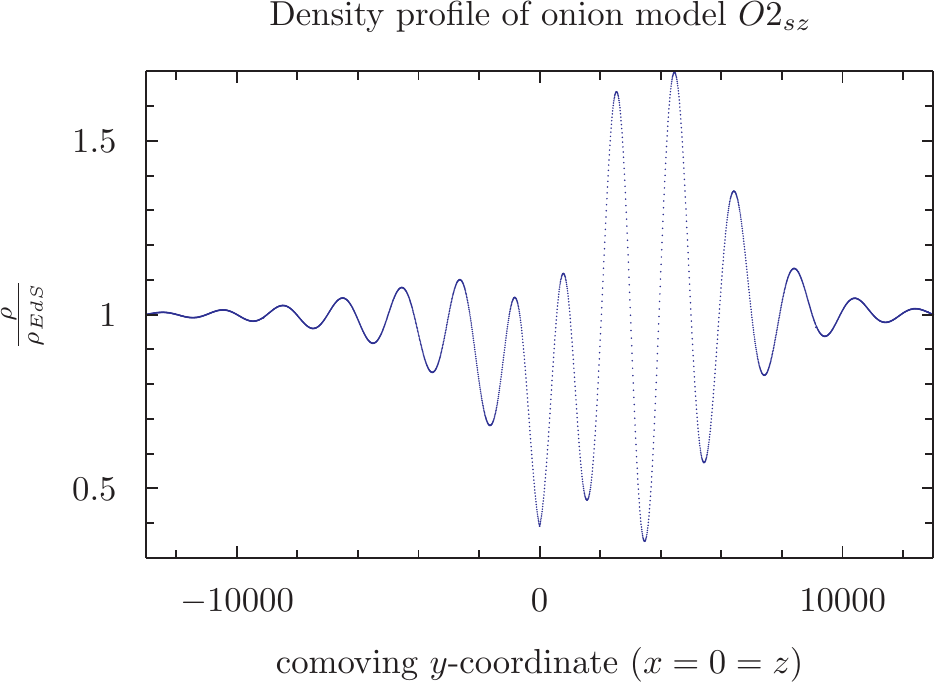}
}
\subfigure[model $O3_{\text{sz}}$]{
\includegraphics[scale = 0.5]{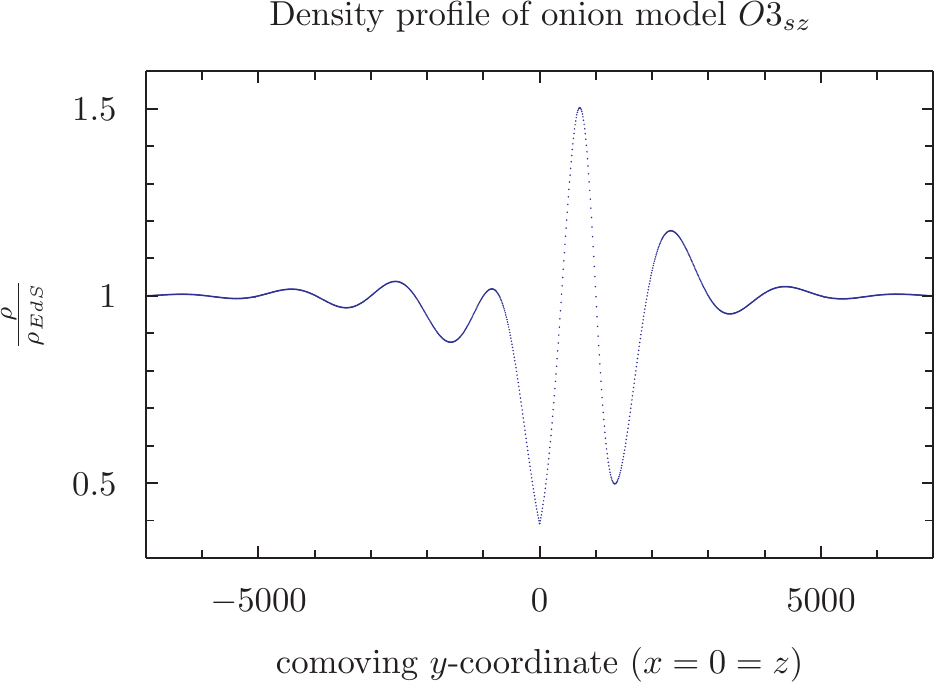}
}
\caption{Present time 1D density profiles of the onion models. The comoving coordinates are normalized at present time in units of $0.1\text{Mpc}$. The non-symmetric models' density profiles are shown along the axis of largest anisotropy.}
\label{fig:onion_densities}
\end{figure*}

Aside from swiss cheese models, also onion models (see {\em e.g.} \cite{onion}) will be used. Onion models have an oscillating density field in the radial direction. In the models studied here, the oscillations are damped and will be referred to as truncated onion models. The damping is achieved by combining $k(r)$ with an exponentially decaying function as shown in equation (\ref{eq:k_onion}). The radial coordinate where the density fluctuations become negligible will be referred to as the truncation radius.
\newline\newline
The final specification of the LTB onion models studied here is made by specifying $k(r)$ as:
\begin{equation}\label{eq:k_onion}
k(r) =
-k_{\text{max}} r^2 \cos^2\left( \frac{\pi r}{r_{b}} \right) e^{-\alpha \frac{r}{r_{b}}}
\end{equation}
All models will have $k_{\text{max}} = 6.2\cdot 10^{-10}$ and $r_{b} = 200\text{Mpc}$. The parameter $\alpha$ is varied. For $\alpha = 0$, the model is a true onion model with an oscillating density field for all values of the radial coordinate. In this particular true onion model, the density becomes negative and hence that model will not be used. Truncating the model with $\alpha \gtrsim 0.25$ removes the occurrence of negative densities. The density fields of the onion models used here are shown in figure \ref{fig:onion_densities} and specified in table \ref{table:onion}. The spherical symmetry is removed by using the dipole function $Q = ne^{-\sigma r}$, with the values of $n$ and $\sigma$ specified in table \ref{table:onion}.

\begin{table}[]
\centering
\begin{tabular}{c c c c}
\hline\hline
Model & $\alpha$ & $n$ & $\sigma$ \\
\hline
$O1_{\text{sz/ltb}}$ & 0.5 & 1 & 0.0002 \\
$O2_{\text{sz/ltb}}$ & 1 & 2 & 0.0002 \\
$O3_{\text{sz/ltb}}$ & 2 & 2.3 & 0.001 \\
\hline
\end{tabular}
\caption{Specification of non-symmetric Szekeres onion models and their underlying LTB models. The model names are subscripted with "sz" or "ltb" in order to distinguish between the non-symmetric and spherically symmetric versions of the models. $\alpha$ alone specifies the LTB models, while $n$ and $\sigma$ refer to the dipole function $Q$ used to remove their spherical symmetry.}
\label{table:onion}
\end{table}

\subsection{Swiss cheese {\em vs.} onion models} \label{sec:cheese_vs_onion}
As semi-realistic models of large scale structure formation, the swiss cheese and onion models each have their advantages and disadvantages. Swiss cheese models have density fields that look very realistic since the inhomogeneities are arranged in voids and surrounding clusters whereas the LTB onion models at best can be considered as models with a density distribution that has been averaged in angular directions - a feature which is only barely removed by adding anisotropy through the dipole functions. On the other hand, the individual single void structures of a swiss cheese model are restricted from interacting with each other, while the inhomogeneities of onion models have coupled dynamics.
\newline\indent
Single void and swiss cheese models have intrinsic backgrounds, namely the backgrounds they reduce to outside their structures. Onion models do not have this feature. This is inconvenient since an FLRW background must be defined in order to use standard ray tracing schemes. It is also an advantage of the onion models since the real universe does not have a background either. Newtonian N-body simulations have implicit backgrounds that determine the overall dynamics of the simulation box. It is not clear to what extent these backgrounds affect the structure formation of N-body simulations. However, structures in typical N-body simulations are not separated by extended patches where the density corresponds to the background value. Thus, the structures in N-body simulations interact freely with each other similar to in onion models and unlike in swiss-cheese models.

\section{Constructing mock N-body data}\label{subsec:mockNbody}
In order to compare ray tracing through N-body data with exact light propagation, the Szekeres/LTB models are reproduced as mock N-body data. A detailed description of constructing mock N-body data from Szekeres models was given in \cite{dig_selv} (see also \cite{Troels}) and only a brief review will be given below.
\newline\indent
In this section, subscripts will be used to distinguish between coordinates in the exact Szekeres spacetime and the underlying FLRW spacetime of the mock N-body data. In all other sections, these subscripts are suppressed as it should be clear from the context which spacetime the coordinates refer to. It is emphasized that the exact models and their N-body counterparts do {\em not} share coordinates and that mapping between the two types of spacetimes is vital for {\em e.g.} obtaining mock N-body velocity fields.
\newline\indent
Tildes are used to indicate fiducial spacetime points.
\newline\newline
Newtonian N-body simulations are based on FLRW backgrounds. In the case of the swiss cheese models studied here, the background model is the EdS model so this will be used as the underlying FLRW model for the mock N-body data. Considerations regarding the choice of background for the onion models are discussed in section \ref{sec:onion_results} and appendix \ref{app:curved}.
\newline\indent 
With the background model being the EdS model, the mock N-body data is constructed by using the following relations between the EdS and Szekeres coordinates:
\begin{equation}\label{stereo_map}
\begin{split}
\tilde t:=\tilde t_{\text{sz}} = \tilde t_{\text{eds}}\\
\tilde p:= \tilde p_{\text{sz}} = \tilde p_{\text{eds}}\\
\tilde q:=\tilde q_{\text{sz}} = \tilde q_{\text{eds}}\\
\int_{0}^{\tilde r_{\text{sz}}}dr_{\text{sz}}\sqrt{g_{rr,\text{sz}}} = \int_{0}^{\tilde r_{\text{eds}}}dr_{\text{eds}}\sqrt{g_{rr,\text{eds}}}
\end{split}
\end{equation}
These four equations make up a map between the Szekeres and EdS/N-body spacetimes. $g_{rr}$ is the $rr$-component of the metric tensor in stereographic coordinates. The subscripts "eds" and "sz" are used to denote EdS and Szekeres coordinates respectively. The four equations making up the map are coupled since $g_{rr,\text{sz}}$ depends on all four coordinates.
\newline\indent
Note that $p$ and $q$ are related to spherical coordinates differently in the Szekeres and FLRW spacetimes (see {\em e.g.} \cite{dig_selv}). Hence, even though $(\tilde p_{\text{sz}},\tilde q_{\text{sz}}) = (\tilde p_{\text{eds}}, \tilde q_{\text{eds}})$, it is generally not true that $(\tilde \theta_{\text{sz}}, \tilde\phi_{\text{sz}}) = (\tilde\theta_{\text{eds}}, \tilde\phi_{\text{eds}})$.
\newline\newline
Equation (\ref{stereo_map}) is used to map the Szekeres density field onto the EdS N-body background. Aside from the density field, the corresponding potential, $\psi$, will also be needed as it enters into the perturbed FLRW metric in the Newtonian gauge which is the basis for the ray tracing scheme
\footnote{In standard perturbation theory and thus standard ray tracing schemes, the coordinates of the perturbed FLRW metric are simply set equal to those of the FLRW background. Thus no map is needed for going between these two spacetimes/coordinate sets.}. This can be obtained from the mapped density field through the Poisson equation $\nabla^2\psi = \frac{4\pi G a^2}{c^2}\delta\rho$ with the overdensity, $\delta\rho$, defined as the difference between the inhomogeneous "N-body" density and the background density, {\em i.e.} $\delta\rho := \rho_{\text{Nbody}} - \rho_{\text{bg}}$.
\newline\indent
The velocity field is needed in order to compute the observable redshift and the Doppler convergence (see subsection \ref{sec:Dopler} for a definition of the latter). The angular velocities are negligible and only the radial velocity, $v_r$, needs to be computed. The radial velocity can {\em e.g.} be computed from the following equation obtained by taking the time derivative of the $r$-equation in the map given above:
\begin{equation}\label{eq:velocity_field}
\begin{split}
\frac{d}{dt}dp_r(\tilde t,\tilde r_{\text{sz}},\tilde p,\tilde q) = \frac{d}{dt}(a(\tilde t)\tilde r_{\text{eds}}) = a_{,t}(\tilde t)\tilde r_{\text{eds}} + a(\tilde t)v_{r}(\tilde r_{\text{eds}})\\
\end{split}
\end{equation}
$dp_r$ denotes the proper radial distance from the origin and the left-hand side of this equation is to be evaluated in the Szekeres spacetime. $v_r$ on the right-hand side is the radial velocity field of the mock N-body data at the EdS spacetime position $(\tilde t,\tilde r_{\text{eds}}, \tilde p, \tilde q)$ corresponding to the Szekeres spacetime position $(\tilde t,\tilde r_{\text{sz}},\tilde p,\tilde q)$.
\newline\newline
Using swiss cheese models implies involving one more coordinate system than the two discussed above. In particular, the standard ray tracing scheme is partially based on the global coordinate system of the swiss cheese model. On the other hand, the individual structures in a swiss cheese model are described using the local coordinate systems of the individual Szekeres single void model spacetimes making up the swiss cheese model. Note that the mapping described above, in principle, pertains to global swiss cheese coordinate systems. The difference between using global and local coordinate systems for the mapping is however negligible.
\newline\indent
In the following, global coordinates will be indicated by a subscripted "$g$". Coordinates without such a subscript are local coordinates. For onion models, local and global coordinates are the same.

\section{Light propagation and weak lensing}
Light rays move along null geodesics which can be determined by solving the geodesic equations obtained from the metric. In order to study weak lensing, the deformation of images along null-geodesics, also {\em bundles} of light rays must be studied. This can be done by solving the screen space projected geodesic deviation equation, where screen space is the two-dimensional Euclidean space orthogonal to the ray direction as seen by the observer. The Sachs formalism (based on \cite{sachs}) gives a convenient form of the formal solution through the Weyl and Ricci tensors. This formalism is widely used and detailed descriptions can be found in {\em e.g.} \cite{arbitrary_spacetime,dallas_cheese}
\footnote{In connection with \cite{arbitrary_spacetime}, see also \cite{correction}.}
. Here, only a brief description of the Sachs formalism and the resulting ray tracing schemes are given.

\subsection{The Sachs formalism}\label{sec:Sachs}
The screen space deviation vector, $\xi^{a}$, describes the deviation between geodesics in a bundle of light rays. The Latin indices $a,b,c \in \left\lbrace 1,2 \right\rbrace  $ denote screen space components while Greek indices $\alpha,\beta,\mu,\nu,... \in \left\lbrace 0,1,2,3 \right\rbrace $ will be used to denote spacetime components. The deviation vector is related to its initial derivative through a Jacobi relation {\em i.e.} $\xi(\lambda) = D(\lambda)\dot{\xi}_0$, where dots denote differentiation with respect to the affine parameter, a subscript "$0$" is used to denote present time, and the screen space matrix $D$ represents the Jacobi map. $D$ is the central quantity for describing deformation of the ray bundle image; $\sqrt{|\det(D)|}$ gives the relation between the areas described by $\dot{\xi^a_0}$ and $\xi^a(\lambda)$. Setting the initial conditions such that $\lambda$ is initially the local proper distance from the observer, $\dot{\xi^a_0}$ becomes the solid angle element and $\sqrt{|\det(D)|}$ becomes the angular diameter distance $ D_A$ along the fiducial ray of the ray bundle (see also {\em e.g.} \cite{arbitrary_spacetime}). Further more, if the cosmological model of interest can be considered as having an FLRW background, $D$ can be related to the distortion/magnification matrix $A$ by the relation $D = D_{A,bg} A$ \cite{dallas_first,mis_interp}, where $D_{A,\text{bg}}$ is the background FLRW angular diameter distance. Neglecting rotation (see {\em e.g.} \cite{dallas_cheese}), the entries of the distortion matrix can be decomposed into a convergence $\kappa$ and two shear components,  $\gamma_1$ and $\gamma_2$:
\begin{equation}
A = 
\begin{pmatrix}
1-\kappa -\gamma_1 & \gamma_2\\ \gamma_2& 1-\kappa + \gamma_1
\end{pmatrix}
\end{equation}
The convergence can thus be computed from the diagonal component of $D$ as follows:
\begin{equation}\label{eq:kappa}
\kappa = 1-\frac{D_{11} + D_{22}}{2D_{A,\text{bg}}}
\end{equation}
The FLRW background models used in this work contain non-relativistic matter and a possible non-vanishing curvature. In such cases, the background angular diameter distance can be computed using the Mattig relation \cite{mattig}:
\begin{equation}\label{eq:Mattig}
\begin{split}
D_{A,\text{bg}} = \frac{2c}{H_0\Omega_{m,0}^2(1+z)^2}\cdot \\
\left[ \Omega_{m,0}z -\left(2-\Omega_{m,0} \right)\left(\sqrt{1+\Omega_{m,0}z}-1 \right)   \right] 
\end{split}
\end{equation} 
The components of $D$ along a fiducial ray in a ray bundle can be obtained by using the transport equation:
\begin{equation}\label{D_dot}
\ddot D^a_b = T^a_cD^c_b
\end{equation}
The indices are arranged in accordance with the Einstein summation convention which will also be used in the following.
\newline\indent
The optical tidal matrix $T$ can be written using $\mathbf{R}: = -\frac{1}{2}\mathcal{R}_{\mu\nu}k^{\mu}k^{\nu}$ and $\mathbf{F}:=-\frac{1}{2}C_{\alpha\beta\mu\nu}(\epsilon^*)^{\alpha}k^{\beta}(\epsilon^*)^{\mu}k^{\nu}$, where $\mathcal{R}_{\mu\nu}$ is the Ricci tensor, $C_{\alpha\beta\mu\nu}$ the Weyl tensor and $\epsilon^{\mu} := E_1^{\mu} + iE_2^{\mu}$ with $E_1^{\mu}, E_2^{\mu}$ spanning screen space:
\begin{equation}
 T_{ab} = 
  \begin{pmatrix} \mathbf{R}- Re(\mathbf{F}) & Im(\mathbf{F}) \\ Im(\mathbf{F}) & \mathbf{R}+ Re(\mathbf{F})  \end{pmatrix} 
\end{equation}
It is convenient to note that $k^{\mu}$ and $\epsilon^{\mu}$ are null-geodesics orthogonal to each other; this implies that the Ricci-part of the Weyl tensor drops out in $\mathbf{F}$ so that $\mathbf{F}=-\frac{1}{2}\mathcal{R}_{\alpha\beta\mu\nu}(\epsilon^*)^{\alpha}k^{\beta}(\epsilon^*)^{\mu}k^{\nu}$ (this simplification is also noted in a footnote in \cite{arbitrary_spacetime}).
\newline\indent
In order to obtain $\kappa$ along null-geodesics in Szekeres models, the transport equation must be solved simultaneously with solving the geodesic equations and the parallel propagation equations for $E_{a}^{\mu}$. These equations are shown in appendix \ref{appendix:sz_Riemann} together with the explicit expressions for $\mathbf{R}$ and $\mathbf{F}$ for the Szekeres metric. The resulting set of ODEs is the same as that in \cite{dallas_first} though some of the components are here given in a slightly simpler form {\em e.g.} because the Riemann tensor is used instead of the Weyl tensor when computing $\mathbf{F}$.
\newline\newline
The convergence and the reduced shear $g_{\gamma} := \frac{\gamma}{1-\kappa}$ are highly important cosmological observables as they affect redshift-distance relations and can be used to trace matter (see {\em e.g.} \cite{only_grav3, Schneider1,Schneider2, lensing_important1, lensing_important3, lensing_important4, lensing_important6,shear_review, only_grav1, only_grav2, only_grav5, weak_survey} for examples of using $\kappa$ and $\gamma$ in relation to observations). Especially with regard to the onion models it is worth mentioning that the quantities $\kappa$ and $\gamma$ are traditionally defined using linear perturbation theory, and their relations to observables is conditional to this regime. Equally important is it that the relation between $A$ and $D$ requires a background so that $D_{A,\text{bg}}$ is defined.
\newline\indent
In settings where the linear perturbation scheme is ill-suited, the algebraic relations between the components of $D$ can still be used to define quantities generalizing $\kappa$, $\gamma_1$ and $\gamma_2$. These quantities' relation to observables is not {\em a priori} as in the linear limit (see also \cite{dallas_first}). In this work, the terms "convergence" and "shear" will refer to the generalized definitions of $\kappa$ and $\gamma$ without restricting to linear perturbation theory.

\subsection{Standard ray tracing}\label{sec:Ray_tracing_theory}
In this section, the standard ray tracing expression for $\kappa$ is introduced. For more details, see {\em e.g.} \cite{Schneider1, lensing_important1}.
\newline\newline
Ray tracing combines the perturbed FLRW metric in the Newtonian gauge with the non-linear density contrasts and velocity fields obtained from N-body simulations. The standard recipe for going between Newtonian N-body results and a relativistic description of these is thus in line with the simple dictionary given in \cite{recipe} which gives a recipe for going between Newtonian N-body data and a relativistic spacetime with the line element:
\begin{equation}\label{line_element_ng}
\begin{split}
ds^2 = -c^2(1+2\psi)dt^2 +\\
a^2(1-2\psi) \left( dr^2 + r^2d\theta^2 + r^2\sin^2(\theta)d\phi^2\right) 
\end{split}
\end{equation}
The scale factor is normalized to 1 at present time, {\em i.e.} $a(t_0) = 1$. Following the recipe, the dimensionless gravitational potential is obtained through the Poisson equation,  $\nabla^2\psi = \frac{4\pi G a^2}{c^2}\delta\rho$, using the non-linear overdensity of the given N-body data. The velocity field is needed {\em e.g} for computing the observable redshift. It is given by $u^{\mu} \propto (1,v^i)$, with $v^i$ the spatial velocity field of the N-body data. In the case of the Szekeres models studied here, the angular velocities are negligible and $u_{\mu} \approx \frac{c}{V}(-c^2(1+2\psi), a^2(1-2\psi)v_r,0,0)$, with $V = \sqrt{c^2(1+2\psi)-a^2v_r^2(1-2\psi)}$.
\newline\indent
The metric corresponding to the line element in equation (\ref{line_element_ng}) combined with the non-linear velocity and density fields will in the following be referred to as the Newtonian gauge metric.
\newline\newline
$\mathbf{R}$ and $\mathbf{F}$ can be computed and used to obtain ODEs for the components of $D$. It is, however, standard instead to write the equations in terms of the deformation matrix $A = \frac{1}{D_{A,\text{bg}}}D$ and use Green's method to obtain implicit integral equations for its components. The resulting equations can be combined with equation (\ref{eq:kappa}) to obtain the following expression for the gravitational convergence:
\begin{equation}\label{eq:kappa_old}
\kappa_{\delta} = \frac{4\pi G}{c^2}\int_{0}^{r_g}dr_g'a^2\delta\rho\frac{(r_g-r_g')r_g'}{r_g}
\end{equation}
This integral expression is obtained by neglecting lens-lens coupling (see {\em e.g.} \cite{millenium}) and by using the Limber approximation (see {\em e.g.} \cite{improve3, lss}). The integrals are along the path of a given light ray and are usually solved by introducing the Born approximation, where rays are traced as radial null-geodesics in the background model. 
\newline\indent
The convergence gives the most important contribution to alterations of redshift-distance relations in N-body simulation models where it is standard to use the approximation $D_A \approx D_{A,\text{bg}}(1-\kappa_{\delta})$. $(1-\kappa_{\delta})$ is the lowest order approximation to the determinant of $D$ since the shear is usually several orders of magnitude smaller than $\kappa_{\delta}$.

\subsubsection{Doppler convergence}\label{sec:Dopler}
The total first order convergence consists of several other contributions than just the gravitational convergence. These other contributions {\em e.g.} include the well-known Sachs-Wolfe and integrated Sachs-Wolfe effects which are generally subdominant to the gravitational convergence. There is also a Doppler contribution, $\kappa_v$, which arises because peculiar velocities affect the redshift along light rays. This Doppler contribution to the convergence and redshift-distance relations has been known for several decades (see {\em e.g.} \cite{first_doppler} for an early derivation), and has been included in several more recent studies ({\em e.g.} \cite{bright_side, use_doppler1,use_doppler2, use_doppler3, use_doppler4,use_doppler5,use_doppler6, use_doppler7}). The Doppler convergence is typically not included when ray tracing through N-body simulations. This is presumably because the effect is important mostly at small redshifts, and because the effect should become negligible when averaged over an infinite number of observations in a universe satisfying the Copernican principle. The Doppler contribution will be included here since it is automatically included in the geodesic deviation equation and thus will be portrayed in the exact convergence.
\newline\newline
The Doppler convergence can be computed as (see \cite{use_doppler1}):
\begin{equation}\label{eq:Doppler}
\begin{split}
\kappa_vc =   \left(1-\frac{1}{a_{,t}(\eta_0-\eta)} \right) \mathbf{n}\cdot(\mathbf{v}-\mathbf{v}_0) + \mathbf{n}\cdot \mathbf{v}_0 \\
\end{split}
\end{equation}
The vector $\mathbf{n}$ is the photon direction vector in the source frame {\em i.e.} it is the spatial part of the unit $4$-vector $n^{\mu} \propto k^{\mu}+\frac{k^{\nu}u_{\nu}}{c^2}u^{\mu}$ evaluated along the null geodesics. $\eta$ denotes conformal time. 
\newline\indent
The velocity field $\mathbf{v}$ is the spatial part of the normalized 4-velocity $u^{\mu}$. The sign of $\mathbf{v}$ is determined from the global coordinate system and can in swiss cheese models differ from the local single void sign of $u^{\mu}$.
\newline\newline
Note that in equation (\ref{eq:Doppler}), the scale factor etc. are evaluated at the spacetime position corresponding to the {\em observed} redshift, while equation (\ref{eq:kappa_old}) for the gravitational convergence is evaluated in accordance with the background redshift.

\subsubsection{Summary of the ray tracing method}\label{sec:ray_trace_summary}
In the following two sections, a simple and popular approximate ray tracing method is compared with exact light propagation and with an extended ray tracing method which includes the Doppler convergence. For convenience, the approximations included in the ray tracing methods are summarized here.
\newline\newline
The convergence is computed by adding the lowest order gravitational convergence, $\kappa_{\delta}$, to the Doppler convergence, $\kappa_v$. The gravitational convergence is computed using the Born and Limber approximations and by neglecting lens-lens coupling. In standard ray tracing schemes, the Doppler convergence is neglected. Thus, in standard ray tracing schemes, the angular diameter distance is computed as $D_A = D_{A,\text{bg}}(1-\kappa_{\delta})$ while in the extended ray tracing scheme included here, it is computed as $D_A = D_{A,\text{bg}}(1-\kappa_{\delta}-\kappa_v)$.
\newline\indent
The convergence will be portrayed as a function of the observable redshift given by $z_{\text{obs}}+1 = \frac{(k^{\alpha}u_{\alpha})_e}{(k^{\alpha}u_{\alpha})_0}$. The subscript "$e$" denotes time of emission.
\newline\newline
The potential $\psi$ turns out to be insignificant for the results. This is only the case since $\kappa_{\delta}$ is computed using the Limber approximation and thus is computed independently of $\psi$. The only place $\psi$ plays a role is thus when lowering indices in the computations of $\kappa_v$ and $z_{\text{obs}}$. The most significant "first order" quantity modifying these from their background values is the velocity field, not $\psi$. By using a few of the studied LTB models, it has been tested that simply neglecting $\psi$ altogether is of no significance for the results obtained here. The results presented below have thus all been obtained under the approximation of setting $\psi = 0$.

\section{Swiss cheese model results}\label{sec:cheese_results}
In order to compare results obtained using the exact formalism and the approximate ray tracing schemes, these methods will be used to study the convergence along individual rays in swiss cheese and onion models. A direct comparison between the exact and approximate results requires that the exact and ray traced rays are initialized equivalently. Since the exact ODEs are solved backwards in time, this implies that the observer position and line of sight in the two spacetimes must be equivalent according to the map in equation (\ref{stereo_map}).
\newline\newline
When the purpose is studying light propagation, swiss cheese models are most simply constructed on the fly. In such construction schemes, a ray will be moved from one single void model to another once it has reached a specific coordinate distance from the given local structure (see also \cite{dallas_cheese}). In general, the exact and ray traced rays do not follow the same paths. Since the swiss cheese models are made on the fly, this implies that the swiss cheese models that the exact and ray traced rays propagate through are unrelated. Clearly, this leads to a rather contrived comparison between the exact and ray tracing results. To avoid this, when a ray traced ray is moved to a new structure, its position and direction will always be chosen to be the initial position and direction that the corresponding exact ray had when it reached that specific structure
\footnote{Within a given redshift interval, the ray traced ray may traverse more structures than the exact ray. When traversing such final structures, the ray traced ray is moved to a "new" structure simply by turning it on the spot and tracing it radially back towards the structure it came from.}
. This way, the exact and ray traced rays will move through the "same" swiss cheese model, but the rays will generally traverse each structure along different paths - depending on how good an approximation the Born approximation is. The swiss cheese results are thus to be considered indicative of accumulative effects of local inhomogeneities.
\newline\indent
The standard ray tracing scheme described in section \ref{sec:Ray_tracing_theory} is adapted to radial rays. In practice, ray tracing through swiss cheese models thus becomes much simpler if rays are always initialized as radial at the beginning of a new single void structure. The ray tracing results shown below have therefore been obtained by following the procedure of moving a ray to a new structure by turning it around  and sending it radially back towards the structure it just left. Especially the part of initializing rays as radial at each "new" structure clearly makes the paths very special. The consequent bias of the results induced by studying only special rays implies that it is inappropriate to compare with background results. This does not affect the integrity of the study as the point is not to compare with background results anyway. The point here is to study the precision of the standard ray tracing scheme with the added Doppler convergence as well as to study implications of neglecting the Doppler convergence. Thus, for the results obtained from the swiss cheese models based on non-symmetric Szekeres models, the special paths do not compromise the study much; if the ray traced rays cannot reproduce exact rays that are initialized as radial then surely they cannot reproduce more general rays that are initialized with arbitrary impact parameters at each structure.
\newline\indent
The consequence of only studying radial rays is much more significant in the case of the spherically symmetric LTB models; in these models, rays initialized as radial will stay so. This implies that the exact and ray traced rays will automatically move along the same spatial path. In order to see how (dis-)similar the exact and ray traced paths along non-radial rays in LTB models are, the Sachs formalism has been used to obtain an approximation scheme for computing the convergence along such non-radial rays. The scheme employs the Born approximation and should yield results that are no more precise than the regular ray tracing scheme described in section \ref{sec:Ray_tracing_theory}. The details of the scheme are explained in appendix \ref{app:sachs_born}. 
\newline\newline
Unless otherwise stated, the observer is placed at the center of a void. The results are always plotted against the observable redshift and light rays are propagated/traced through swiss cheese models until they reach a redshift of $z_{\text{obs}} = 0.6$.

\subsection{LTB results} \label{sec:LTB_results}
In this section, the results from studying light propagation in a swiss cheese model based on LTB single void models are shown. As discussed above, the results shown here pertain to rays that propagate {\em radially} through consecutive LTB single void structures. This makes the rays very special since is guarantees that the exact and ray traced rays move along the same spatial paths. Non-radial rays in LTB models will be discussed in subsection \ref{sec:nonradial}.
\newline\newline

\begin{figure*}[htb!]
\subfigure[]{
\includegraphics[scale = 0.88]{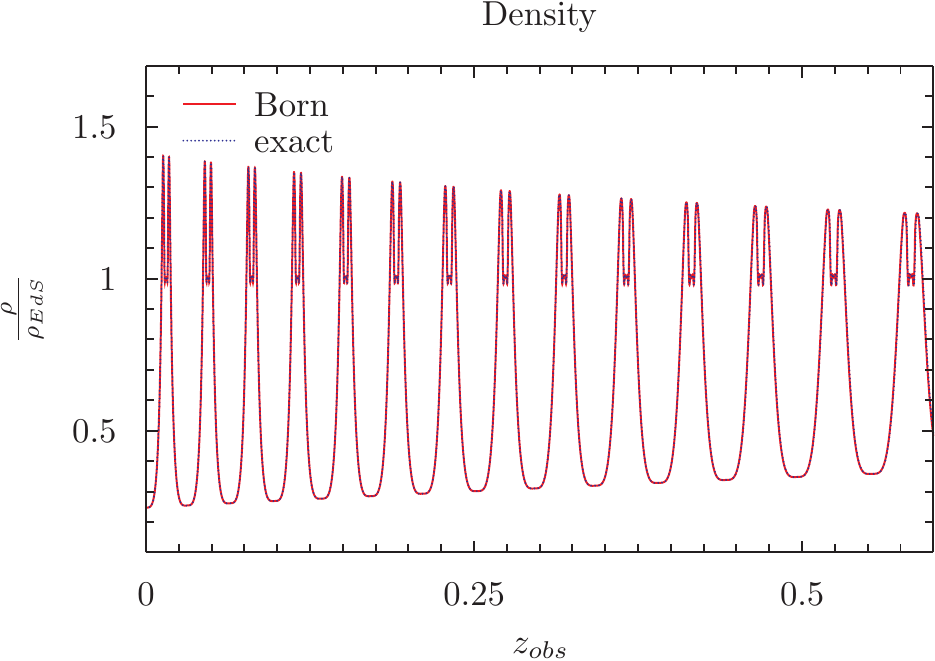}
\label{fig:LTBcheese_rho}
}
\subfigure[]{
\includegraphics[scale = 0.88]{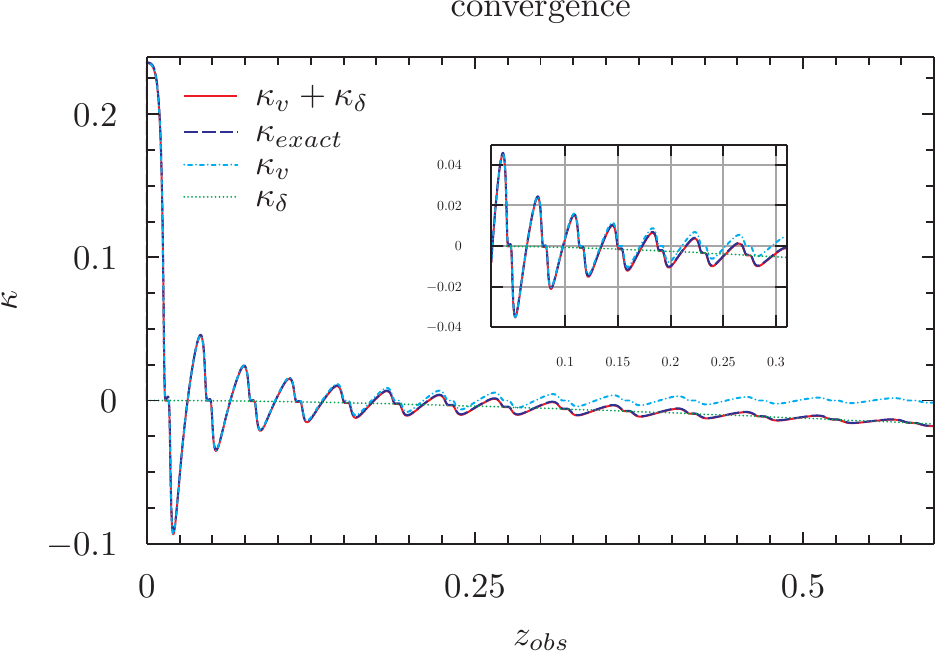}
\label{fig:LTBcheese_kappa}
}
\caption{Density and convergence along a light ray in a swiss cheese model based on compensated LTB single void models with $m = 4$ and $r_b = 60\text{Mpc}$. The ray path has been computed using both the Born approximation and the geodesic equations of the exact LTB metric. The two resulting density distributions are indistinguishable in the figure. Four convergences are shown. The line labeled "$\kappa_v + \kappa_{\delta}$" shows the sum of the Doppler convergence and the gravitational convergence obtained using the simple ray tracing scheme. The two components are also shown individually. The line labeled "$\kappa_{\text{exact}}$" is the exact convergence as obtained from the exact set of ODEs. The figure includes a close-up of the region where the gravitational convergence begins to be non-negligible.}
\end{figure*}

\begin{figure*}[htb!]
\centerline{\includegraphics[scale = 0.88]{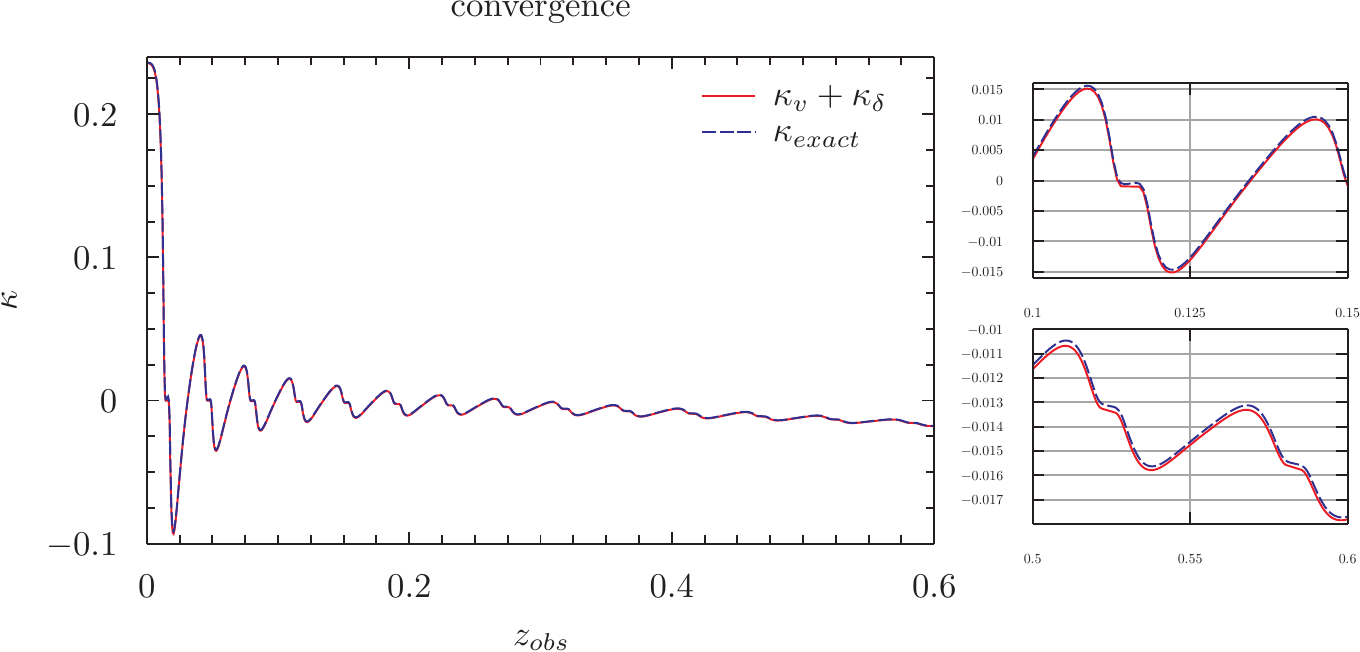}}
\caption{Convergence along a radial ray in the swiss cheese model based on the LTB single void model specified by $m = 4$ and $r_b = 60\text{Mpc}$. The line labeled "$\kappa_v + \kappa_{\delta}$" shows the sum of the Doppler convergence and the gravitational convergence obtained using the simple ray tracing scheme. The line labeled "$\kappa_{\text{exact}}$" is the exact convergence as obtained from the exact set of ODEs. The figure includes close-ups at different redshift intervals.}
\label{fig:LTBcheese_zoom}
\end{figure*}

As seen in figure \ref{fig:LTBcheese_rho}, the local density distributions along the approximate and exact geodesics are the same. This indicates that the combined (mock) N-body data and Newtonian gauge metric yield a good approximation of the exact observable redshift. Note that it is important to use the approximate observable redshift; simply using the background redshift leads to clear deviations between the exact and ray tracing results.
\newline\indent
The convergence along the ray is shown in figure \ref{fig:LTBcheese_kappa}. There is an overall good agreement between the exact convergence and the sum of the Doppler and gravitational convergences. The figure also shows that the Doppler convergence clearly dominates the convergence at low redshifts while the gravitational convergence dominates at higher redshifts. The former is in agreement with the results obtained in \cite{bright_side} where the Doppler and gravitational contributions to the convergence were studied in Szekeres single void models. In \cite{bright_side}, mock N-body data was not employed when computing the two contributions to the convergence. Instead, regular linear perturbation theory was used. It was found that, at late times when the Szekeres structures had gone non-linear, linear perturbation theory lead to an underestimation of the Doppler convergence by approximately $20 \%$. Here, it is found that this underestimation vanishes when linear perturbation theory is combined with non-linear Newtonian N-body velocity fields. This is emphasized in figure \ref{fig:LTBcheese_zoom} which includes close-ups along the convergence curves. Not until approximately $z>0.5$ do the exact and ray tracing convergences begin to differ notably from each other - and even then, only by a few percent. The difference between the exact and approximate convergences grows {\em slightly} with $r_b$ and $m$ (defined in section \ref{sec:cheese_models}). It is virtually nonexistent in the $r_b = 30\text{Mpc}$ cases, and for the $m = 2$ cases also when $r_b$ is larger. The figures shown here are for LTB models with $m = 4$ and $r_b = 60\text{Mpc}$.
\newline\indent
The difference between the exact and ray traced convergences are at least partially numerical artifacts.
\newline\indent
The distinct oscillations of the Doppler convergence appear as a manifestation of the Doppler convergence being a measure of the {\em local} velocity field. The gravitational convergence is an accumulated effect of the matter distribution along the given geodesic. The gravitational convergence is thus much less affected by individual local variations than the Doppler convergence is, though close-ups (not shown) of the gravitational convergence do reveal small ripples accounting for the local fluctuations in the density field.
\newline\newline

\begin{figure*}[htb!]
\centering
\subfigure[]{
\includegraphics[scale = 0.8]{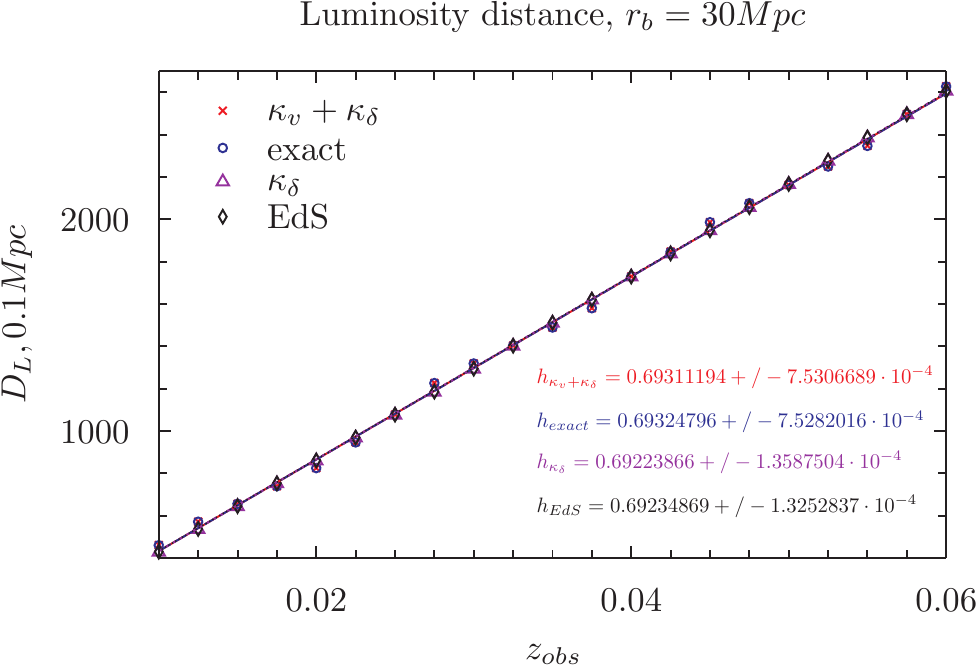}
}
\subfigure[]{
\includegraphics[scale = 0.8]{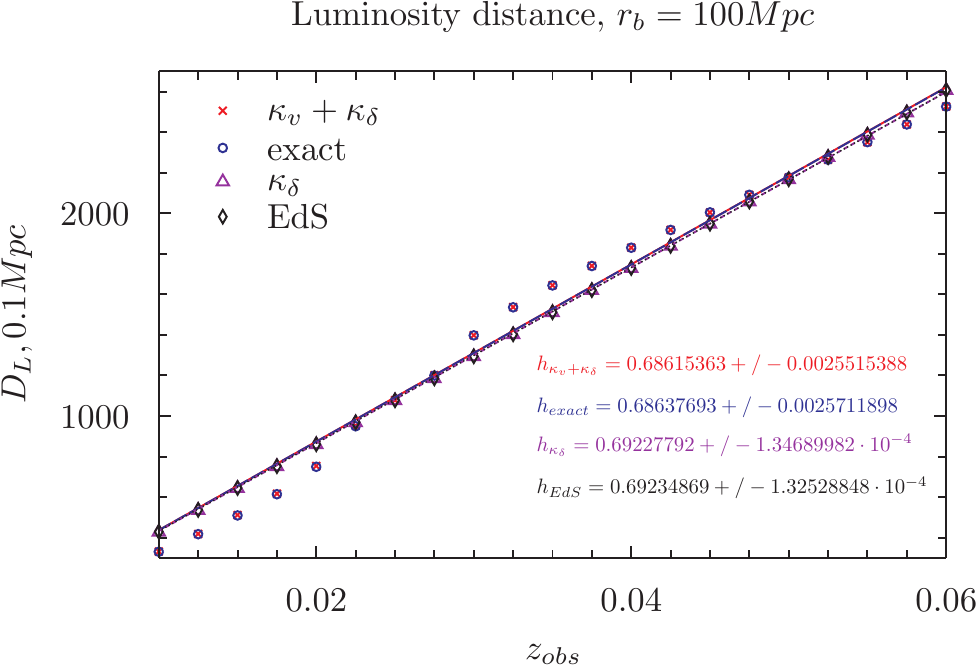}
\label{fig:h_voidsize_100}
}
\par
\subfigure[]{
\includegraphics[scale = 0.8]{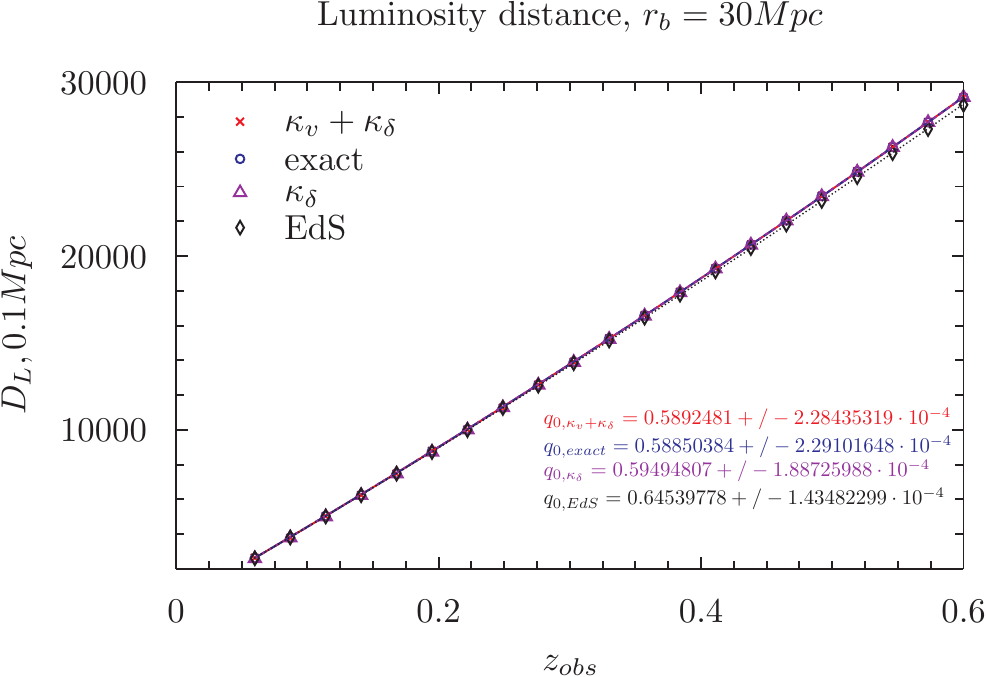}
}
\subfigure[]{
\includegraphics[scale = 0.8]{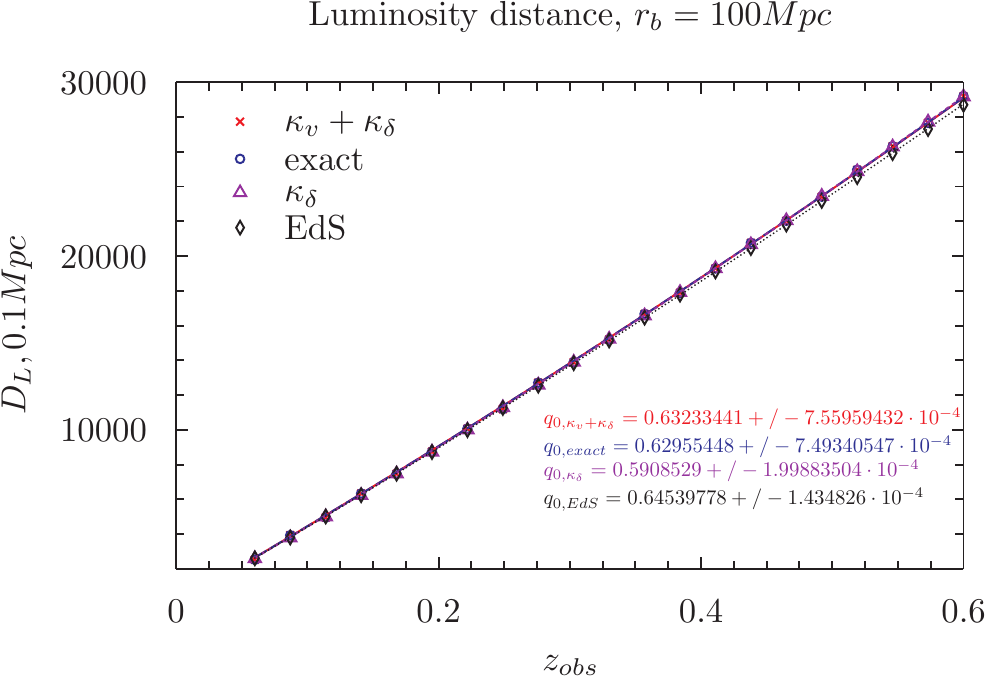}
}
\caption{Luminosity distance plotted as a function of the observed redshift and fitted to low-redshift FLRW approximations of $D_L(z)$. The results from four luminosity distances are shown corresponding to the exact convergence, the sum of the gravitational and Doppler convergences, the gravitational convergence and the background solution. The results are shown for LTB swiss cheese models specified by $m = 4$ and both $r_b = 30\text{Mpc}$ and $r_b = 100\text{Mpc}$.}
\label{fig:h_voidsize}
\end{figure*}

\begin{figure*}[!htb] 
\subfigure[]{
\includegraphics[scale = 0.7]{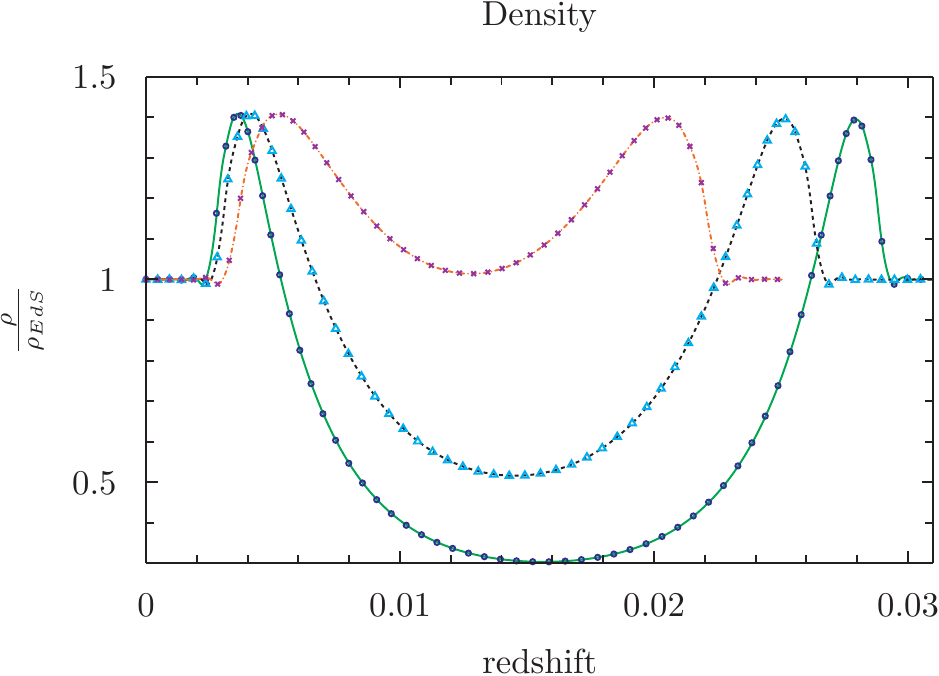}
}
\subfigure[]{
\includegraphics[scale = 0.7]{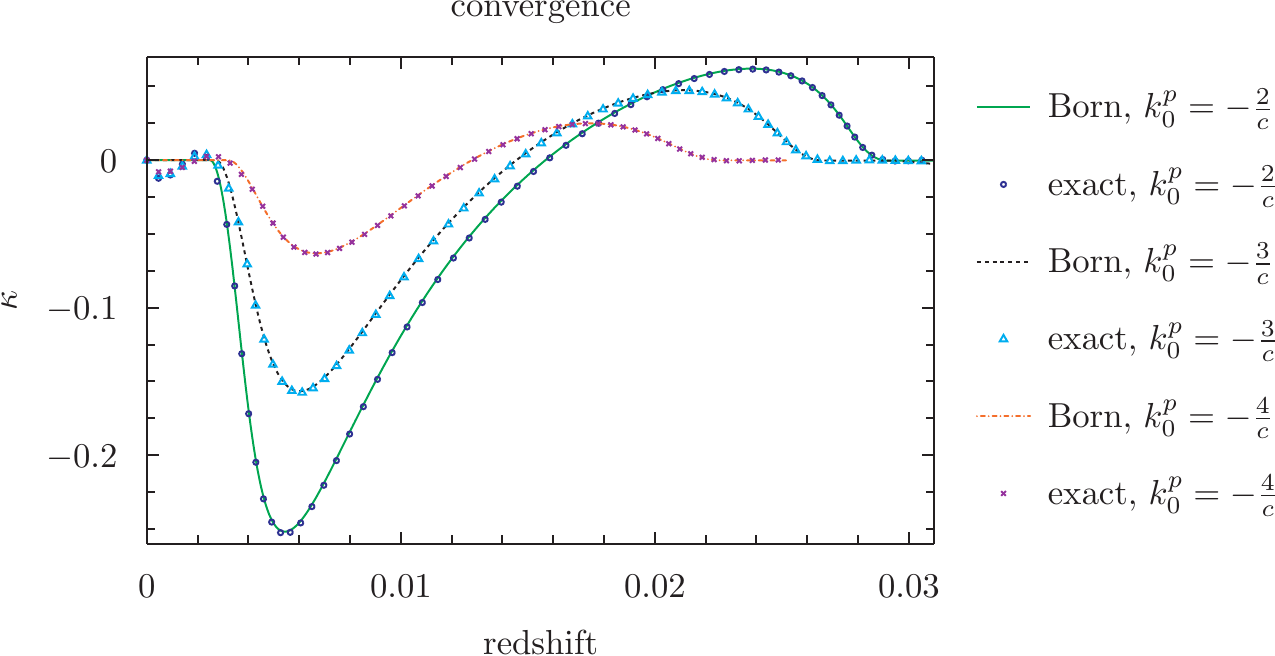}
}
\caption{Density and convergence along non-radial light rays in an LTB single void model specified by $m = 4$ and $r_b = 60\text{Mpc}$. The observer has been placed at $(t,r,\theta,\phi) = (t_0,70\text{Mpc}, 8/5\pi, 3/5\pi)$, looking in various directions determined by $(k^t,k^r,k^p,k^q)=(-\frac{1}{c}, k^r_0, k^p_0, 0)$, where $k^p_0$ is varied as  $k^p_0 = -\frac{2}{c}, -\frac{3}{c}, -\frac{4}{c}$ and $k^r_0$ is determined from the null-condition. Note that $k^p_0$ does not have the same units as $\frac{1}{c}$ so it is only the numerical value of $c$ that is implied in the values of $k^p_0$. Since the LTB model reduces exactly to the EdS model outside the structure, these initial conditions are for both the exact Szekeres coordinate system and the ray tracing coordinate system. The exact and ray traced rays are labeled by "exact" and "Born" respectively.}
\label{fig:ltb_nonrad}
\end{figure*}

It is clear from figure \ref{fig:LTBcheese_kappa} that the Doppler convergence, in principle, will affect observations. As a simple study of the significance of these effects, the different convergences have been used to compute the luminosity distance as a function of redshift which has then been used to obtain estimates of the reduced Hubble parameter, $h$, and the deceleration parameter, $q_0$. The luminosity distance is computed as $D_L = D_A(1+z)^2 = D_{A,\text{bg}}(1-\kappa)(1+z)^2$. In the low redshift interval of $0.01<z<0.06$
\footnote{This interval is chosen based on figure 1 of \cite{Io}.}
, 200 points along the redshift-distance relations are fitted to the low-redshift FLRW approximation $D_L(z) \approx \frac{c}{H_0}z$ with $h$ as the fitting parameter. In the remaining redshift interval of $0.06<z<0.6$, 1000 points are fitted to the FLRW approximation $D_L(z) \approx \frac{c}{H_0}z\left( 1 + \frac{1-q_0}{2}z \right) $ with $q_0$ as the fitting parameter, and using the value of $H_0$ obtained from the low-redshift fit.
\newline\indent
To illustrate the significance of void size, the fitting results are shown for two void sizes, namely $r_b = 30\text{Mpc},100\text{Mpc}$. The fitting was done with pyxplot\footnote{http://pyxplot.org.uk/} and the best-fit values together with 1 standard deviation intervals are shown together with the plots in figure \ref{fig:h_voidsize}.
\newline\newline
There is a distinct difference in the estimates of $h$ based on the $\kappa_{\delta}$ fits compared to the fits based on $\kappa_{\text{exact}}$. This again affects the estimates of $q_0$. The difference in the two estimates of $h$ is only approximately 0.1\%, which is quite small compared to uncertainties of observational results. For instance, in \cite{H_0} the Hubble constant is estimated to be $H_0 = 73.8 \pm 2.4 \text{km/s/Mpc}$ {\em i.e.} the uncertainty is $3.3\%$, and the authors express no hope of their procedure in the future to yield results with a precision below approximately 1\%.
\newline\indent
The fits obtained using the sum $\kappa_{\delta}+\kappa_v$ agree well with the exact results. In particular, the two best-fit values of $h$ agree within approximately 0.02\% for the $r_b = 30\text{Mpc}$ model and 0.03\% for the $r_b = 100\text{Mpc}$ model. The void size clearly affects the determination of both $h_{\text{exact}},h_{\kappa_{\delta}+\kappa_v}$ and $q_{0,\text{exact}},q_{0,\kappa_{\delta}+\kappa_v}$ both regarding best-fit values and their uncertainties.
\newline\indent
A notable feature of the fits is that the uncertainties are markedly higher for the fits obtained using $\kappa_{\text{exact}}$ and the sum $\kappa_v + \kappa_\delta$ than for those obtained using just $\kappa_{\delta}$ and background values of $D_L$. This finding agrees with what should be expected from figure \ref{fig:LTBcheese_kappa}; the oscillating Doppler convergence leads to larger uncertainties in the low redshift estimates and to deviations of the best-fit values from those obtained from the background $D_L$ and $\kappa_{\delta}$. At higher redshifts, the Doppler convergence becomes less important and the uncertainties in the best-fit values become more similar.
\newline\indent
As is seen in especially figure \ref{fig:h_voidsize_100}, the imprints of the Doppler convergence oscillations on observables can be very prominent. In fact, the Doppler convergence is so prominent at low redshifts that it in \cite{use_doppler2} was studied to what extent measures of the Doppler convergence can be obtained observationally with the prospect of using it to {\em e.g.} study peculiar velocity fields.

\subsubsection{Non-radial rays}\label{sec:nonradial}
\begin{figure*}[htb!] 
\centering
\subfigure[]{
\includegraphics[scale = 0.8]{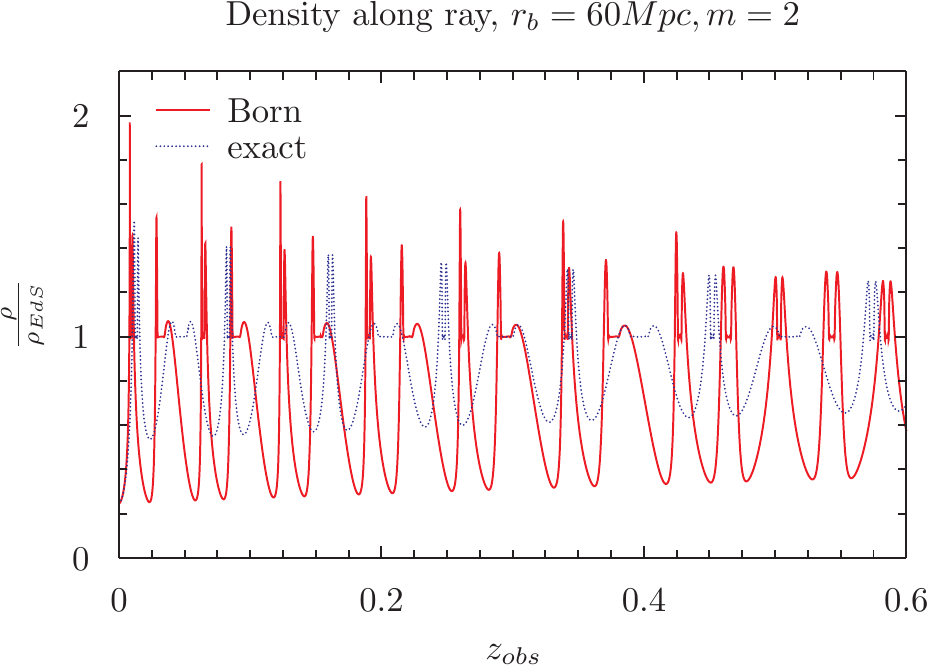}
}
\subfigure[]{
\includegraphics[scale = 0.8]{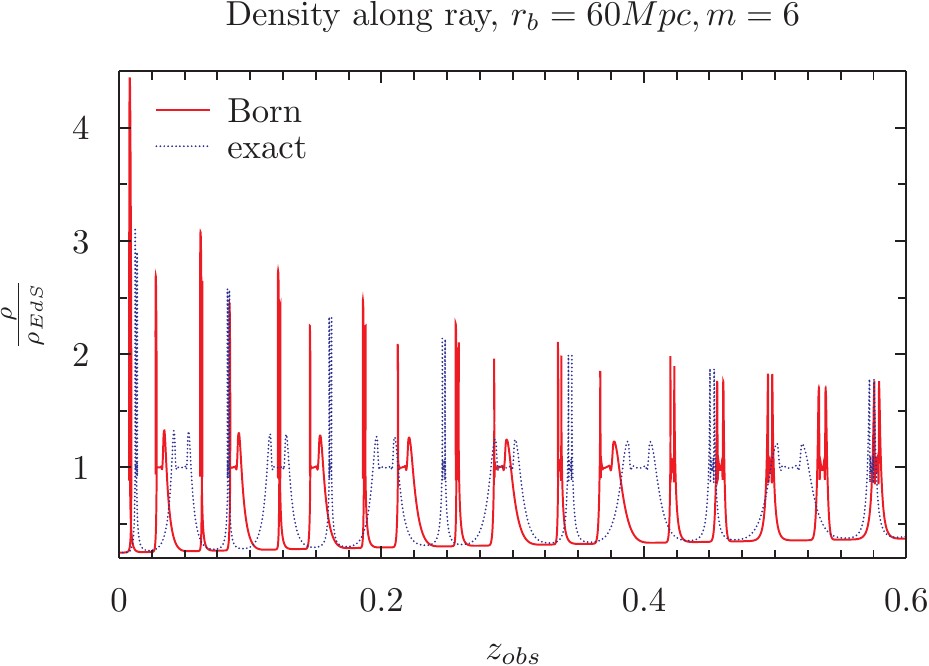}
}
\caption{Density along geodesics in swiss cheese model based on compensated non-symmetric Szekeres single void models. The geodesics have been computed using both the Born approximation and the exact geodesic equations according to the exact Szekeres metric.}
\label{fig:szcheese_rho}
\end{figure*}
It was shown above that the simple ray tracing scheme with the convergence computed as $\kappa_v+\kappa_{\delta}$ yields an almost exact reproduction of the exact results. This exact reproduction does not depend on the rays being radial. To show this, the Born approximated Sachs formalism described in appendix \ref{app:sachs_born} has been used to compute an approximate convergence along non-radial rays in different LTB models. Results are shown in figure \ref{fig:ltb_nonrad} for the LTB single void model specified by $m = 4$ and $r_b = 60\text{Mpc}$. As seen, the exact paths are exactly traced out by the Born paths. Aside from a minor oscillation at $z_{\text{obs}}\approx 0$, the convergences are also reproduced very precisely. These oscillations are due to changes in the velocity field which have, for simplicity, been neglected in the ray tracing method described in appendix \ref{app:sachs_born}. The oscillations are, however, described very well by the standard ray tracing method described in section \ref{sec:Ray_tracing_theory}. 
\newline\newline
Before moving on to study non-symmetric Szekeres models, it should be pointed out that the results of this section show that the LTB models can generally be described extremely well using the standard "perturbation" schemes used for studying N-body data. The quotation marks around "perturbation" emphasize that the ray tracing scheme is not strictly based on linear perturbation theory as the velocity and density fields are non-linear. This result is in agreement with the results of \cite{dig_selv} where it was shown that the Newtonian gauge metric could reproduce light propagation in a specific uncompensated LTB single void model, while it was insufficient for describing light propagation in non-symmetric models.

\subsection{Non-symmetric Szekeres results}\label{sec:szekeres_2}
\begin{figure*}[]
\centering
\subfigure[]{
\includegraphics[scale = 0.8]{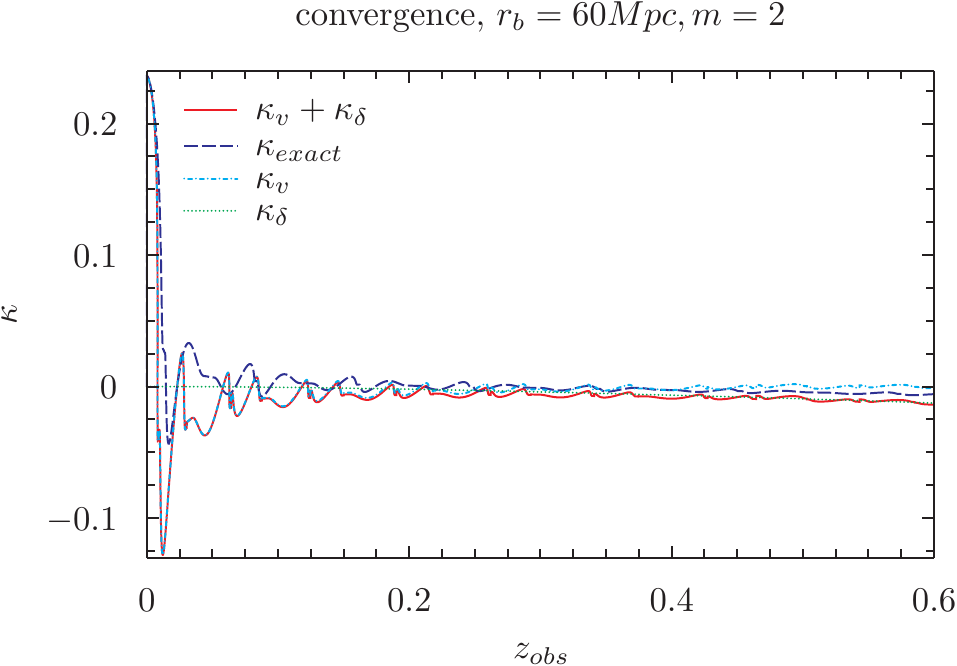}
}
\subfigure[]{
\includegraphics[scale = 0.8]{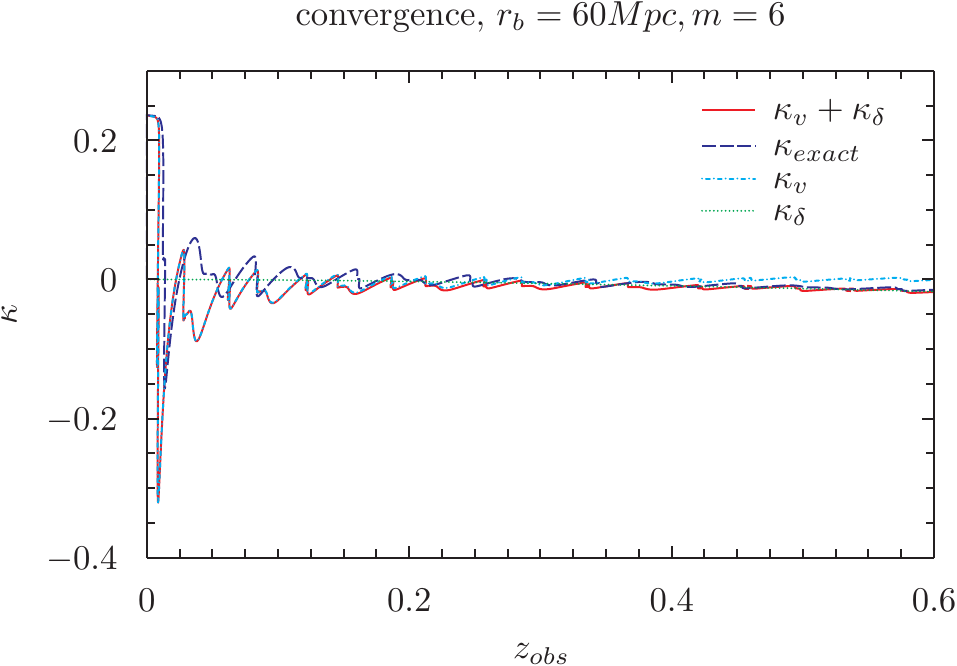}
}
\caption{Convergence along rays in swiss cheese models based on non-symmetric Szekeres single void models. Four convergences are shown in each figure. The line labeled "$\kappa_v + \kappa_{\delta}$" shows the sum of the Doppler convergence and the gravitational convergence. The two components are also shown individually. The line labeled "$\kappa_{\text{exact}}$" is the exact convergence as obtained from the exact set of ODEs.}
\label{fig:szcheese_kappa}
\end{figure*}

In this section, results from studying exact and ray traced rays in swiss cheese models based on the non-symmetric Szekeres models with $r_b = 60\text{Mpc}$ and $m = 2,6$ are shown.
\newline\newline
Because of the anisotropy of the voids, the exact rays are not radial. As seen in figure \ref{fig:szcheese_rho} this results in clearly different densities along the exact and ray traced rays, with the difference being most significant for the $m = 2$ case. In particular, the exact rays seem to completely avoid the central most devoid parts of the voids - even though they are initialized as radial at each single void structure. On the other hand, the ray traced rays are forced through the centers of each void.
\newline\indent
In accordance with this clear difference of local paths, the convergence along the exact and ray traced rays are also different. The convergences along the rays are shown in figure \ref{fig:szcheese_kappa}. Aside from being different locally, the exact and ray traced convergences do not even converge towards each other at high redshifts. This is because the gravitational convergence dominates at high redshifts. Especially in the $m = 2$ case, the very different densities along the exact and ray traced paths apparently lead to gravitational convergences that are different. In the $m = 6$ case, the exact and ray traced convergences have almost converged to each other at $z_{\text{obs}}\approx 0.6$.
\newline\indent
It is also notable that, unlike in the LTB case, the Doppler convergence does not oscillate about the gravitational convergence. This is {\em e.g.} seen by noting that the gravitational convergence is approximately zero at very low redshifts. If the exact convergence oscillated about the gravitational convergence, it would thus oscillate about zero at these lowest redshifts. This is clearly not the case.
\newline\newline

\begin{figure*}[]
\centering
\subfigure[]{
\includegraphics[scale = 0.8]{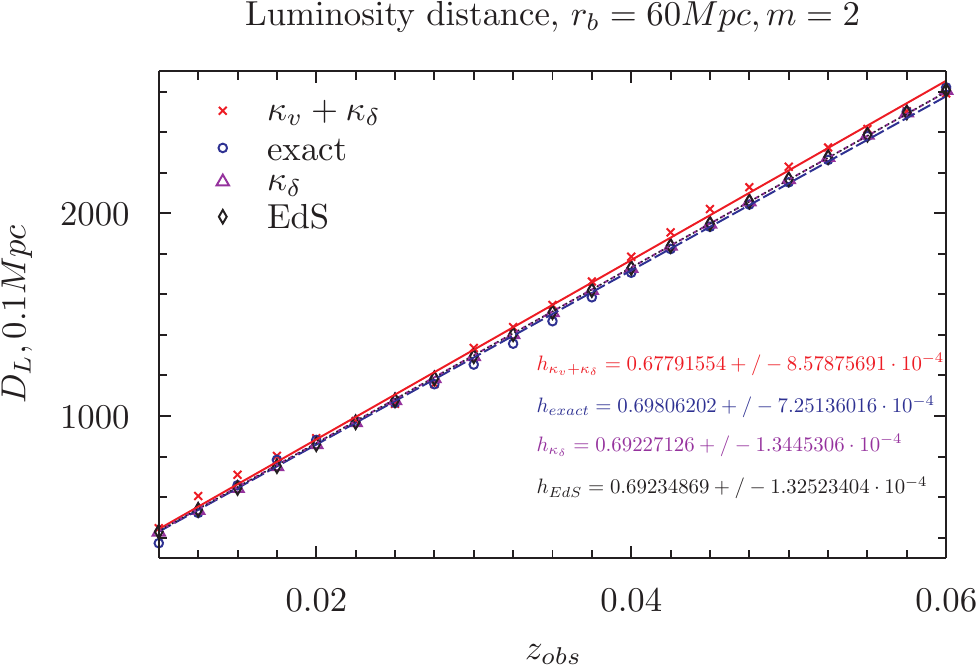}
}
\subfigure[]{
\includegraphics[scale = 0.8]{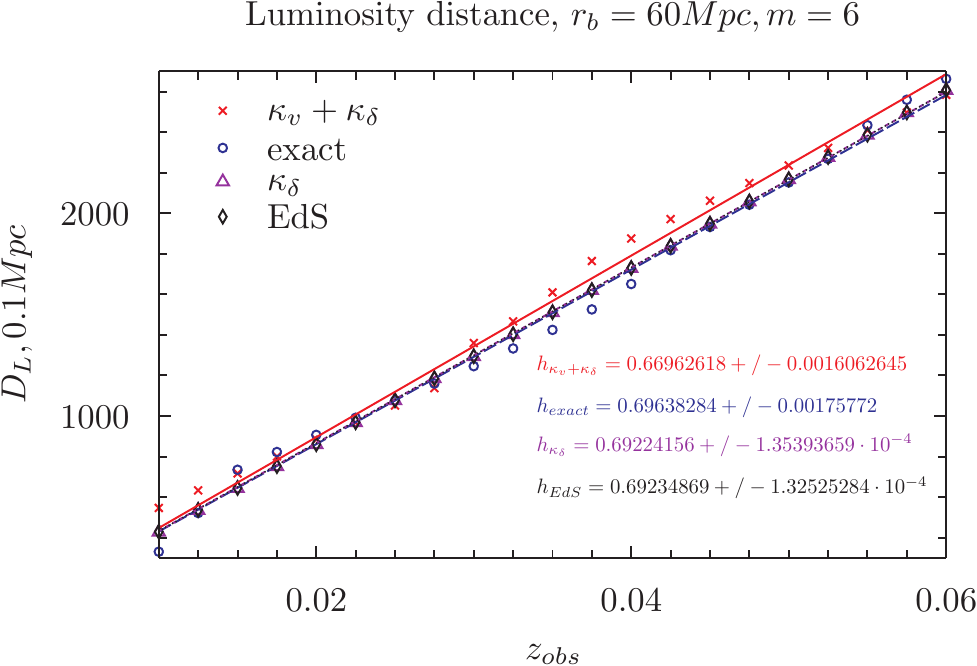}
}\par
\subfigure[]{
\includegraphics[scale = 0.8]{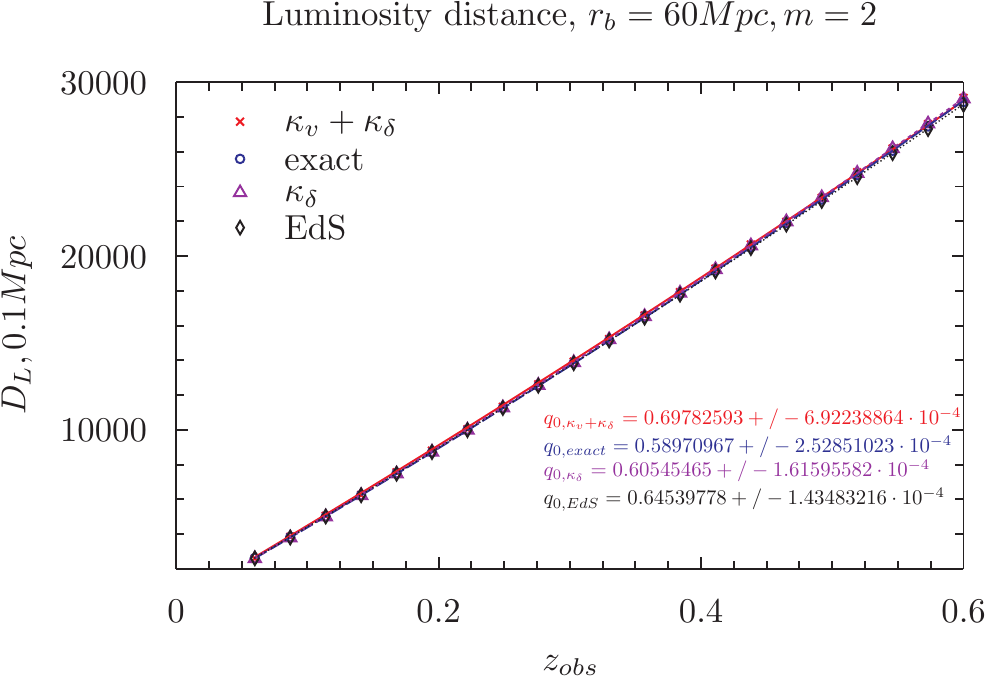}
}
\subfigure[]{
\includegraphics[scale = 0.8]{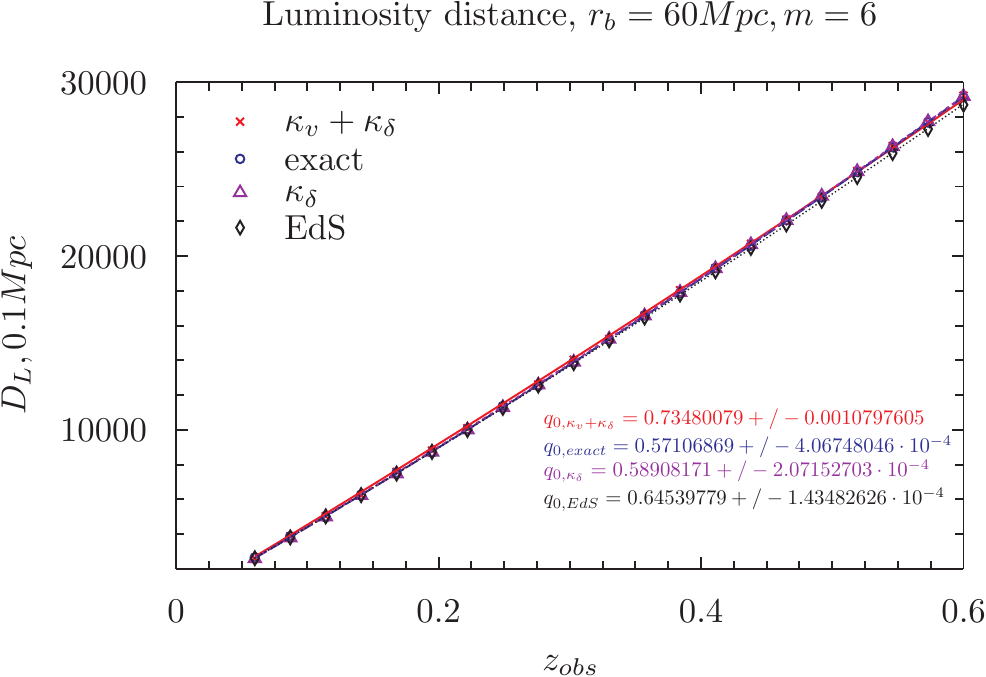}
}
\caption{Luminosity distance plotted as a function of the observed redshift and fitted to low-redshift FLRW approximations of $D_L(z)$. The results from four luminosity distances are shown corresponding to the luminosity distance obtained from the exact equations, the sum of the gravitational and Doppler convergences, the exact convergence and the background solution.}
\label{fig:h_sz}
\end{figure*}

As an illustration of the importance of the differences in the exact and ray traced paths, the luminosity distances corresponding to the exact and ray tracing convergences are used to estimate $h$ and $q_0$. The estimates are obtained using the fitting scheme introduced in section \ref{sec:LTB_results}. The results are shown in figure \ref{fig:h_sz}.
\newline\indent
The parameter values determined from the exact and ray traced rays are clearly in less agreement than in the LTB case. Specifically, the determinations of $h$ based on $\kappa_{\text{exact}}$ and $\kappa_v+\kappa_\delta$ differ by approximately $ 2.9\%$ and $3.9\%$ in the $m = 2$ and $m = 4$ cases respectively. The deviation between the exact and ray tracing estimates of $h$ depends significantly on the line of sight of the observer. The results shown here are for a direction where the difference between the approximations is particularly large. In many other directions the deviation is sub-percent. It should in this connection be emphasized that the point with the estimates made here is not to determine whether or not local inhomogeneities may affect determinations of $h$ (or $q_0$) in the real universe; this would require averaging over many rays with variable impact parameters at each void. This has, for instance, been done in \cite{dallas_cheese} and as mentioned in the introduction, the conclusion of that and other studies is that when averaging over many rays, the distance-redshift relation reduces to that of the average model. Thus, even though there may be differences in the exact and ray tracing estimates of $h$ along individual rays, the average results should be similar. The point with this section is thus that the exact paths of rays are significantly affected by inhomogeneities and to such an extent that it clearly affects observables along the individual rays. It is shown here that this is not described correctly by simple ray tracing schemes. This may be important in relation to interpretation of observations such as shear maps and CMB temperature fluctuation maps, where averaging over large portions of the sky is not viable.
\newline\indent
The differences in the estimates of $q_0$ are also quite large in the example shown here, but this is mainly due to the differences in the estimates of $h$. 
\newline\newline
The large deviation between exact and Born approximated paths is highly dependent on the level of anisotropy. If the anisotropy is lowered to approximately 25\% of that used here, {\em i.e.} if $Q$ defined in section \ref{sec:models} is multiplied by 0.25, then the exact and Born approximated rays are almost indistinguishable.

\section{Onion model results}\label{sec:onion_results}
The swiss cheese studies indicate that, in as far as the Born approximation is a good approximation, the simple ray tracing scheme studied here is very precise. The swiss cheese studies also indicate that the Born approximation generally cannot be assumed to be a good approximation. In the swiss cheese studies, the global paths of the exact and ray traced rays were the same since a ray traced ray was initialized at the same position and in the same direction as the corresponding exact ray at each new structure. The onion models are multiple structured models {\em ab initio} and not only after an on-the-fly construction scheme. Whereas the swiss cheese models were used to study the local precision of the Born approximation, onion models can thus also be used to study the global precision of the Born approximation. In addition, the convergence along radial rays in LTB onion models will be studied in order to examine the validity of using the ray tracing scheme in models without intrinsic backgrounds.

\subsection{LTB results} \label{sec:onion_LTB_results}
\begin{figure*}[]
\centering
\subfigure[]{
\includegraphics[scale = 0.59]{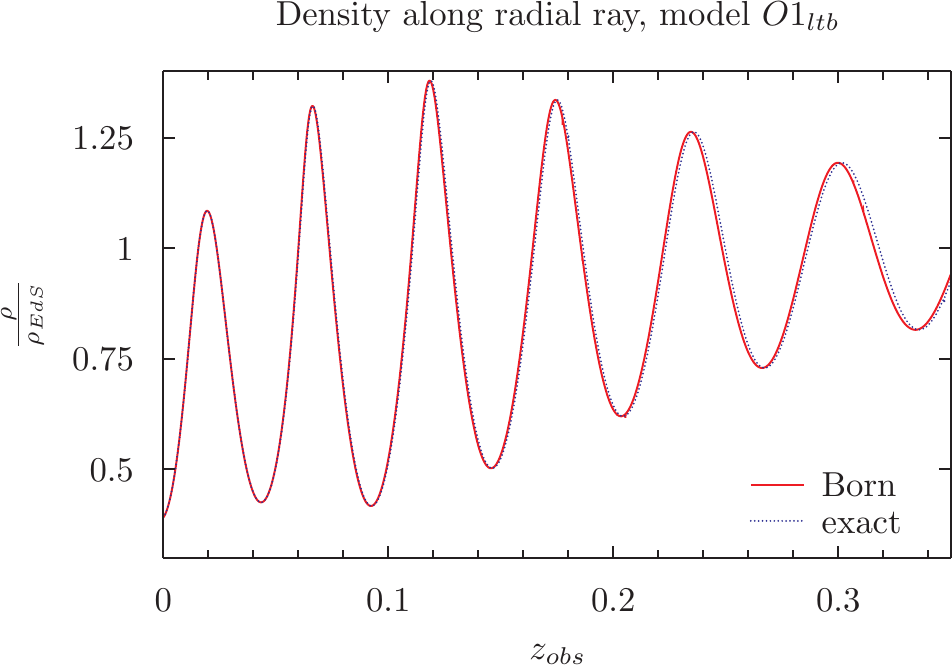}
\label{fig:onion1}
}
\subfigure[]{
\includegraphics[scale = 0.59]{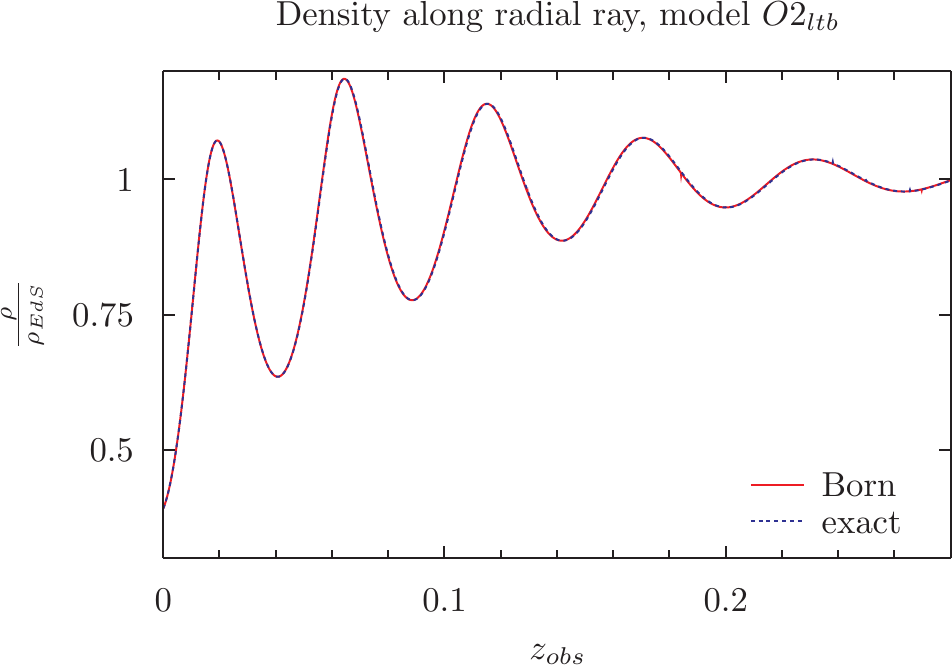}
}
\subfigure[]{
\includegraphics[scale = 0.59]{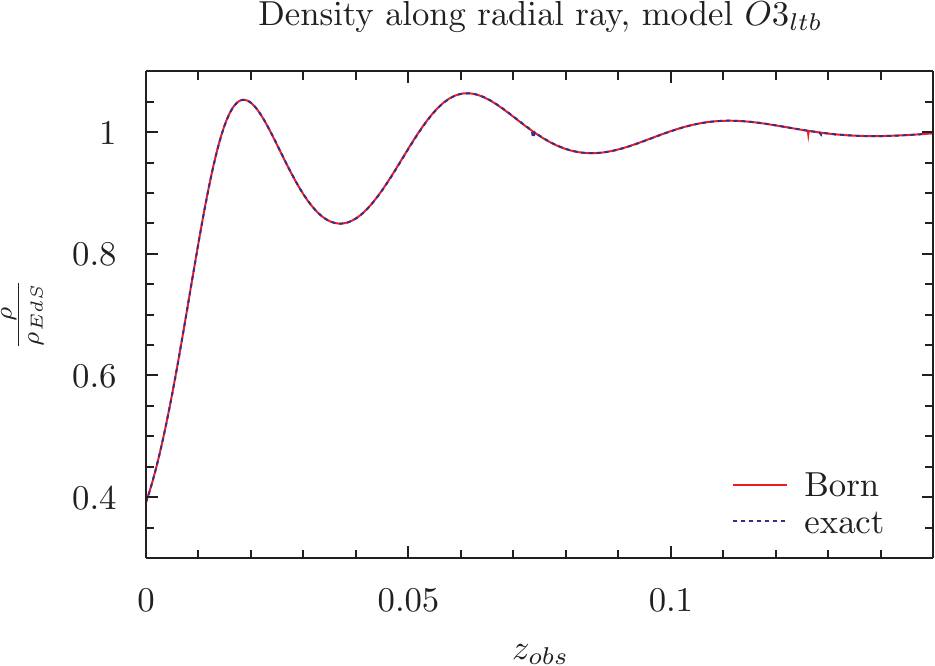}
}\par
\subfigure[]{
\includegraphics[scale = 0.59]{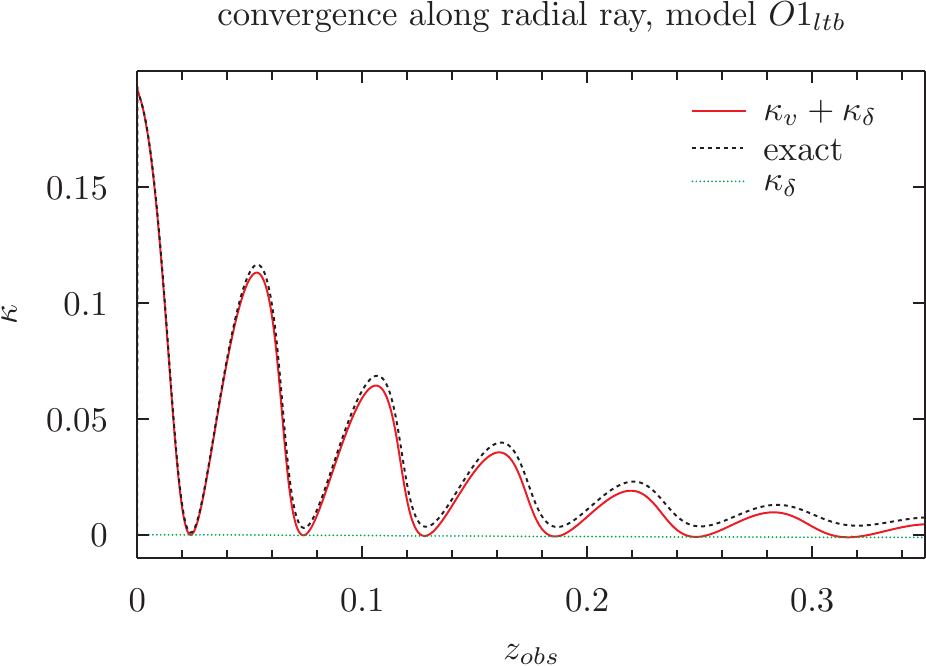}
}
\subfigure[]{
\includegraphics[scale = 0.59]{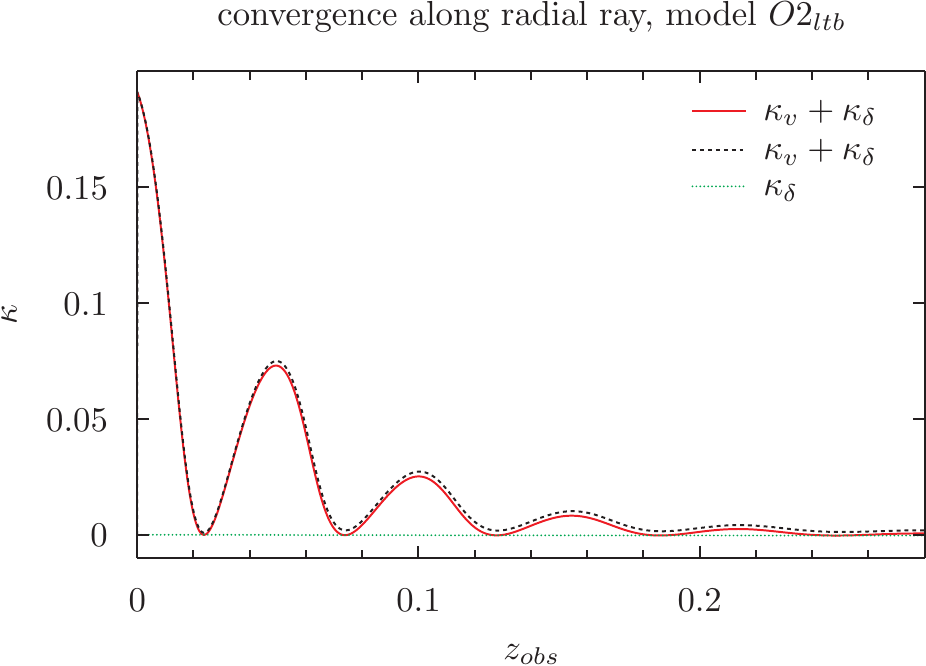}
}
\subfigure[]{
\includegraphics[scale = 0.59]{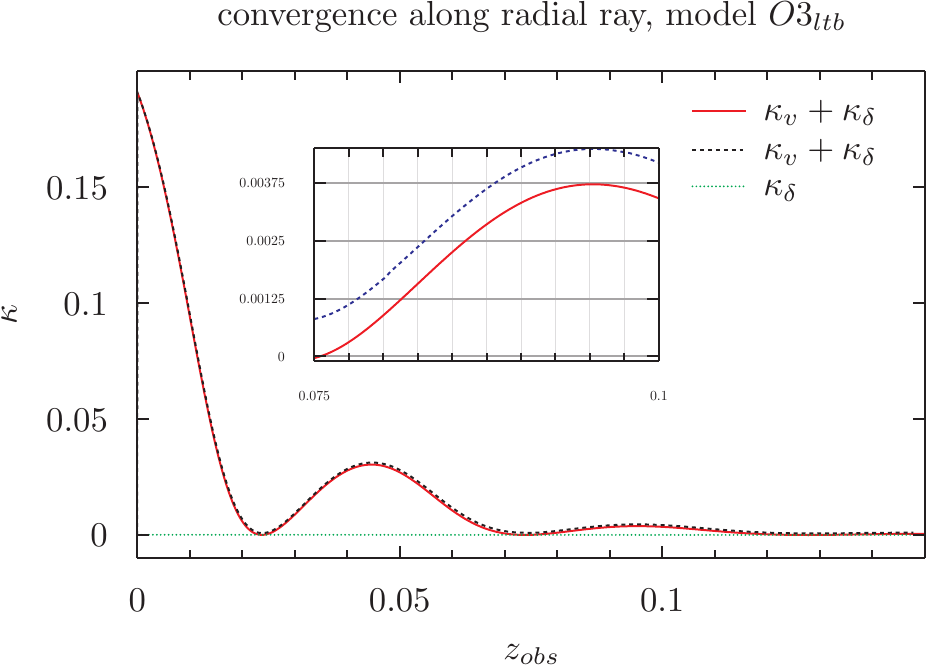}
}
\caption{Density and convergence along exact and ray traced radial rays in the onion models $O1_{\text{ltb}}$, $O2_{\text{ltb}}$ and $O3_{\text{ltb}}$. The exact convergence is shown together with the sum of the Doppler and gravitational convergences. The gravitational convergence is also shown alone but its deviation from zero is too small to be visible in the figure. For model $O3_{\text{ltb}}$, a close-up of the convergence has been included to highlight that the exact and approximate convergences are not exactly identical.}
\label{fig:onion12}
\end{figure*}

The convergence along light rays in LTB swiss cheese models was very precisely reproduced by the simple ray tracing scheme that included the Doppler convergence. It is here tested whether or not this conclusion also holds true for radial rays in LTB onion models. The results are shown in figure \ref{fig:onion12}.
\newline\indent
The density fields along the rays indicate that the redshift is approximated well by the ray tracing scheme. An exception is in model $O1_{\text{ltb}}$ which has the largest truncation radius of the studied onion models. In figure \ref{fig:onion1} it is seen that the densities along the exact and ray traced rays in model $O1_{\text{ltb}}$ diverge slightly from each other at the highest part of the studied redshift interval. Since the exact and ray traced rays automatically move along the same spatial path, this indicates a deviation between the exact and approximate redshifts. The figures also reveal that the convergence is not reproduced quite as precisely in the LTB onion models as in the swiss cheese case. The deviation between the exact and ray tracing convergences grows with the truncation radius {\em i.e.} the deviation is most prominent in model $O1_{\text{ltb}}$ while it almost vanishes in model $O3_{\text{ltb}}$.
\newline\newline

\begin{figure*}[htb!]
\centering
\subfigure[]{
\includegraphics[scale = 0.75]{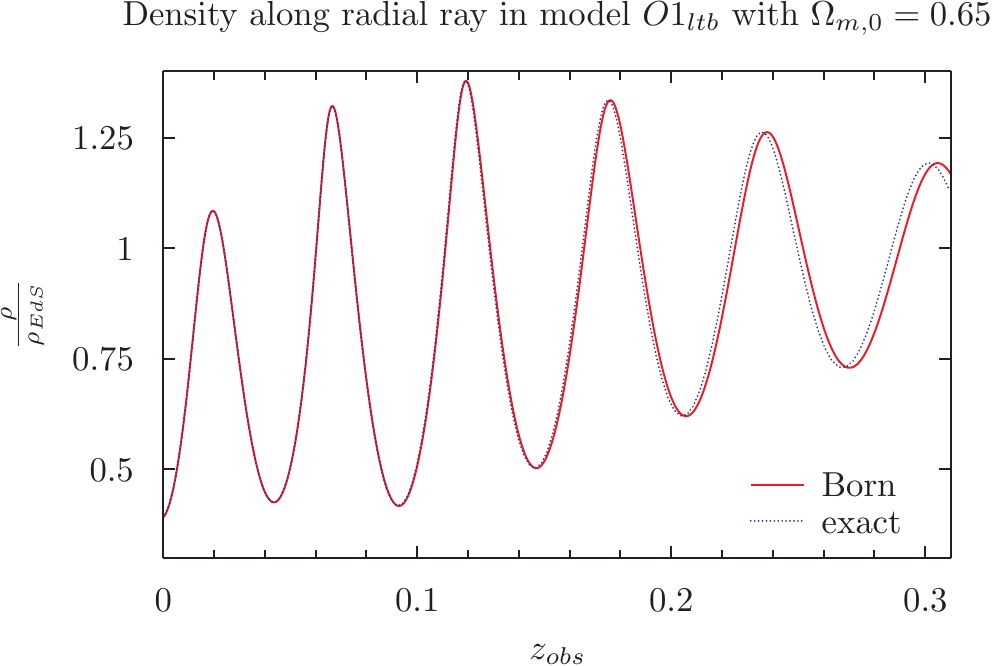}
}
\subfigure[]{
\includegraphics[scale = 0.75]{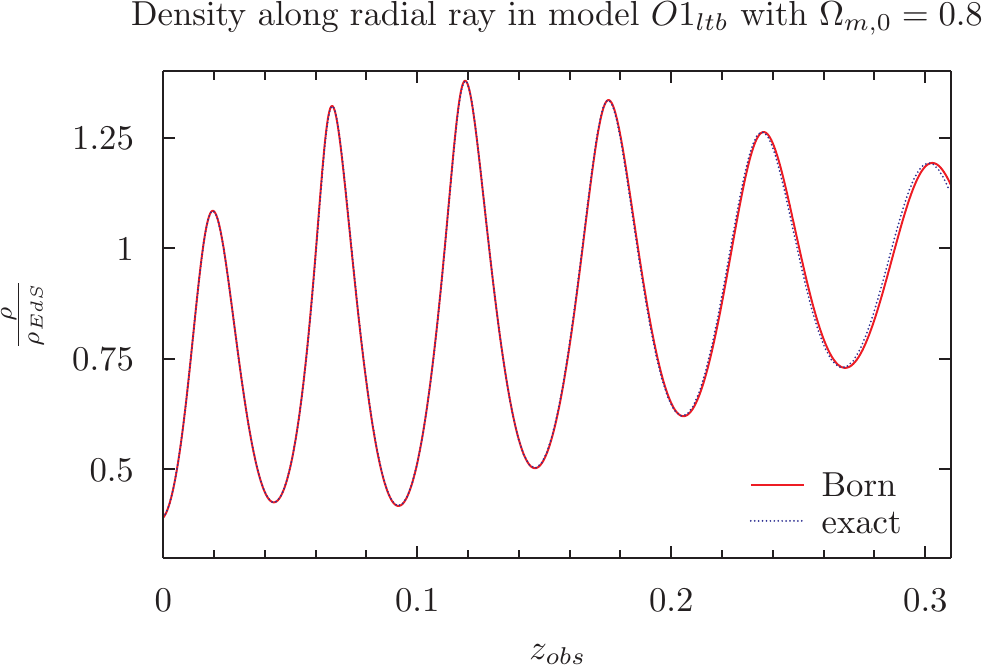}
}\par
\subfigure[]{
\includegraphics[scale = 0.75]{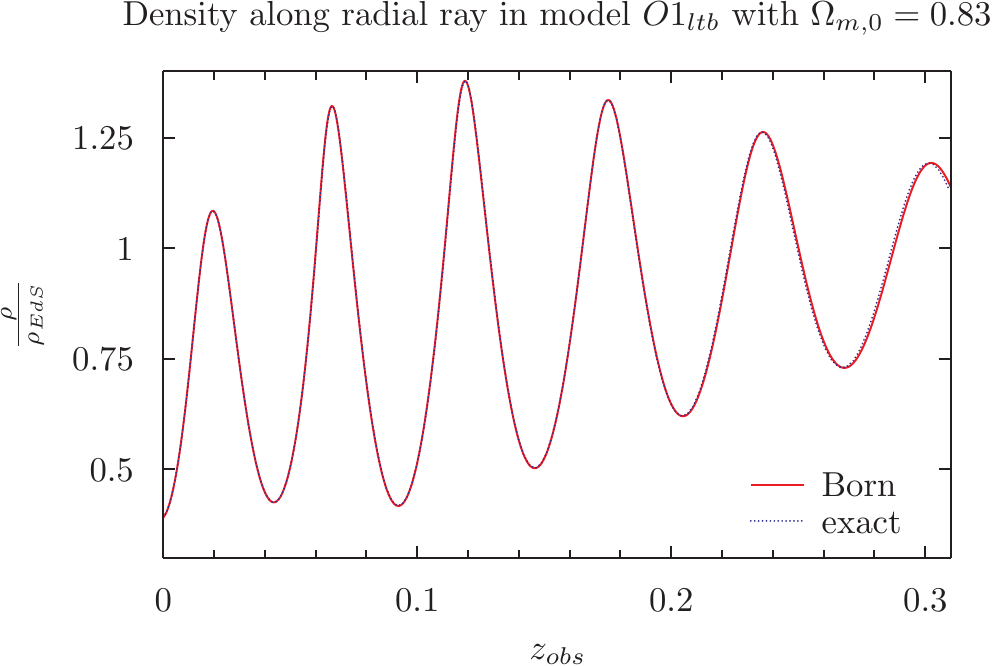}
}
\subfigure[]{
\includegraphics[scale = 0.75]{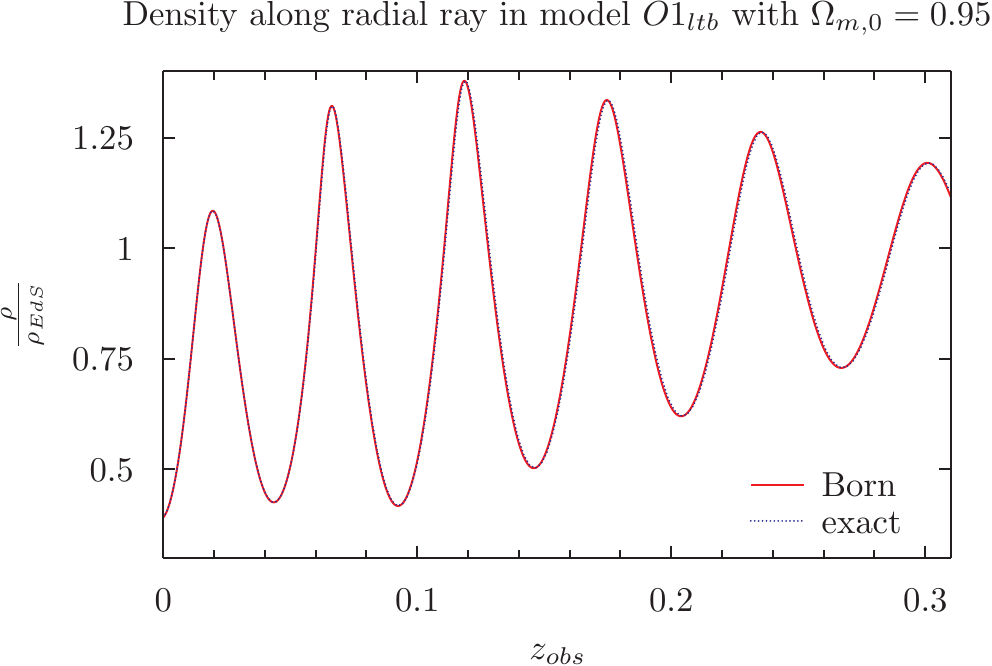}
}\par\subfigure[]{
\includegraphics[scale = 0.75]{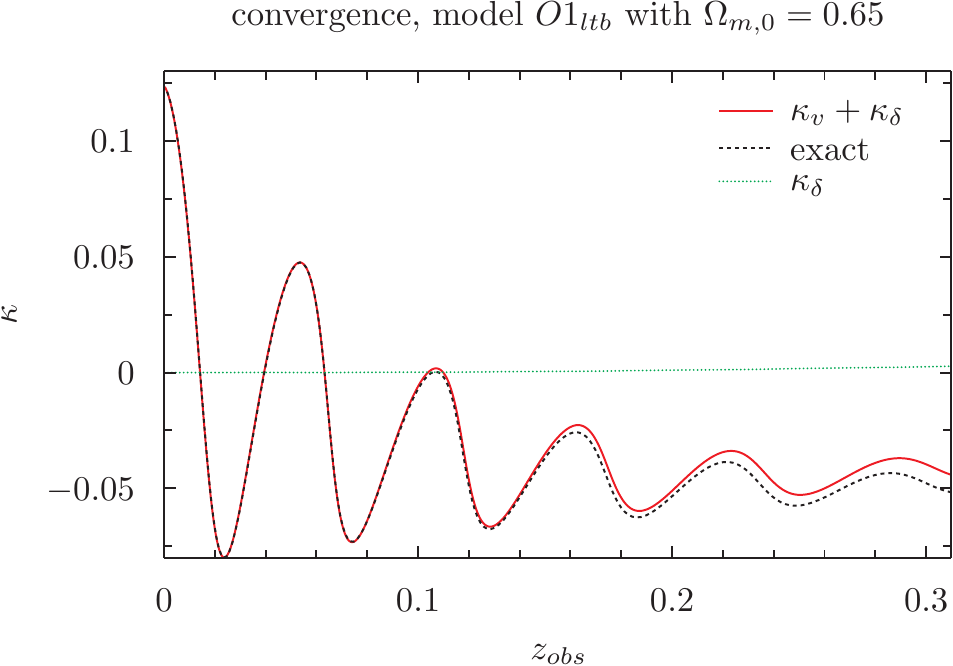}
}
\subfigure[]{
\includegraphics[scale = 0.75]{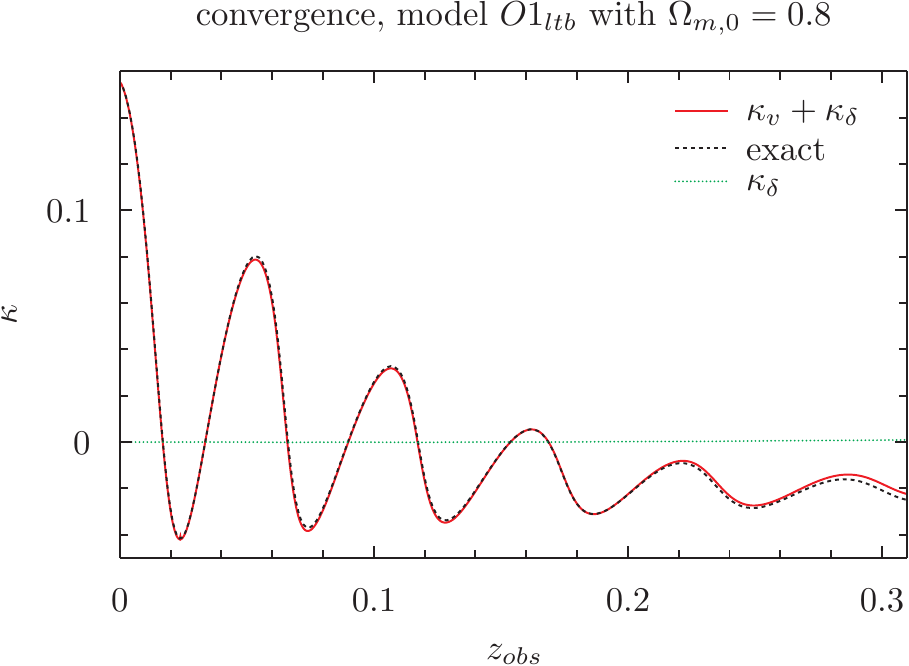}
}\par
\subfigure[]{
\includegraphics[scale = 0.75]{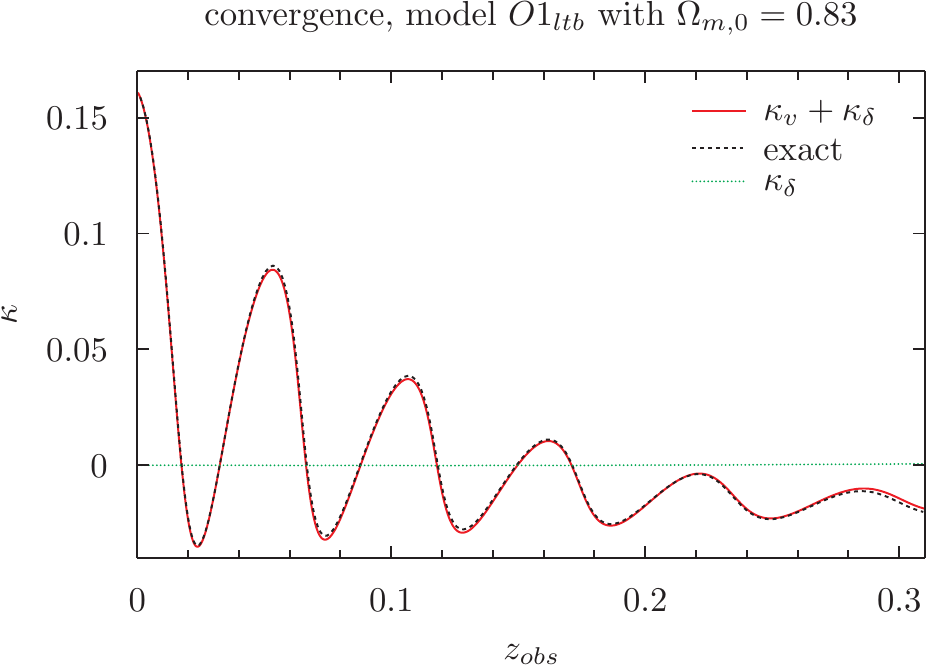}
}
\subfigure[]{
\includegraphics[scale = 0.75]{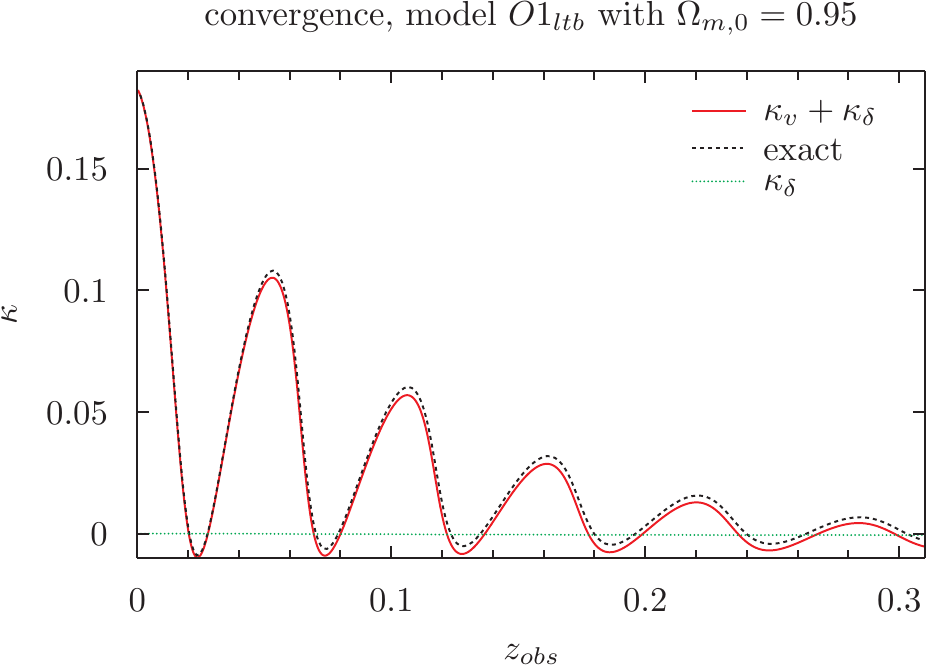}
}
\caption{Density and convergence along exact and ray traced radial rays in the model $O1_{\text{ltb}}$. The ray tracing has been done on different backgrounds as indicated in the individual figure titles. The average density of model $O1_{\text{ltb}}$ along the ray corresponds approximately to $\Omega_{m,0} = 0.8$.}
\label{fig:curved_density}
\end{figure*}

It may be that a better reproduction of the exact results can be obtained if the ray tracing is done on another global background than the EdS model. The theory presented in sections \ref{subsec:mockNbody} and \ref{sec:Ray_tracing_theory} is based on the assumption of a flat background. Modifications needed in order to adapt the equations to curved backgrounds is described in appendix \ref{app:curved} together with discussions and examples regarding the use of curved FLRW backgrounds. Using these modified equations, the ray tracing through model $O1_{\text{ltb}}$ has been done on different backgrounds. The results are shown in figure \ref{fig:curved_density}. As seen, the accuracy of the reproduction of the convergence depends on the background. The best reproduction is obtained when using a background defined by a present time density parameter $\Omega_{m,0}:=\frac{\rho(t_0)}{\rho_{crit,0}} \approx 0.83$. The average density along the ray corresponds to $\Omega_{m,0}\approx 0.8$. Using a background with a small negative curvature clearly yields a better reproduction of the exact results although the reproduction is still not quite as good as along rays in swiss cheese and single void models. It seems likely that this is simply a result of the density fluctuations in the onion model oscillating about a "mean" that changes in the radial direction. In particular, the mean starts below, but converges towards, the critical density. This feature is not particularly realistic and it seems likely that the reproduction of the convergence and redshift would be better in more realistic models or if the studies of the models used here were restricted to redshift intervals where the oscillations occur approximately about the same mean. For instance, figure \ref{fig:curved_density} shows that at redshifts below approximately 0.15, the change in this mean is very small. In accordance with this, the convergence in the same redshift interval is well described by ray tracing with a constant background corresponding to $\Omega_{m,0} = 0.8$.

\subsection{Non-symmetric Szekeres results}\label{sec:onion_sz}

\begin{figure*}[htb!]
\centering
\subfigure[]{
\includegraphics[scale = 0.52]{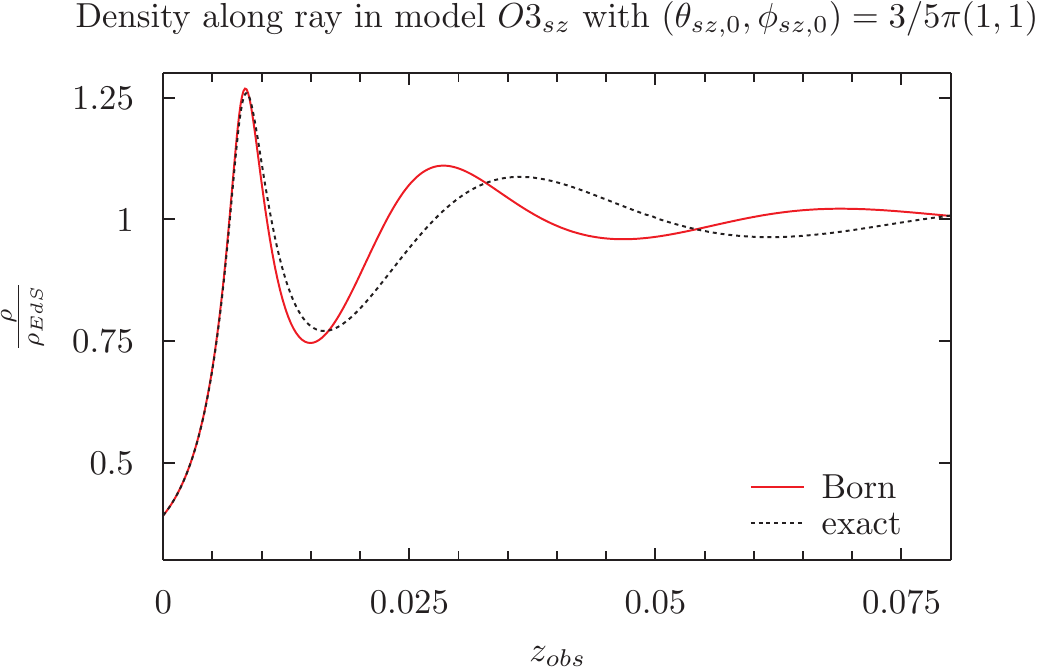}
}
\subfigure[]{
\includegraphics[scale = 0.52]{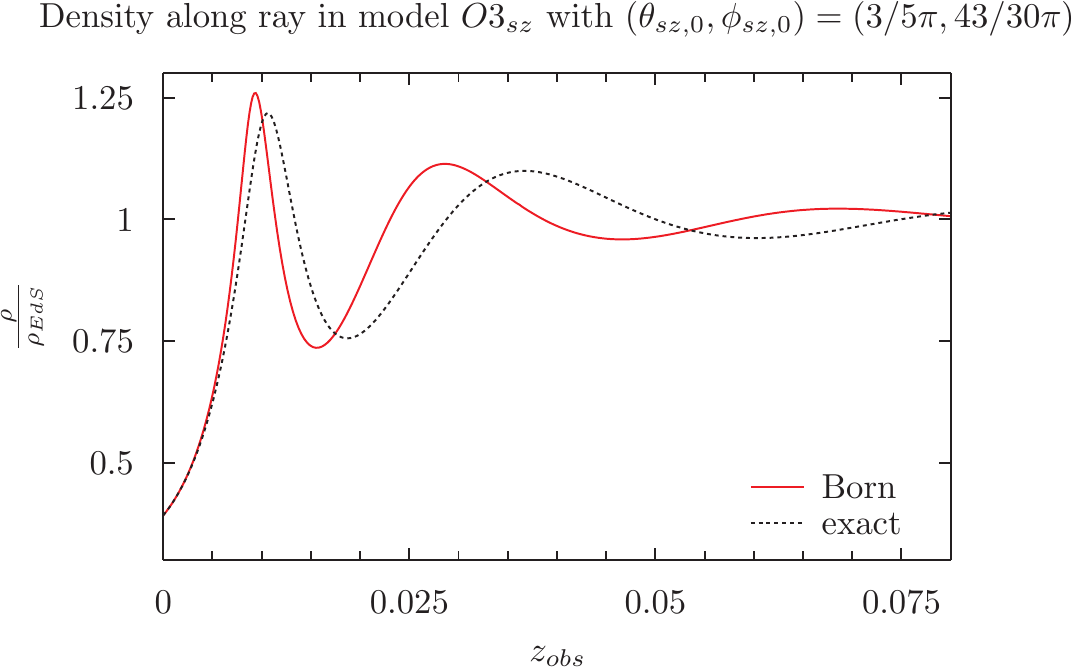}
}
\subfigure[]{
\includegraphics[scale = 0.52]{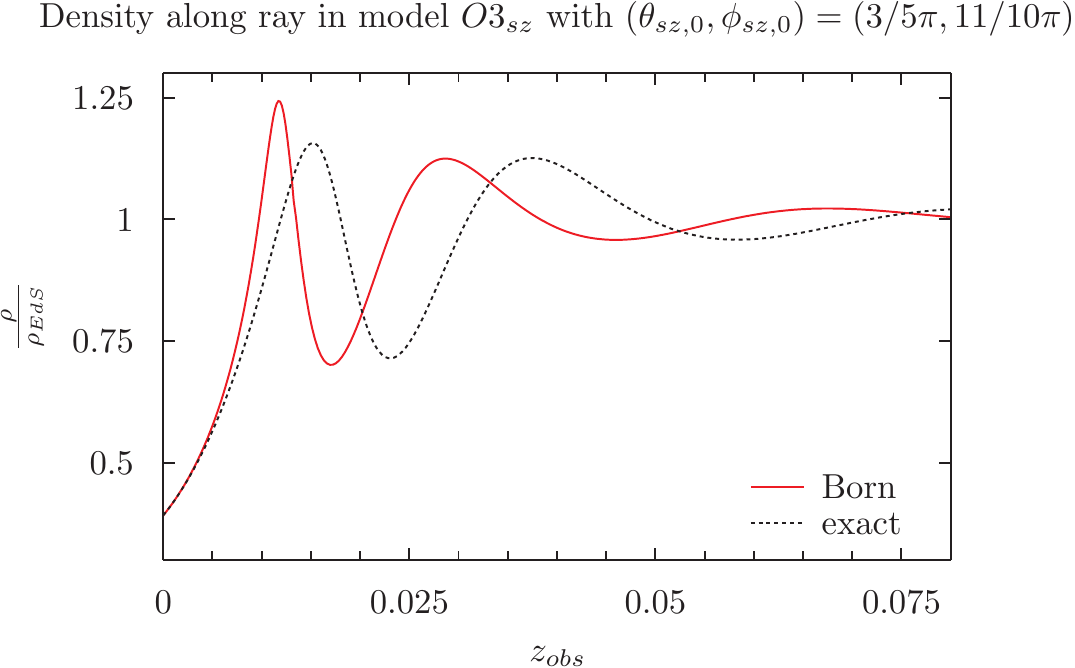}
}\par\subfigure[]{
\includegraphics[scale = 0.52]{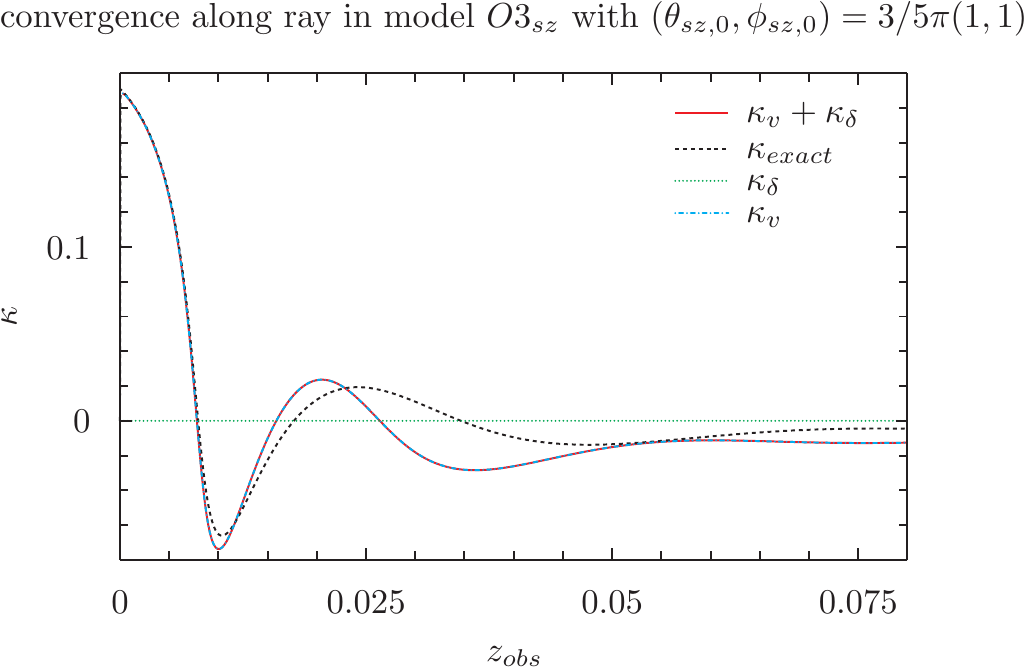}
}
\subfigure[]{
\includegraphics[scale = 0.52]{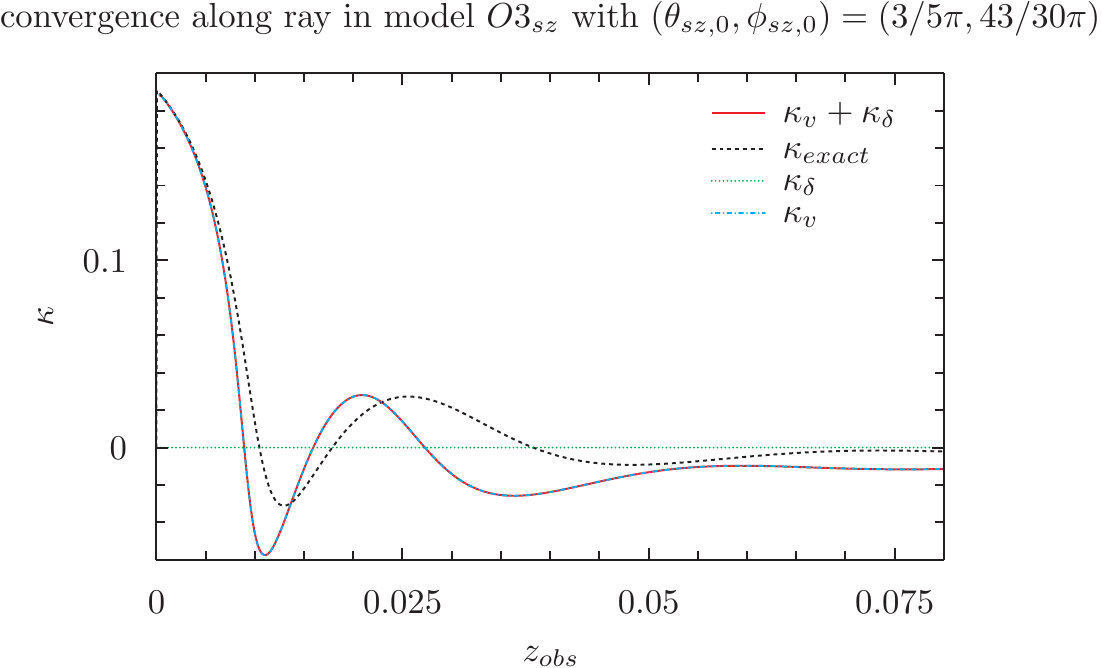}
}
\subfigure[]{
\includegraphics[scale = 0.52]{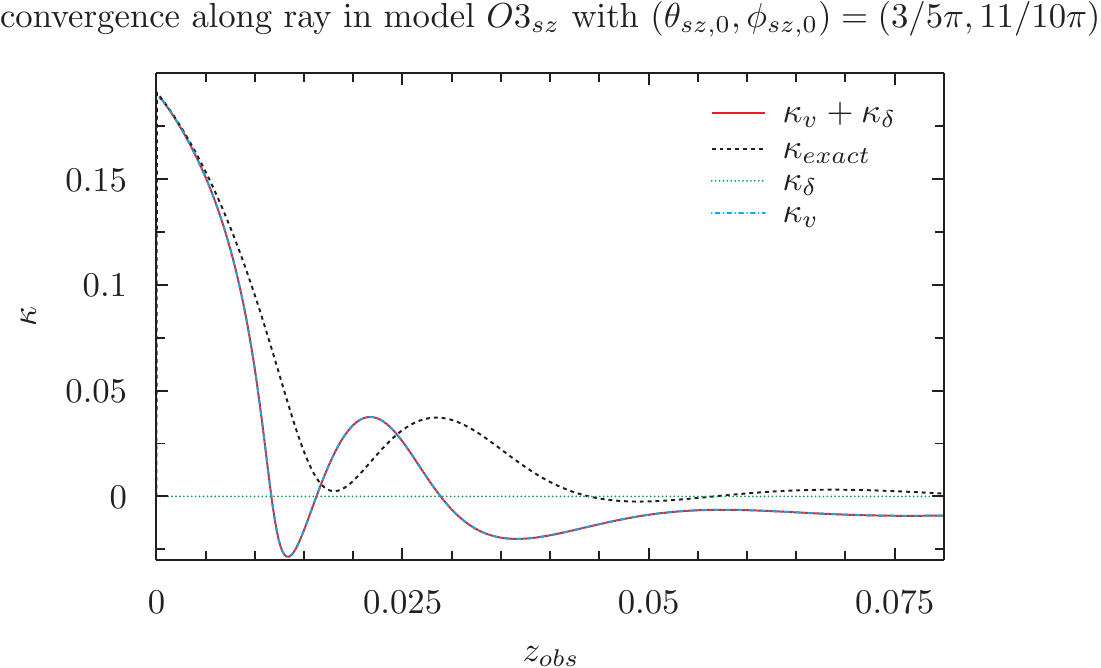}
}
\caption{Density and convergence along exact and ray traced rays in onion model $O3_{\text{sz}}$. Four convergences are shown: Doppler, gravitational, sum of these two, and the exact. The gravitational convergence is negligible in the low-redshift interval studied here, so the lines showing the Doppler convergence and the sum of the Doppler convergence and the gravitational convergence are indistinguishable.}
\label{fig:onion_rho_sz}
\end{figure*}
\begin{figure*}[htb!]
\centering
\subfigure[]{
\includegraphics[scale = 0.52]{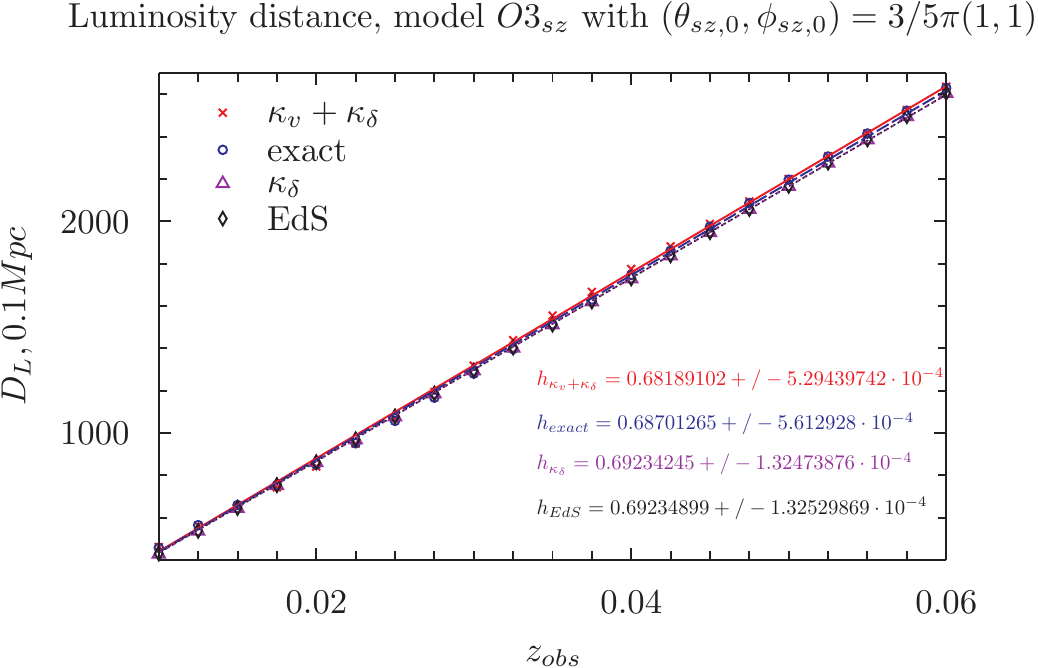}
}
\subfigure[]{
\includegraphics[scale = 0.52]{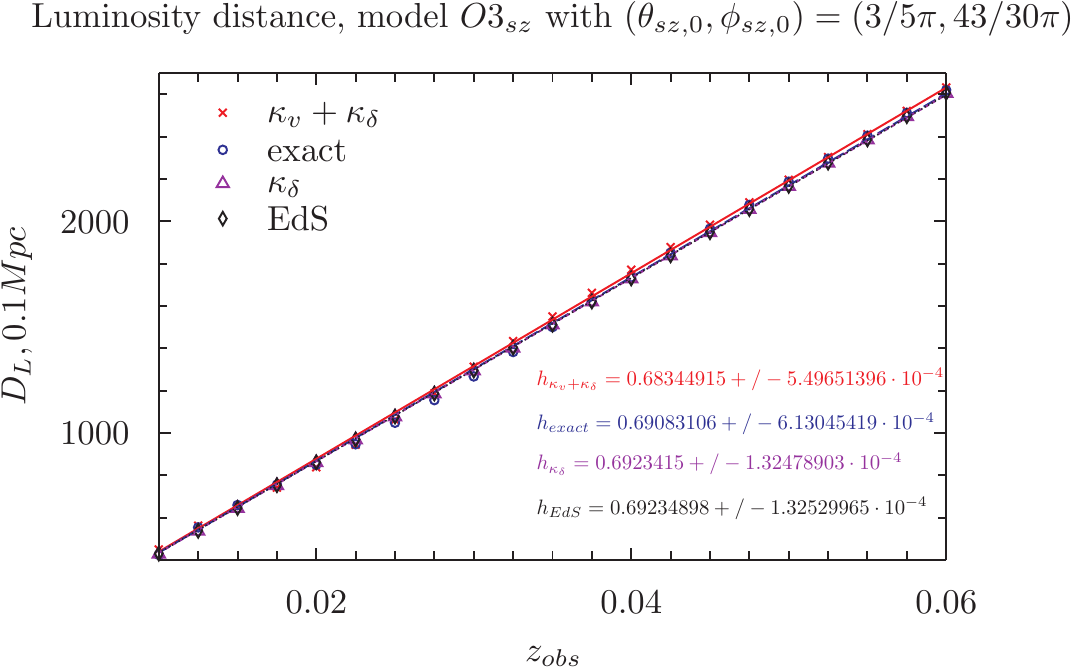}
}
\subfigure[]{
\includegraphics[scale = 0.52]{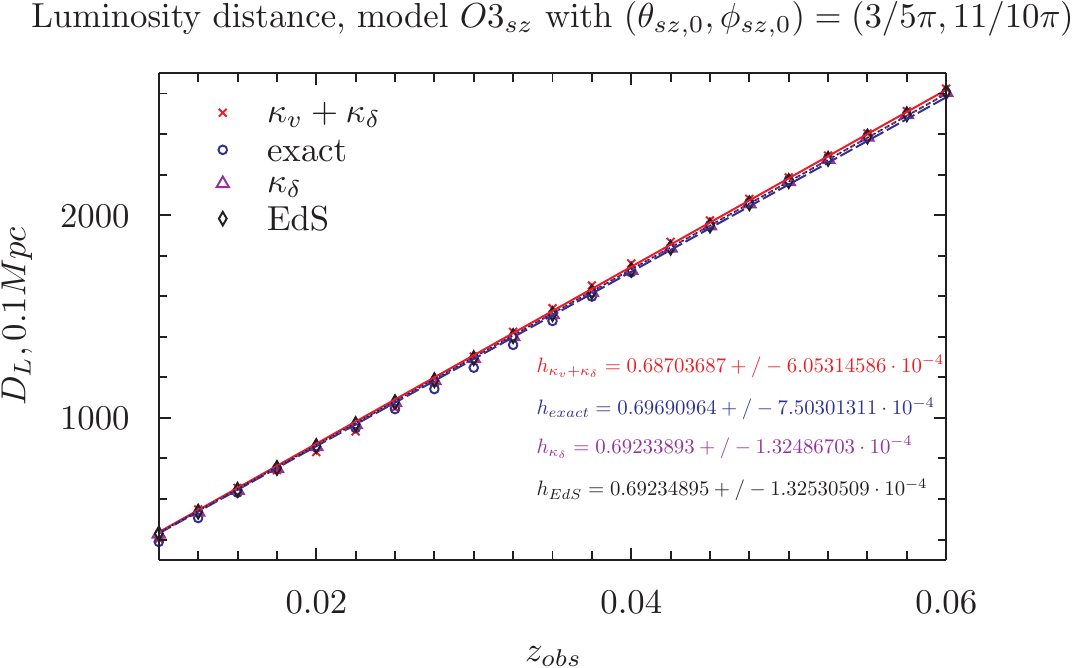}
}
\caption{Luminosity distance plotted as a function of the observed redshift and fitted to the low-redshift FLRW approximation for $D_L(z)$. The results from four luminosity distances are shown corresponding to the exact convergence, the sum of the gravitational and Doppler convergences, the gravitational convergence and the background solution. The results are shown along three different lines of sight for a central observer in onion model $O3_{sz}$.}
\label{fig:onion_h_sz}
\end{figure*}

The convergence is not correctly reproduced along the radial rays in the LTB onion models when an EdS background is used for the ray tracing. Instead, a curved background must be introduced. The background necessary for obtaining accurate reproductions depends on the density fluctuations along the given ray. Even for a central observer in the nonsymmetric Szekeres models, these density fluctuations depend on the line of sight. It is thus not entirely clear which background to choose as the global background for the ray tracing. In addition, as discussed in section \ref{sec:onion_LTB_results}, the introduction of the damping of the density fluctuations makes the models unsuitable for studies where FLRW backgrounds are introduced. However, a somewhat precise reproduction of the exact results is obtained along the entire rays when ray tracing on an EdS background in models $O2_{\text{ltb}}$ and $O3_{\text{ltb}}$. Thus, as an initial consideration, the exact and approximate convergences along individual rays in models $O2_{\text{sz}}$ and $O3_{\text{sz}}$ will be compared here while model $O1_{\text{sz}}$ will not be studied. A more thorough, future study of the effects of nonsymmetric, interacting structures when ray tracing with fixed FLRW backgrounds, could {\em e.g.} be based on the models introduced in \cite{Sussman1, Sussman2}.
\newline\newline
The density distributions and convergences along individual rays initialized at an observer at the origin in model $O3_{\text{sz}}$ are shown in figure \ref{fig:onion_rho_sz}. The results are shown along the three lines of sight defined by $(\theta_{\text{sz},0}, \phi_{\text{sz},0}) = 3/5\pi(1,1)$, $(\theta_{\text{sz},0}, \phi_{\text{sz},0}) = (3/5\pi, 11/10\pi)$ and $(\theta_{\text{sz},0}, \phi_{\text{sz},0}) = (3/5\pi, 43/30\pi)$. The first direction is close to the direction of highest anisotropy while the second is approximately orthogonal to it. The last line of sight is in a direction in between the first two. Note in figure \ref{fig:onion_rho_sz} that the exact and ray traced rays deviate less from each other the closer the line of sight is to the line of highest anisotropy. This is not surprising since the density is close to being symmetrically distributed about this line.
\newline\newline
The three sets of results have been used to estimate the reduced Hubble parameter, $h$, as in earlier sections. The resulting estimates of $h$ and the corresponding uncertainties are shown in figure \ref{fig:onion_h_sz}. The estimate of $h$ is approximately $1 \%$ higher when based on $\kappa_{\text{exact}}$ than when using $\kappa_{v} + \kappa_{\delta}$.
\newline\newline

\begin{figure*}[]
\centering
\subfigure[]{
\includegraphics[scale = 0.7]{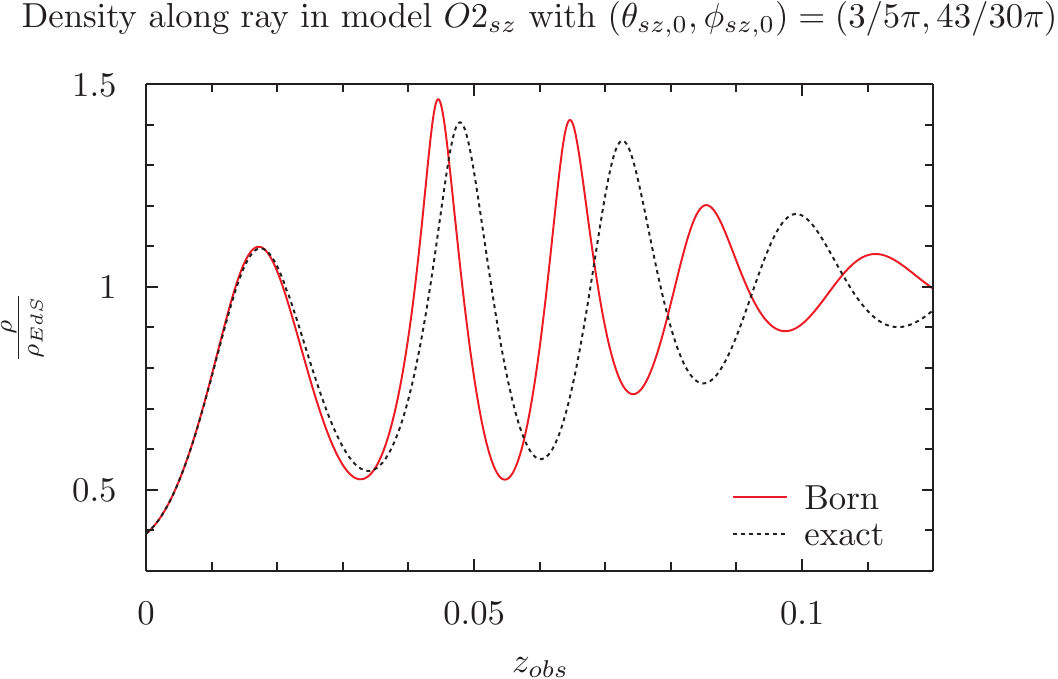}
}
\subfigure[]{
\includegraphics[scale = 0.7]{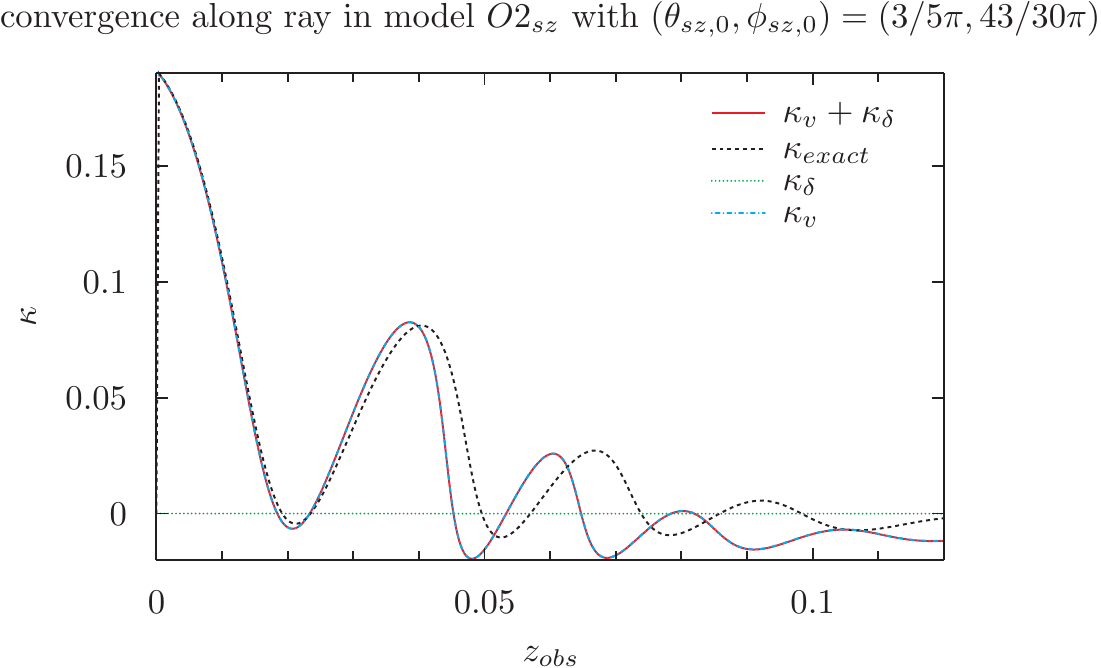}
}
\caption{Density and convergence along exact and ray traced rays in onion model $O2_{\text{sz}}$. Note that $\kappa_{\delta}\approx 0$ so that the lines representing $\kappa_{v}$ and the sum $\kappa_v + \kappa_{\delta}$ are indistinguishable in the figure.}
\label{fig:onion_sz1}
\end{figure*}

\begin{figure*}[]
\centering
\subfigure[]{
\includegraphics[scale = 0.7]{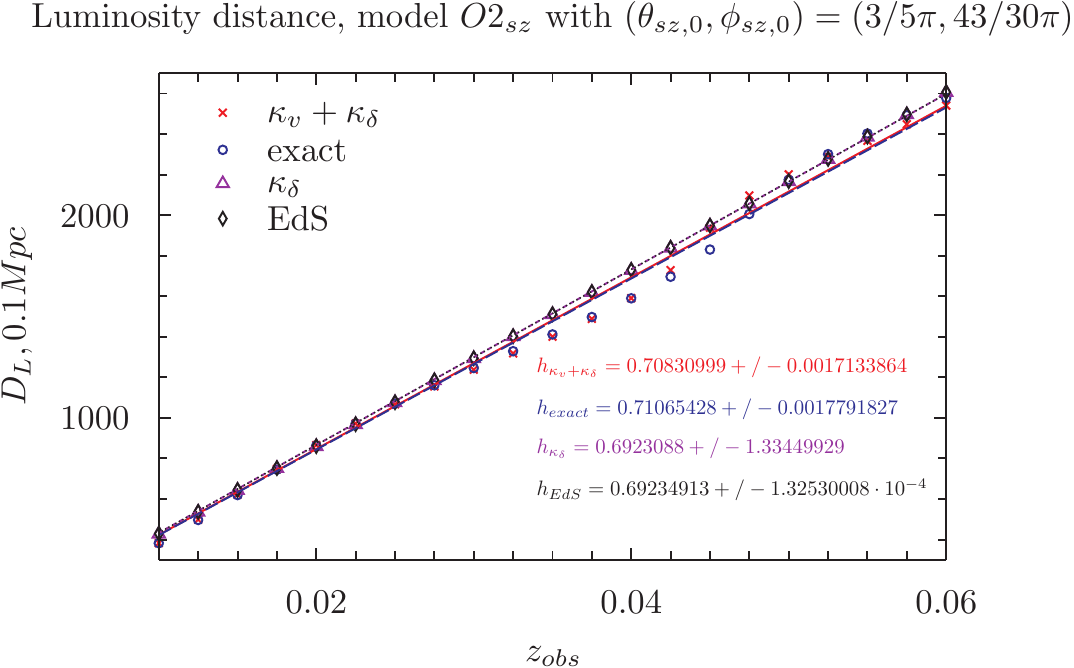}
}
\subfigure[]{
\includegraphics[scale = 0.7]{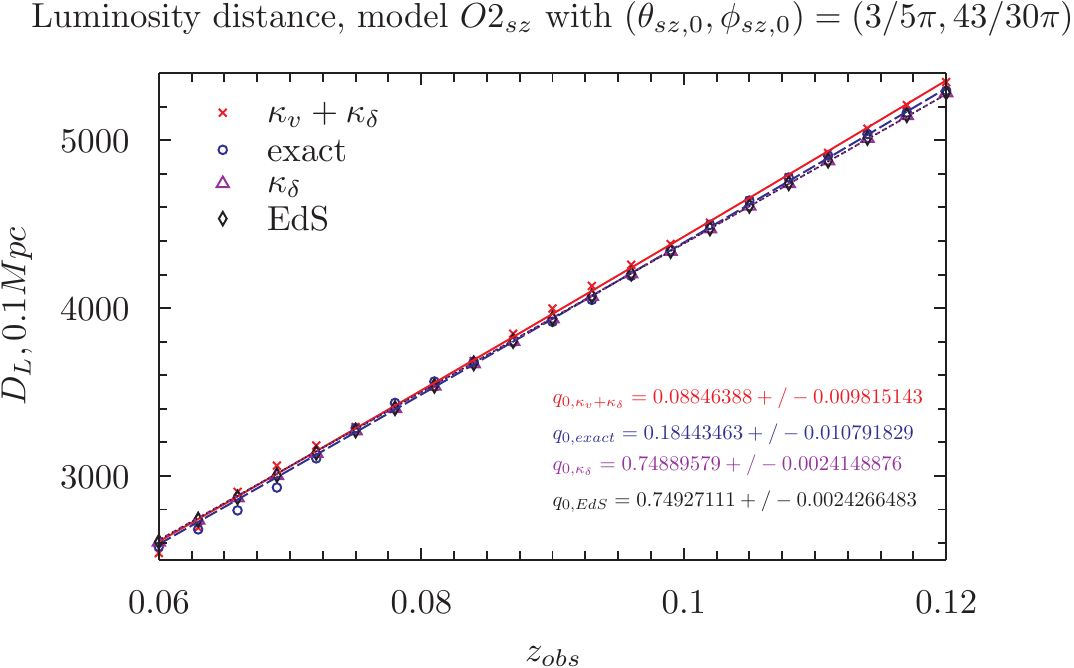}
}
\caption{Luminosity distance plotted as a function of the observed redshift and fitted to low-redshift FLRW approximations for $D_L(z)$. The results from four luminosity distances are shown corresponding to the exact convergence, the sum of the gravitational and Doppler convergences, the gravitational convergence and the background solution.}
\label{fig:onion_hq_sz1}
\end{figure*}

A single ray has been studied in model $O2_{\text{sz}}$. The ray is initialized along the line of sight $(\theta_{\text{sz},0}, \phi_{\text{sz},0}) = (3/5\pi, 43/30\pi)$, {\em i.e.} in between the lines of highest and lowest anisotropy. The convergence and density field along the exact and ray traced rays are compared in figure \ref{fig:onion_sz1}. As with the $O3_{\text{sz}}$ results, the exact and ray traced rays clearly do not move along equivalent spacetime paths.
\newline\indent
In model $O2_{\text{sz}}$, the inhomogeneities in the density distribution are truncated at a large enough distance from the origin for it to be reasonable to make an estimate of the present time deceleration parameter, $q_0$. $h$ is estimated from 200 data points in the redshift interval $0.01<z<0.06$ while $q_0$ is estimated from 200 points along the redshift interval $0.06<z<0.12$. Here, the interval and number of data points used to estimate $q_0$ is much less than in the swiss cheese cases which explains the larger uncertainties in the results shown in figure \ref{fig:onion_hq_sz1}. The estimates of $h$ based on $\kappa_{\text{exact}}$ and the sum $\kappa_v + \kappa_{\delta}$ are quite similar with a difference of approximately 0.3\%. The difference between the two corresponding estimates of $q_0$ are significantly different; the estimate based on $\kappa_{\text{exact}}$ is more than twice the estimate based on the sum $\kappa_{v} + \kappa_{\delta}$. This result is very different from that obtained by studying the swiss cheese models where the different estimates of $q_0$ were primarily due to significant differences in the estimates of $h$. The difference in the estimates obtained here is in line with the convergences along the two rays shown in figure \ref{fig:onion_sz1} where $\kappa_v+\kappa_{\delta}$ is seen to be more negative than $\kappa_{\text{exact}}$. The difference in the $q_0$ estimates may merely be a manifestation of the differences in the density distributions along the exact and ray traced rays. Considering the results found in section \ref{sec:onion_LTB_results} it seems likely that the differences are partially due to the apparent insufficiency of the ray tracing scheme to describe light propagation in onion models.
\newline\newline\newline
Lastly it is noted that light propagation through onion models has its limits regarding how realistic it is; as was seen in figure \ref{fig:szcheese_rho}, real light rays will avoid the most underdense regions of a void. In onion models, the light rays are less free to bend around the most underdense regions since these are spread out over large angular regions.

\section{Conclusions}
By studying both radial and non-radial rays in LTB swiss cheese and single void models it was shown that the sum of the gravitational and Doppler convergences reproduces the exact convergence very precisely. The Doppler convergence is not included in standard ray tracing schemes, but in accordance with other studies, it is here found that the Doppler convergence is a very significant contribution at low redshifts. Its significance for parameter determinations was here illustrated through estimates of the reduced Hubble and deceleration parameters. It was found that the Doppler convergence has a distinct yet rather small effect on especially the uncertainty of these estimates.
\newline\newline
In the non-symmetric Szekeres models, exact rays and ray traced rays do not propagate through equivalent portions of spacetime. The difference between exact and ray traced paths is very dependent on the void profile. The more square-like the void profiles are, the closer the Born approximation comes to being exact. In the case where the void profiles are very cone-like, the densities along the ray traced rays and the exact rays are very different because the exact rays avoid the central part of the void. Even in this case, the consequence for {\em e.g.} determinations of the reduced Hubble and deceleration parameters generally seems to be limited. However, along some particular lines of sight, the differences in the estimates of $h$ based on the exact $D_A$ and that obtained using $\kappa_v+\kappa_\delta$ can become $\sim3-4 \%$. The effect might be more significant for other quantities. It would, for instance, be interesting to see if there are any significant differences between CMB temperature fluctuation maps obtained using exact rays and Born approximated rays.
\newline\indent
Many weak lensing studies employ more sophisticated ray tracing schemes than the one used here. In particular, the Born approximation is often replaced by multiple-lensing techniques where the ray traced rays are bent at a discrete number of lensing planes. It would be interesting to see to what extent multiple-lensing techniques increase the agreement between the exact and ray tracing light paths. The study in \cite{dig_selv} indicates that even if the full Newtonian gauge metric is used to propagate rays, the resulting ray paths will generally be different from the corresponding exact ray paths. It is thus not clear that using  multiple-lensing techniques would result in significantly more correct ray paths than the Born approximation.
\newline\newline
The structure formation in swiss cheese models is very special since each single void structure is inhibited from interacting with the other structures in the swiss cheese. To test if this has any significance for the precision of the ray tracing scheme, onion models have also been studied.
\newline\indent
In the case of LTB onion models, the accuracy and precision of the ray tracing results  are not quite as good as in the case of the LTB single void and swiss cheese models, and for one of the studied models, obtaining a somewhat good reproduction requires introducing a curved FLRW background. The study based on radial rays in LTB onion models indicates that the damping of the density fluctuations in the models compromises the study. Generally though, the results are i overall agreement with those found for the swiss cheese models. A more thorough study of effects of interacting, nonsymmetric, dynamical structures on ray tracing on fixed FLRW backgrounds may be in order and could for instance be done using the models introduced in \cite{Sussman1, Sussman2}.

\section{Acknowledgments}
We thank the anonymous referee for comments and suggestions that have improved the presentation of our work.
\newline\indent
Parts of the numerical work for this project have been done using computer resources from the Center for Scientific Computing Aarhus. The numerical codes written for the work employ the GNU Scientific Library
\footnote{http://www.gnu.org/software/gsl/}
.

\appendix
\section{The Sachs formalism specified to Szekeres models}\label{appendix:sz_Riemann}
In this appendix, the Sachs formalism described in section \ref{sec:Sachs} is specialized to the Szekeres model. The goal is to obtain $\kappa$ along light bundles. This can be achieved by simultaneously solving the geodesic equations, the parallel propagation equations for the components of $E_a^{\mu}$ and the equations for the components of the Jacobi matrix $D$. ODEs for this were obtained in \cite{dallas_first} and later used in \cite{dallas_cheese}. The ODEs given below are the same as these, with the notation used here being slightly different.
\newline\newline
To simplify the notation, the metric components will be written using functions defined according to the line element written as:
\begin{equation}\label{eq:simple_line}
\begin{split}
ds^{2} = -c^2dt^{2} +\frac{\left(\Phi_{,r}(t,r)-\Phi(t,r)\frac{E_{,r}(r,p,q)}{E(r,p,q)}\right)^2}{1-k(r)}dr^2 +\\
\frac{\Phi(t,r)^2}{E(r,p,q)^2}(dp^2+dq^2) \\
 = -c^2dt^2 + R(t,r,p,q)dr^2 + F(t,r,p,q)\left(dp^2 + dq^2 \right) 
\end{split}
\end{equation}
\newline\newline
The geodesic equations corresponding to this line element are:
\begin{equation}
\dot k^t = -\frac{1}{2c^2}\left( R_{,t}(k^r)^2 + F_{,t}\left[ (k^p)^2 + (k^q)^2 \right]  \right) 
\end{equation}
\begin{equation}
\dot k^r = \frac{1}{2R}\left(-2\dot Rk^r + R_{,r}(k^r)^2 + F_{,r}\left[ (k^p)^2 + (k^q)^2 \right]  \right) 
\end{equation}
\begin{equation}
\dot k^p = \frac{1}{2F}\left(-2\dot F k^p + R_{,p}(k^r)^2 + F_{,p}\left[ (k^p)^2 + (k^q)^2 \right]  \right) 
\end{equation}
\begin{equation}
\dot k^q = \frac{1}{2F}\left(-2\dot F k^q + R_{,q}(k^r)^2 + F_{,q}\left[ (k^p)^2 + (k^q)^2 \right]  \right) 
\end{equation}
The parallel propagated screen space basis vectors, $E_a^{\mu}$, are needed since they enter into the ODEs for $D$. ODEs describing this parallel propagation are simply obtained by substituting $\frac{1}{2}\left(k^{\mu}E_a^{\nu} + k^{\nu}E_a^{\mu}  \right) $ for $k^{\mu}k^{\nu}$ in the geodesic equations with the resulting equations being:
\begin{equation}
\dot E^t_a = -\frac{1}{2c^2}\left(R_{,t}E_a^rk^r + F_{,t}\left[E_a^pk^p + E_a^qk^q \right]  \right) 
\end{equation}
\begin{equation}
\begin{split}
 2R\dot E_a^r = -R_{,t}\left[E_a^tk^r + E_a^rk^t \right] - R_{,r}k^rE_a^r 
-\\ R_{,p}\left[E_a^rk^p + E_a^pk^r \right] -
R_{,q}\left[ E_a^rk^q +E_a^qk^r  \right] + F_{,r}\left[E_a^pk^p + E_a^qk^q \right] 
\end{split}
\end{equation}
\begin{equation}
\begin{split}
 2F\dot E_a^p = -F_{,t}\left[E_a^tk^p + E_a^pk^t \right] + R_{,p}k^rE_a^r -\\
F_{,r}\left[k^rE_a^p + k^pE_a^r \right]-
F_{,p}E_a^pk^p + F_{,p}E_a^qk^q -F_{,q}\left[E_a^pk^q + E_a^qk^p \right] 
\end{split}
\end{equation}
\begin{equation}
\begin{split}
 2F\dot E_a^q = -F_{,t}\left[E_a^tk^q + E_a^qk^t \right] + R_{,q}k^rE_a^r -\\
 F_{,r}\left[k^rE_a^q + k^qE_a^r \right]
-F_{,p}\left[k^pE_a^q + k^qE_a^p \right] - F_{,q}k^qE_a^q + F_{,q}k^pE_a^p 
\end{split}
\end{equation}
The last ODEs needed are those for the components of $D$ which by equation (\ref{D_dot}) are seen to be:
\begin{equation}
\ddot D_{11} = \left(\mathbf{R}-Re(\mathbf{F}) \right) D_{11} + Im(\mathbf{F})D_{12}
\end{equation}
\begin{equation}
\ddot D_{12} = \left( \mathbf{R}-Re(\mathbf{F})\right)  D_{12} + Im(\mathbf{F})D_{22}
\end{equation}
\begin{equation}
\ddot D_{21} =  Im(\mathbf{F})D_{11} + \left( \mathbf{R} + Re(\mathbf{F}) \right) D_{21}
\end{equation}
\begin{equation}
\ddot D_{22} = Im(\mathbf{F})D_{12} + \left( \mathbf{R} + Re(\mathbf{F}) \right) D_{22}
\end{equation}
Explicit expressions for $\mathbf{R}$ and $\mathbf{F}$ need to be obtained. The expression for $\mathbf{R}$ is easy to obtain; the Einstein equation can be written as $\mathcal{R}_{\mu\nu} = \frac{1}{2}\mathcal{R}g_{\mu\nu} + \frac{8\pi G_N}{c^4}T_{\mu\nu}$. Combining this with the null condition $k^{\mu}k_{\mu} = 0$ and noting that the Szekeres spacetime is given in a comoving spacetime foliation such that $u^{\mu} = (1,0,0,0)$, this leads to:
\begin{equation}
\mathbf{R} =-4\pi G_N\rho(k^t)^2
\end{equation}

To obtain an explicit expression for $\mathbf{F}$, the Riemann components are needed. Only a few of the components of the Riemann tensor are non-vanishing. These are shown at the end of this appendix in subsection \ref{appendix_subsection:sz_Riemann}. With the knowledge of the exact expressions of the Riemann components, it is seen that $\mathbf{F}$ can be written as:
\begin{widetext}
\begin{equation}
\begin{split}
2\mathbf{F} =
\mathcal{R}_{trtr}\left( 2(\epsilon^*)^t(\epsilon^*)^rk^tk^r -
\left[ (\epsilon^*)^r \right]^2(k^t)^2 - \left[ (\epsilon^*)^t \right]^2(k^r)^2 \right) + \\
\mathcal{R}_{tptp}\left(2 (\epsilon^*)^tk^t \left\lbrace (\epsilon^*)^pk^p + (\epsilon^*)^qk^q \right\rbrace -(k^t)^2\left\lbrace\left[(\epsilon^*)^p \right]^2 + \left[(\epsilon^*)^q \right]^2   \right\rbrace -\left[(\epsilon^*)^t \right]^2\left\lbrace  (k^p)^2 + (k^q)^2 \right\rbrace       \right) + \\
\mathcal{R}_{rprp}\left( 2(\epsilon^*)^r(\epsilon^*)^p k^r k^p - \left[(\epsilon^*)^r \right]^2(k^p)^2 - \left[(\epsilon^*)^p \right]^2(k^r)^2 \right)  +\\
\mathcal{R}_{rqrq}\left(2(\epsilon^*)^r(\epsilon^*)^qk^qk^r - \left[ (\epsilon^*)^r \right]^2(k^q)^2 - \left[(\epsilon^*)^q\right]^2(k^r)^2  \right)+\\
\mathcal{R}_{pqpq}\left(2(\epsilon^*)^p(\epsilon^*)^qk^pk^q - \left[(\epsilon^*)^p \right]^2(k^q)^2 - \left[(\epsilon^*)^q \right]^2(k^p)^2   \right)  
\end{split}
\end{equation}

In order to solve the set of ODEs given above, initial conditions need to be determined. Most of the geodesics are initialized as radial. Since $u^{\mu} =(1,0,0,0)$, this implies that the initial conditions for the screen space vectors are $E_1^{\mu} = \frac{1}{\sqrt{F}}(0,0,1,0)$ and $E_2^{\mu} = \frac{1}{\sqrt{F}}(0,0,0,1)$. In the more general case of rays that are not initially radial, the screen space basis unit vectors are initialized according to:

\begin{equation}
\begin{split}
E_1^{\mu} \propto \left( 0,\frac{\sqrt{F}}{\sqrt{R}}\sqrt{(k^p_0)^2 + (k^q_0)^2}, -\frac{\sqrt{R}}{\sqrt{F}}\frac{k^r_0k^p_0}{\sqrt{(k^p_0)^2 + (k^q_0)^2}}, -\frac{\sqrt{R}}{\sqrt{F}}\frac{k^r_0k^q_0}{\sqrt{(k^p_0)^2 + (k^q_0)^2}}\right) 
\end{split}
\end{equation}
\begin{equation}
E_2^{\mu} \propto \left( 0,0,\frac{1}{\sqrt{F}}\frac{k^q_0}{\sqrt{(k^p_0)^2 + (k^q_0)^2}},-\frac{1}{\sqrt{F}}\frac{k^p_0}{\sqrt{(k^p_0)^2 + (k^q_0)^2}} \right) 
\end{equation}
\end{widetext}

These expressions are obtained by requiring that the two basis vectors are orthogonal to the velocity field and $k^{\mu}$.
\newline\newline
In order for $\sqrt{|\det(D)|}$ to be the angular diameter distance, $k^t$ must be initialized as $k^t_0 =  -\frac{1}{c}$. The initial values of $k^r, k^p,k^q$ are arbitrary though $k^{\mu}$ must fulfill the null-condition.
\newline\indent
By differentiating the equation $\xi = D\dot{\xi_0}$ it can be seen that $D$ must be initialized as the zero-matrix and $\dot D$ as the identity matrix.

\begin{widetext}
\subsection{Riemann components of Szekeres models} \label{appendix_subsection:sz_Riemann}
The Riemann components are computed using the definition $\mathcal{R}_{\alpha\beta\mu\nu} = \frac{1}{2}\left( g_{\beta\mu,\nu\alpha} - g_{\alpha\mu,\nu\beta} + g_{\alpha\nu,\mu\beta} - g_{\beta\nu,\mu\alpha}\right)  - g^{\sigma\gamma} \left( \Gamma_{\sigma\alpha\mu}\Gamma_{\gamma\beta\nu} - \Gamma_{\sigma\alpha\nu}\Gamma_{\gamma\beta\mu}  \right)  $. This equation contains the covariant Christoffel symbols symmetric in their two last indices, {\em i.e.}:
\begin{equation}
\begin{split}
\Gamma_{\alpha\beta}^{\gamma} = \frac{1}{2}g^{\gamma\sigma}\left(g_{\alpha\sigma,\beta} + g_{\beta\sigma,\alpha} - g_{\alpha\beta,\sigma} \right) \\
=> \Gamma_{\gamma\alpha\beta} = \frac{1}{2}\left(g_{\alpha\gamma,\beta} + g_{\beta\gamma,\alpha} - g_{\alpha\beta,\gamma} \right)
\end{split}
\end{equation}
The non-vanishing covariant Christoffel symbols of the quasi-spherical Szekeres models with the metric corresponding to the line element of equation (\ref{eq:simple_line}) are:

\begin{equation}
\begin{split}
\Gamma_{rrt} =\Gamma_{rtr} =  -\Gamma_{trr} =\frac{R_{,t}}{2}\\
\Gamma_{rrr} = \frac{R_{,r}}{2}\\
\Gamma_{rrp} = \Gamma_{rpr}=-\Gamma_{prr} =\frac{R_{,p}}{2}\\
\Gamma_{rrq} = \Gamma_{rqr}=-\Gamma_{qrr} =\frac{R_{,q}}{2}\\
\Gamma_{ppt} = \Gamma_{ptp} = -\Gamma_{tpp} = \Gamma_{qqt} = \Gamma_{qtq} = -\Gamma_{tqq} =\frac{F_{,t}}{2}\\
\Gamma_{ppr} = \Gamma_{prp} = -\Gamma_{rpp} = \Gamma_{qqr} = \Gamma_{qrq} = -\Gamma_{rqq} =\frac{F_{,r}}{2}\\
\Gamma_{qqq} =\Gamma_{ppq} = \Gamma_{pqp} = -\Gamma_{qpp} = \frac{F_{,q}}{2}\\
\Gamma_{ppp} =\Gamma_{qqp} = \Gamma_{qpq} = -\Gamma_{pqq} = \frac{F_{,p}}{2}\\
\end{split}
\end{equation}
The corresponding non-vanishing Riemann component are:
\begin{equation}
\begin{split}
\mathcal{R}_{trtr} = \mathcal{R}_{rtrt}=  -\mathcal{R}_{trrt} = -\mathcal{R}_{rttr}=  -\frac{1}{2}R_{,tt} + \frac{R_{,t}^2}{4R}\\
\mathcal{R}_{tptp} = \mathcal{R}_{ptpt}=- \mathcal{R}_{tppt} = -\mathcal{R}_{pttp} = -\frac{1}{2}F_{,tt}+ \frac{F_{,t}^2}{4F}\\
\mathcal{R}_{tqtq} =\mathcal{R}_{qtqt} =- \mathcal{R}_{tqqt} = -\mathcal{R}_{qttq} = -\frac{1}{2}F_{,tt}+ \frac{F_{,t}^2}{4F}\\
\mathcal{R}_{rprp} = \mathcal{R}_{prpr}=-\mathcal{R}_{prrp} = - \mathcal{R}_{rppr} = -\frac{1}{2}\left(R_{,pp} + F_{,rr} \right) + \frac{1}{4c^2}R_{,t}F_{,t} + \frac{1}{4R}\left( R_{,r}F_{,r} + R_{,p}^2 \right) + \frac{1}{4F}\left( R_{,p}F_{,p} + F_{,r}^2 - R_{,q}F_{,q}\right)\\ 
\mathcal{R}_{rqrq} =\mathcal{R}_{qrqr} =-\mathcal{R}_{qrrq} = -\mathcal{R}_{rqqr} = -\frac{1}{2}\left( R_{,qq} +F_{,rr}  \right) + \frac{1}{4c^2}R_{,t}F_{,t} + \frac{1}{4R}\left(R_{,r}F_{,r} + R_{,q}^2 \right) + \frac{1}{4F}\left(-R_{,p}F_{,p} + R_{,q}F_{,q} + F_{,r}^2 \right)\\
\mathcal{R}_{pqpq} = \mathcal{R}_{qpqp} = -\mathcal{R}_{pqqp} = -\mathcal{R}_{qppq}  = -\frac{1}{2}\left( F_{,qq} + F_{,pp} \right) + \frac{1}{4c^2}F_{,t}^2 - \frac{1}{4R}F_{,r}^2 + \frac{1}{2F}\left( F_{,p}^2 + F_{,q}^2  \right)\\
\end{split}
\end{equation}
\end{widetext}

\section{The Sachs formalism combined with the Born approximation} \label{app:sachs_born}
The ray tracing scheme described in section \ref{sec:Ray_tracing_theory} reproduces the exact results very precisely in the special case where the exact rays are radial rays in an LTB model. Unfortunately, that ray tracing scheme is, in practice, difficult to use for studying rays initialized as non-radial in the local single void models of the swiss cheese. Thus, in order to test if the good reproduction is purely a consequence of the exact rays themselves being radial, the Sachs formalism has been used to devise a simple scheme for "ray tracing" non-radial rays in LTB models. In this ray tracing scheme, rays are propagated according to the background null-geodesic equations, {\em i.e.} in accordance with the Born approximation.
\newline\indent
The convergence can be obtained using the Sachs formalism described in section \ref{sec:Sachs}. Using $\mathcal{R}_{\mu\nu} = \frac{1}{2}\mathcal{R}g_{\mu\nu} + \frac{8\pi G_N}{c^4}T_{\mu\nu}$, it is easily seen that $\mathbf{R}=-\frac{1}{2}\mathcal{R_{\mu\nu}}k^{\mu}k^{\nu} =  -\frac{4\pi G_N}{c^2V^2}\rho\left(-c^2k^t + a^2v_rk^r \right)^2$. Since the ray tracing will only be done through single void models, the gravitational convergence is, however, negligible. This implies that the result will not depend on whether it is the background density or mapped Szekeres density that is used when computing $\mathbf{R}$. In fact, the term $a^2v_rk^r$ can also be neglected, and $\mathbf{F}$ is set equal to zero, corresponding to its lowest order value. The only place where the computations deviate from background computations is thus when computing the observable redshift which corresponds to computing the Doppler convergence.
\newline\indent
In order for $\sqrt{|\det(D)|}$ to be the angular diameter distance, $k^t$ must be initialized as $k^t_0 =  -\frac{1}{c}$ while the initial values of $k^r,k^p,k^q$ are arbitrary though the null-condition for $k^{\mu}$ must be fulfilled. $D$ must be initialized as the zero-matrix and $\dot D$ as the identity matrix.
\newline\indent
The formalism described here has only been used with $k^q$ initially set to zero. In this case, the initial conditions of the screen space unit vectors are $E_1^{\mu} \propto (\frac{v_r}{c^2}, \frac{1}{a^2},-\tilde E^2\frac{\frac{v_r}{c}+k^r}{a^2r^2k^p},0)$ and $E_2^{\mu} = (0,0,0,\frac{\tilde E^2}{a^2r^2})$, where $\tilde E$ is a metric function defined in appendix \ref{app:curved}. These initial conditions are obtained by using the fact that the screen space basis vectors are orthogonal to each other and to $u_{\mu}\approx \frac{c}{V}(-c^2(1+2\psi), a^2(1-2\psi)v_r,0,0)$ and $k^{\mu}$. Once again, it is actually sufficient only to keep the background terms.
\newline\newline
At low redshifts, the scheme just described has the same precision as the ray tracing scheme described in section \ref{sec:Ray_tracing_theory}; it uses the Born approximation and computes the Doppler convergence. Thus, if the ray tracing scheme described here can accurately reproduce the path and corresponding convergence of a non-radial light ray in an LTB model, then the ray tracing scheme of section \ref{sec:Ray_tracing_theory} will also be able to. As shown in section \ref{sec:nonradial}, this is in fact the case.

\section{Curved FLRW backgrounds}\label{app:curved}
The truncated onion models shown in figure \ref{fig:onion_densities} all converge to an EdS background at large $r$. The structures at smaller values of the radial coordinate are however not separated from each other by artificial background patches but are free to interact. This implies that the structures will generally not be locally compensating. Thus, ray tracing through the inhomogeneous region of the onion models may require the use of a different FLRW model than the EdS model as the background. Some of the equations of sections \ref{subsec:mockNbody} and \ref{sec:Ray_tracing_theory} require the background FLRW model to be flat and thus need to be adapted before they can be used together with a curved background. The necessary modifications are summarized below together with examples.

\subsection{Mock N-body data}
Disregarding notation, the map in equation (\ref{stereo_map}) does not require the background to be flat. With a general background, the right-hand side should refer to the given FLRW model of choice and not specifically the EdS model, {\em i.e.}:

\begin{equation}\label{general_map}
\begin{split}
\tilde t:=\tilde t_{\text{sz}} = \tilde t_{\text{flrw}}\\
\tilde p:= \tilde p_{\text{sz}} = \tilde p_{\text{flrw}}\\
\tilde q:=\tilde q_{\text{sz}} = \tilde q_{\text{flrw}}\\
\int_{0}^{\tilde r_{\text{sz}}}dr_{\text{sz}}\sqrt{g_{rr,\text{sz}}} = \int_{0}^{\tilde r_{\text{flrw}}}dr_{\text{flrw}}\sqrt{g_{rr,\text{flrw}}}
\end{split}
\end{equation}
The coordinate system used for the FLRW model should be the one corresponding to the FLRW limit of the Szekeres spacetime. The coordinate choice should thus correspond to the FLRW line element written as:
\begin{equation}
ds^2 = -c^2dt^2 + a^2\left( \frac{dr^2}{1-Kr^2/R_0^2} + \frac{r^2}{\tilde E^2}(dp^2 + dq^2)\right) 
\end{equation}
$K \in \left\lbrace -1,0,1 \right\rbrace $ gives the sign of the spatial curvature while $R_0  =\frac{c}{H_0\sqrt{|\Omega_{m,0}-1}|}$ is the curvature radius. The function $\tilde E$ is given by $\tilde E = \frac{1}{2}\left( p^2 + q^2 +1 \right) $. 
\newline\newline
The velocity field cannot be obtained using equation (\ref{eq:velocity_field}); the right-hand side of that equation is only correct for a flat FLRW model. The velocity field can be obtained through the same considerations as those that lead to equation (\ref{eq:velocity_field}):
\newline
The point identification map of equation (\ref{general_map}) is time dependent. Hence, keeping the spatial Szekeres position constant while letting the time coordinate vary will lead to a mapping into spatially different FLRW points. This change in the spatial FLRW coordinates as a function of the time coordinate leads to a velocity field that can be computed {\em e.g.} by using a finite difference formula.

\subsection{Ray tracing}\label{app:Doppler_curved}

\begin{figure*}[htb!]
\centering
\subfigure[]{
\includegraphics[scale = 0.8]{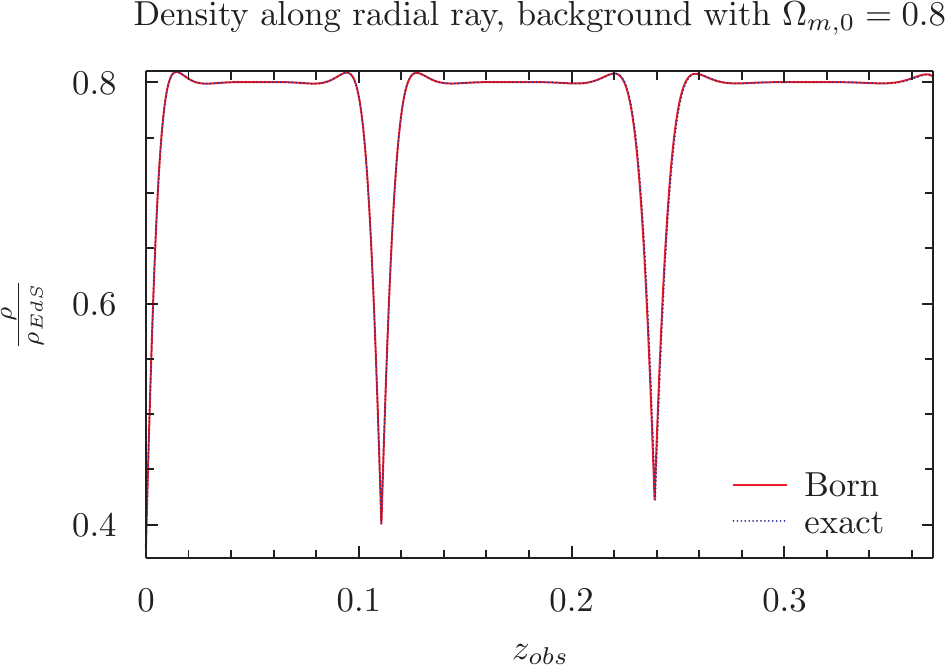}
}
\subfigure[]{
\includegraphics[scale = 0.8]{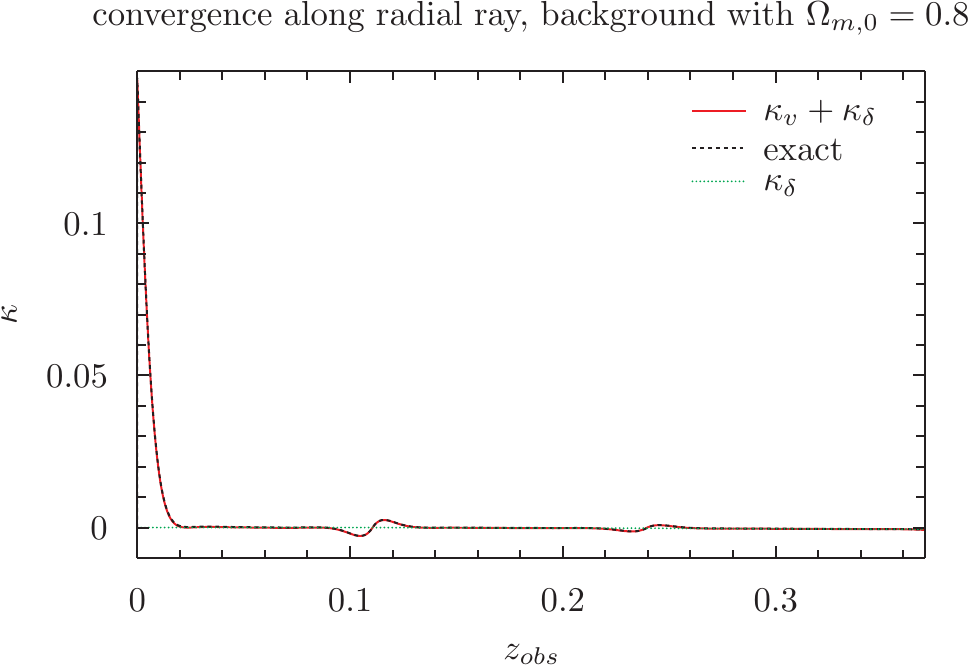}
}
\caption{Density and convergence along a radial geodesic in an LTB swiss cheese model. The density profiles along the exact and ray traced rays are indistinguishable. The same is true for the exact convergence and the sum of the gravitational and Doppler convergence.}
\label{fig:08}
\end{figure*}

The expression for the gravitational convergence for a general FLRW background is well known  (see {\em e.g.} \cite{Schneider1,lensing_important1}):
\begin{equation}\label{eq:kappa_curved}
\kappa_{\delta} =  \frac{4\pi G}{c^2}\int_{0}^{x_g}dx_g'a^2\delta\rho\frac{f_K(x_g-x_g') f_K(x_g')}{f_K(x_g)}
\end{equation}
The relation between $x$ and the radial coordinate is given by $r = f_K(x)$, where $f_K(x)$ is the curvature function given by:
\begin{equation}
f_K(x) = \left\{ \begin{array}{rl}
R_0\sin(x/R_0)  & \text{if} \,\, K = +1 \\
x & \text{if} \,\, K = 0 \\
R_0\sinh(x/R_0) & \text{if} \,\, K = -1
\end{array} \right.
\end{equation}
Equation \eqref{eq:kappa_curved} requires the scale factor to be normalized at the time of observation in the FLRW background coordinates. According to the map in equation (\ref{general_map}), this time coordinate is the same as the time coordinate of the observer in the exact Szekeres spacetime. The Hubble constant of a given background must thus be chosen such that the age of the background model is the same as that of the exact Szekeres model being studied.
\newline\indent
The expression for the Doppler convergence computed on a curved FLRW background can be written as:
\begin{equation}
\begin{split}
\kappa_vc =   \left(1-\frac{c}{a_{,t}r}\sqrt{1-Kr^2/R_0^2} \right) \mathbf{n}\cdot(\mathbf{v}-\mathbf{v}_0) + \mathbf{n}\cdot \mathbf{v}_0 \\
\end{split}
\end{equation}
The radial coordinate $r$ is the coordinate corresponding to a point along a radial null-geodesic in the FLRW background at $z = z_{\text{obs}}$.
\newline\indent
The expression can {\em e.g.} be obtained by following the procedure presented in section 2.2 of \cite{use_doppler2}, modified to a curved background. The main consequences of introducing a curved background are due to changes in the expression for the background angular diameter distance and in particular its dependence on the redshift.
\newline\indent
For the studies presented here, the factor $\sqrt{1-Kr^2/R_0^2}$ is approximately equal to 1 and may thus be neglected both in the expression for the Doppler convergence and in the relation between $r$ and the conformal time $\eta$. Consequently, the expression for the Doppler convergence shown above may be approximated by the expression for the Doppler convergence on a flat FLRW background.
\newline\newline
The ray tracing method on a curved background has been tested using different swiss cheese models. An example is shown in figure \ref{fig:08}. The figure shows the density and convergence along a ray moving radially through consecutive LTB voids. The particular single void model used to construct the swiss cheese model has been specified similarly to the models in section \ref{sec:models} but with $k(r)$ given by:
\begin{equation}\label{eq:k_curved_cheese}
\begin{split}
k(r) = r^2\left( \frac{H_0}{c}\right)^2\left(\Omega_{m,0} -1 \right) \cdot  \\
\left( 1+
4\cos^2\left( \frac{\pi r}{200\text{Mpc}} \right) e^{-8 \frac{r}{200\text{Mpc}}}\right) 
\end{split}
\end{equation}
The results depicted in figure \ref{fig:08} correspond to setting $\Omega_{m,0} = 0.8$. This gives a model that converges asymptotically to an open FLRW model with a present day density parameter of $\Omega_{m,0} = 0.8$. The ray tracing results have been obtained by using this as the background model. As indicated in the figure, the ray tracing results reproduce the exact results very precisely.
\newline\indent
The model defined by equation (\ref{eq:k_curved_cheese}) has an unrealistic, pointy void shape indicating that the model is not well behaved at the origin. Since the results are reproduced well by the ray tracing scheme this is not an issue for the current purpose of the model. The large patches of nearly homogeneous background density along the ray are necessary to avoid velocity artifacts.
\newline\newline
When changing the background in the ray tracing scheme, the background angular diameter distance computed following equation (\ref{eq:Mattig}) and used in equation (\ref{eq:kappa}) must also be changed accordingly.

\subsection{Ray tracing through swiss cheese models with improper backgrounds}
\begin{figure*}[]
\centering
\subfigure[$\Omega_{m,0} = 0.7$]{
\includegraphics[scale = 0.78]{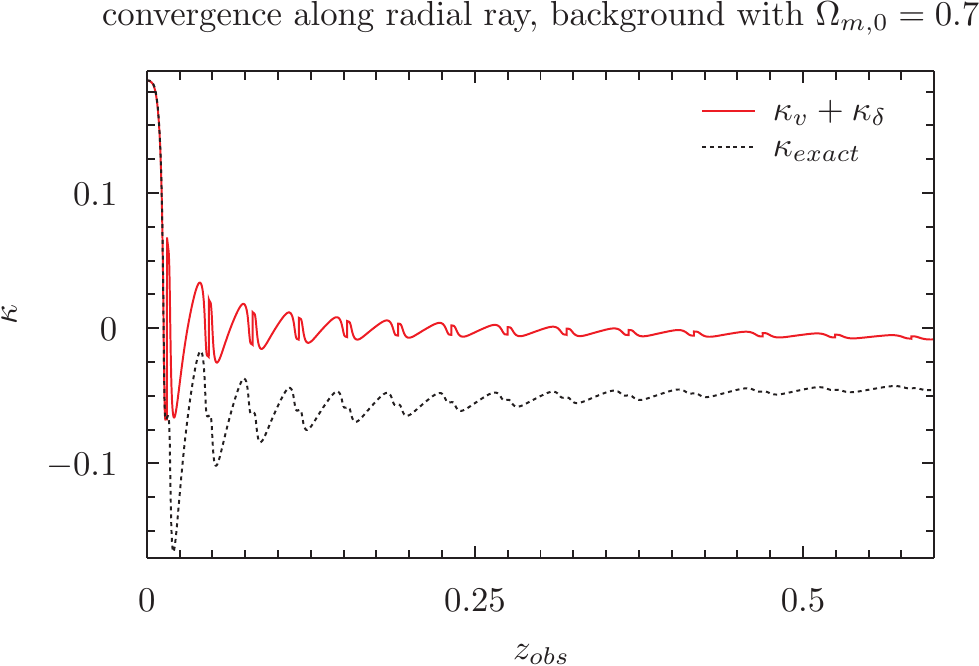}
}
\subfigure[$\Omega_{m,0} = 0.9$]{
\includegraphics[scale = 0.78]{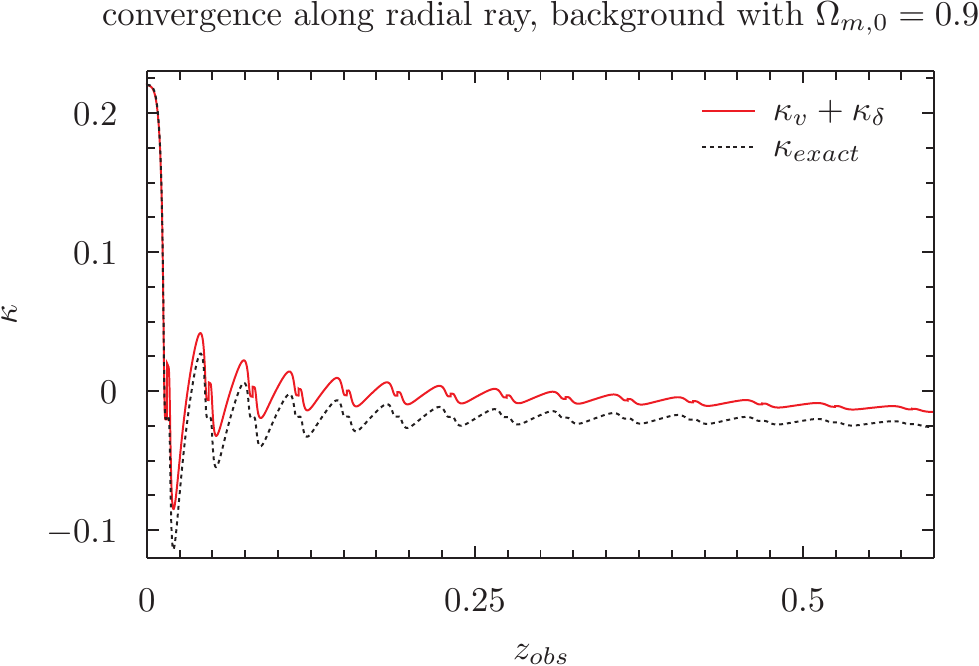}
}\par
\subfigure[$\Omega_{m,0} = 0.95$]{
\includegraphics[scale = 0.78]{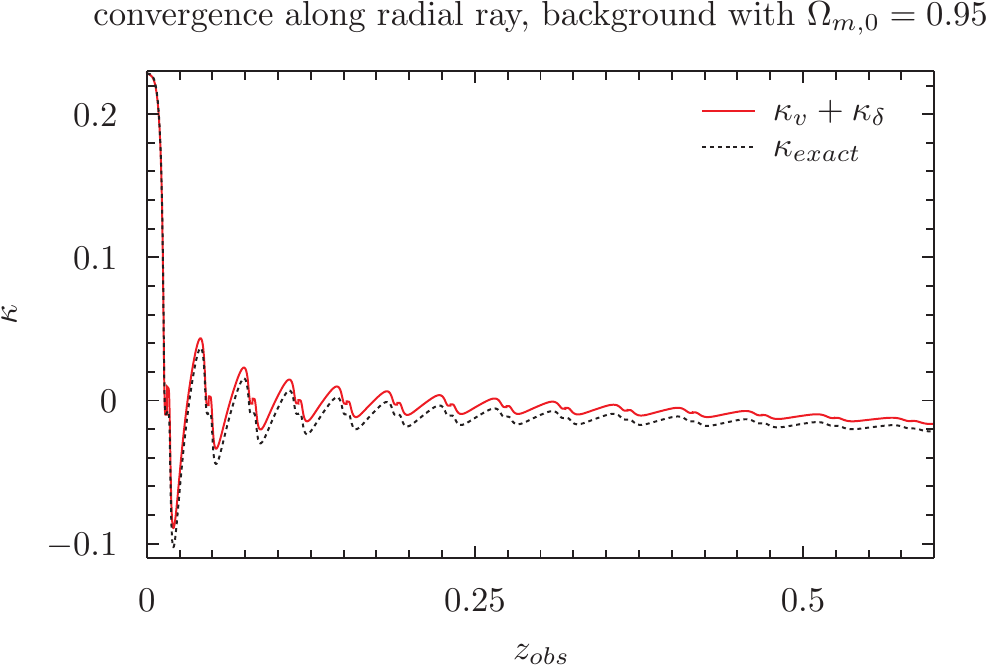}
}
\subfigure[$\Omega_{m,0} = 0.99$]{
\includegraphics[scale = 0.78]{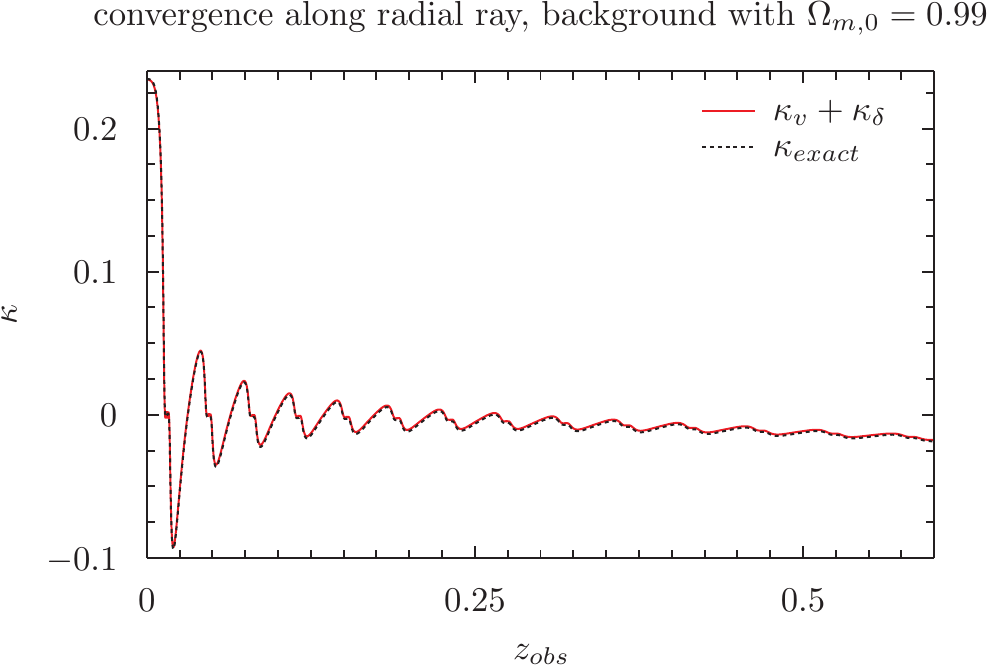}
}\par
\subfigure[$\Omega_{m,0} = 1.01$]{
\includegraphics[scale = 0.78]{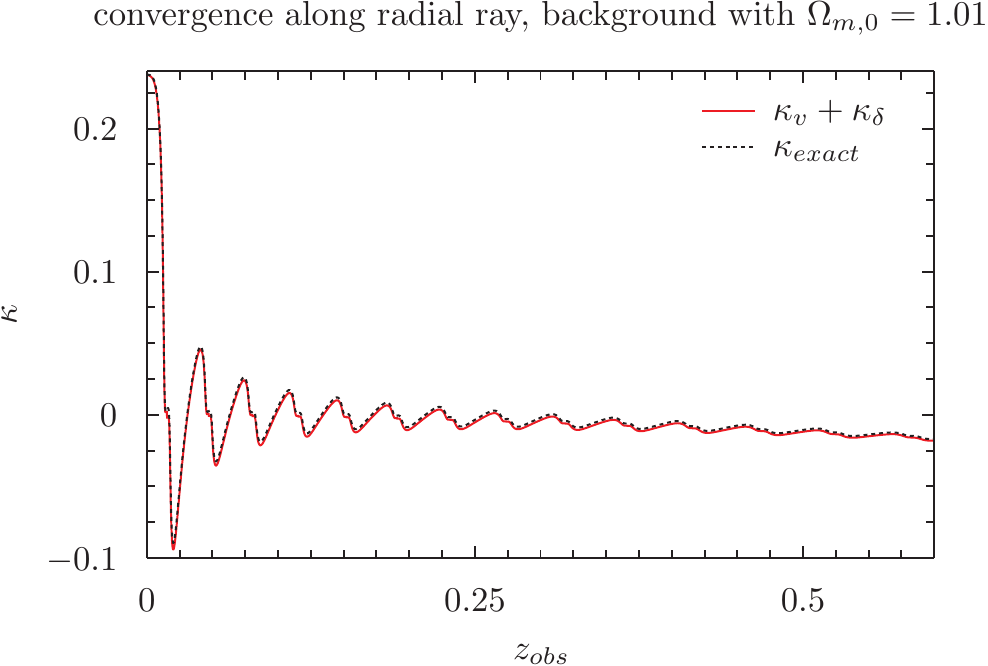}
}
\subfigure[$\Omega_{m,0} = 1.05$]{
\includegraphics[scale = 0.78]{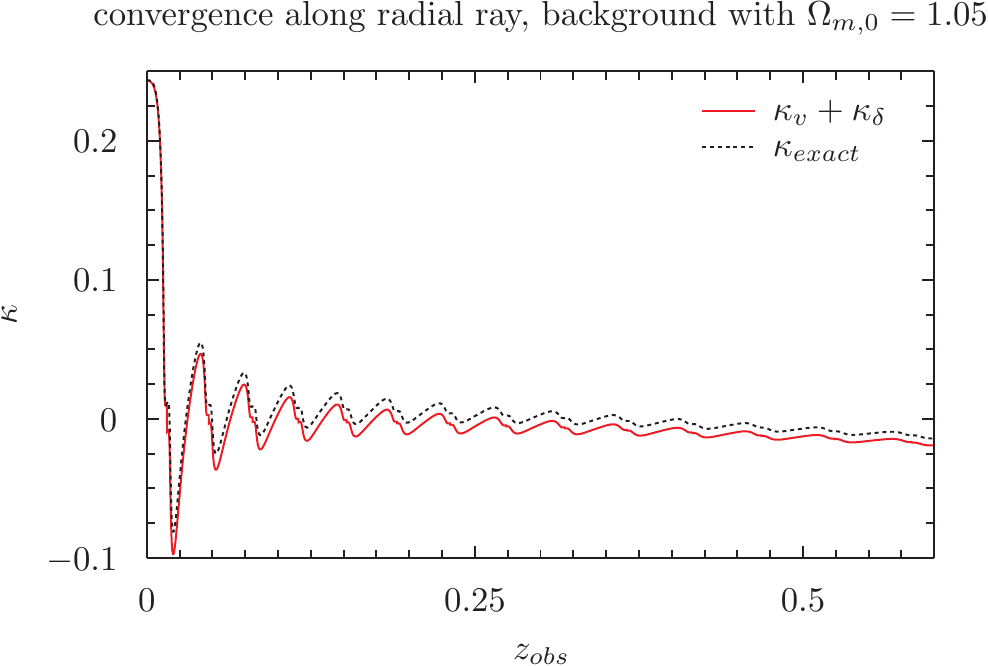}
}\par
\subfigure[$\Omega_{m,0} = 1.1$]{
\includegraphics[scale = 0.78]{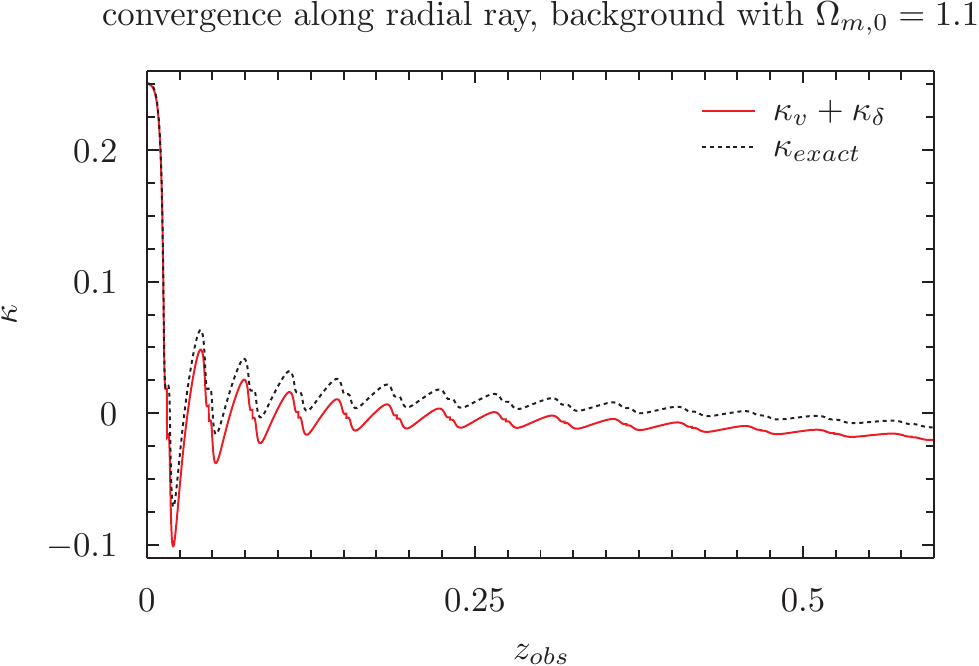}
}
\subfigure[$\Omega_{m,0} = 1.3$]{
\includegraphics[scale = 0.78]{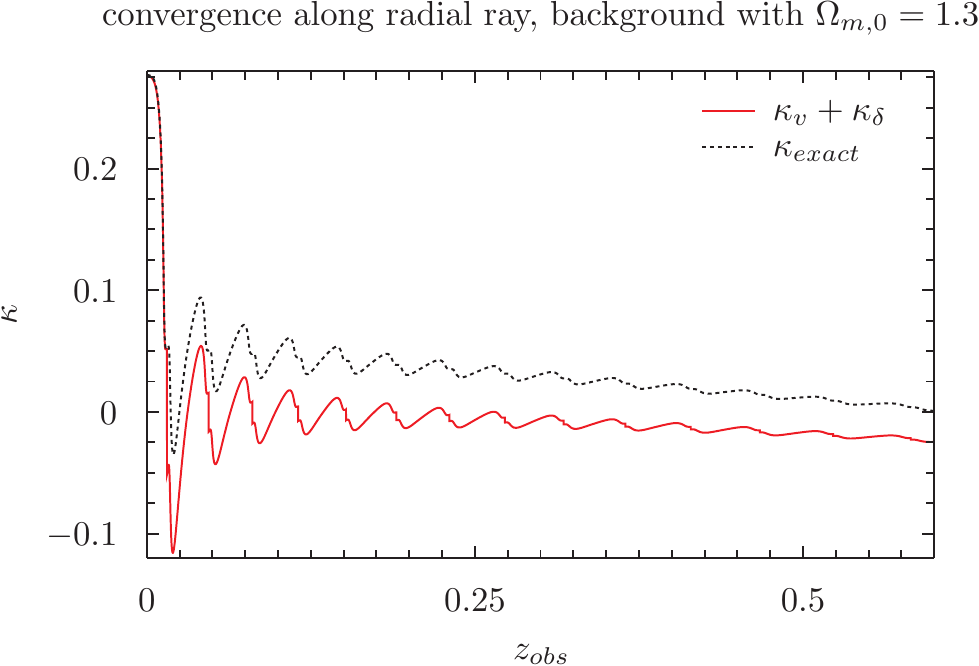}
}
\caption{Convergence along a radial ray in the LTB swiss cheese model based on the LTB single void model specified by $m = 4$ and $r_b = 60\text{Mpc}$. The exact and ray traced convergences are computed using various FLRW backgrounds specified by their values of $\Omega_{m,0}$. The proper background is the EdS model.}
\label{fig:curved_cheese}
\end{figure*}

As shown in section \ref{sec:LTB_results}, the ray tracing scheme very precisely reproduces the exact convergence along rays moving radially through consecutive LTB voids. It is interesting to see to what extent this result requires using the proper background. By "proper background" is here meant the FLRW patches between the compensated single void structures of the given swiss cheese model. For the study in section \ref{sec:LTB_results}, the proper background was the EdS model. Figure \ref{fig:curved_cheese} shows the convergence obtained by using different curved backgrounds for the ray tracing approximation and in equation (\ref{eq:kappa}). The swiss cheese model is the same as the one studied in section \ref{sec:LTB_results}.
\newline\indent
As seen in figure \ref{fig:curved_cheese}, the background {\em is} important in order to obtain precise results with the ray tracing scheme. It is as expected that the ray tracing should yield an incorrect result when using an improper background; the Doppler convergence becomes incorrect when computed from an improper background since the peculiar velocity field is supposed to describe the difference between the local exact expansion of the LTB model and the global expansion of the average model. This is only the case if it is the proper background that is used. Note in this connection that when improper backgrounds are used, the mock N-body velocity field does not reduce to zero outside the single void structures of the cheese.
\newline\indent
Using an improper background for computing the exact convergence from equation (\ref{eq:kappa}) should lead to a shift so that the exact convergence no longer oscillates about zero. This is clearly seen in figure \ref{fig:curved_cheese}. This same does not happen in the ray tracing scheme where the Doppler convergence keeps oscillating about zero. While the gravitational convergence is negligible, the ray tracing convergence will thus oscillate about zero regardless of the background.
\newline\newline
The LTB onion models have also been studied using curved backgrounds. This was done as an attempt to obtain a better reproduction of the exact convergence. This turned out not to be possible. In fact, the difference between exact and ray traced convergence does not appear to depend much on the used background; the exact and approximate convergences are simply raised or lowered by approximately the same amount when the background is changed. This is very different from the swiss cheese case illustrated in figure \ref{fig:curved_cheese} where the difference between the exact and ray tracing convergence clearly depends on the background. The cause for this could be related to the very different Doppler convergence of the onion models compared to that of the LTB swiss cheese models.

\end{document}